{
\documentclass[%
twocolumn,
superscriptaddress,
 amsmath,amssymb,
 aps,
 pre,
]{revtex4}
\usepackage{dcolumn}
\usepackage{bm}
\usepackage{graphicx}
\usepackage{comment}

\begin{document}


\title{Universal Properties of a Run-and-Tumble Particle in Arbitrary Dimension}
\author{Francesco Mori }
\affiliation{LPTMS, CNRS, Univ. Paris-Sud, Universit\'e Paris-Saclay, 91405 Orsay, France}
  \author{Pierre Le Doussal}
  \affiliation{Laboratoire de Physique de l'Ecole Normale Sup\'erieure, PSL University, CNRS, Sorbonne Universit\'es, 24 rue Lhomond, 75231 Paris, France}
\author{Satya N. Majumdar}
\affiliation{LPTMS, CNRS, Univ. Paris-Sud, Universit\'e Paris-Saclay, 91405 Orsay, France}
\author{Gr\'egory Schehr }
\affiliation{LPTMS, CNRS, Univ. Paris-Sud, Universit\'e Paris-Saclay, 91405 Orsay, France}
\date{\today}

\begin{abstract}
We consider an active run-and-tumble particle (RTP) in $d$ dimensions, starting from the origin and evolving  over a time interval $[0,t]$. We examine three different models for the dynamics of the RTP: the standard RTP model with instantaneous tumblings, a variant with instantaneous runs and a general model in which both the tumblings and the runs are non-instantaneous. For each of these models, we use the Sparre Andersen theorem for discrete-time random walks to compute exactly the probability that the $x$ component does not change sign up to time $t$, showing that it does not depend on $d$. As a consequence of this result, we compute exactly other $x$-component properties, namely the distribution of the time of the maximum and the record statistics, showing that they are universal, i.e. they do not depend on $d$. Moreover, we show that these universal results hold also if the speed $v$ of the particle after each tumbling is random, drawn from a generic probability distribution. Our findings are confirmed by numerical simulations. Some of these results have been announced in a recent Letter [Phys. Rev. Lett. {\bf 124}, 090603 (2020)].
\end{abstract}

\maketitle

\newpage

\section{Introduction}

Random walks (RWs) are certainly among the most relevant and studied stochastic processes, with applications in a large number of disciplines, ranging from finance \cite{BP2000,yor01} to climate studies \cite{WK10,RC2011}. Due to their versatility, RWs can be used to study an impressive number of natural and artificial phenomena. One of the simplest examples of RWs are discrete-time walks on a line with independent increments, which can be defined as follows. Let $X_k$ be the position of the random walker at discrete time $k\geq 0$, starting from $X_0=0$ and evolving via
\begin{equation}\label{eq:RW}
X_k=X_{k-1}+\eta_k\,,
\end{equation}
where $\eta_k$ are independent identically distributed (i.i.d.) random variables, drawn from the probability density function (PDF) $f(\eta)$. Even though the increments are uncorrelated, the positions $X_k$'s are strongly correlated. Despite the presence of strong correlations, many observables such as extreme value statistics \cite{EVS_review} can be computed exactly for these random walks even when the noise has a heavy tailed distribution, such as in L\'evy flights. Therefore these random walks models can be used as simple, yet nontrivial, toy examples of strongly correlated systems where new ideas can be tested. 

Moreover, it turns out that several properties of these random walks models, such as
the survival probability and the record statistics, are completely universal, i.e. independent of the jump distribution $f(\eta)$ as long as it is continuous and symmetric. This universality has been traced to the so-called Sparre Andersen theorem \cite{SA_54}. 
For example, one central quantity which has many applications in chemistry \cite{HTB90}, astronomy \cite{Ham61, BF_2005} and finance \cite{BLMV11,Redner_book, SM_review, Persistence_review, fp_book_2014,Masoliver_book}, is the so-called survival probability $q_n$. This is  the probability that the walker, starting initially at the origin, remains on the positive (or negative) side up to step $n$. Sparre Andersen proved, using combinatorial arguments, that for all $n \geq 0$ \cite{SA_54}
\begin{equation}\label{eq:SA}
q_n= {{2n}\choose{n}}2^{-2n}\,,
\end{equation} 
independently of the jump distribution $f(\eta)$ as long as it is symmetric and continuous. Remarkably, Eq. (\ref{eq:SA}) is valid even for heavy-tailed distributions $f(\eta)$, such as the Cauchy distribution $f(\eta)=\pi/(1+\eta^2)$. In particular, for large $n$, $q_n \sim 1/\sqrt{\pi n}$, irrespectively of the jump distribution $f(\eta)$.  Note however that the result in Eq. (\ref{eq:SA}) is universal for any finite $n$ and not just asymptotically for large $n$. Recently, the SA theorem has been generalized also to higher dimensions \cite{Kab1,Kab2}. As a consequence of Eq. (\ref{eq:SA}), many other statistical properties of this class of RWs turn out to be universal. As an example, let $n_1$ be the discrete time at which the RW reaches its global maximum before step $n$. The time of the maximum is one of the key quantities of extreme value statistics and it has been studied for a variety of one-dimensional stochastic processes \cite{EVS_review}. In the case of discrete time RWs with continuous and symmetric jump distribution, one can show that the probability distribution of $n_1$, given the total number $n$ of steps is \cite{Louven_review}
\begin{equation}\label{eq:SA2}
P(n_1|n)=q_{n_1}\,q_{n-n_1}\,,
\end{equation}
where $q_n$ is the survival probability given in Eq. (\ref{eq:SA}). Thus, the distribution of the time $n_1$ is also universal for any $n_1$ and $n$. Another relevant example of the universality of this class of stochastic processes is the record statistics. The statistical properties of records for a stochastic sequence have been extensively studied and have found many applications from hydrology to sports science \cite{Record_review}. However, computing exactly the statistics of records of a correlated sequence is in general challenging, with few known result \cite{Ziff_Satya,mounaix20}. Notably, as a consequence of the SA theorem, the record statistics of a discrete-time RW is also completely universal, if $f(\eta)$ is continuous and symmetric \cite{Ziff_Satya}.

Motivated by recently studied models of non-interacting active self-propelled particles in $d$ dimensions, we showed in a recent Letter \cite{mori20} that some of these universal properties for discrete-time random walks can be transported to study some properties of the
$d$-dimensional run-and-tumble particles (RTP). This involved a nontrivial mapping between the RTP which takes place in continuous time and the discrete-time random walk discussed above \cite{mori20}. For instance, using this mapping, we showed that the survival probability and the record statistics of the $x$-component of the $d$-dimensional RTP of duration $t$ are completely independent of the dimension $d$, as well as of the speed distribution after each tumbling (to be defined more precisely later). The purpose of this long paper is to elucidate this mapping in more detail and show that it can be used further to compute other universal observables, such as the distribution of the time at which the $x$-component reaches its maximum. Moreover, we also introduce two other generalisations of the simple RTP model and show, using again the mapping to discrete-time random walk, that many observables such as the survival probability, the distribution of the time at which the maximum occurs, as well as the record statistics of the $x$-component, become universal and we actually compute them exactly. We also perform extensive numerical simulations to verify our analytical predictions.

Let us recall that the study of RTP has seen a surge of interest in recent times in the context of active matter. This class of stochastic processes describes the motion of self-propelled particles, which are able to absorb energy from the surrounding environment and to convert this energy in directed motion. This is in contrast with the classical passive processes, e.g. Brownian motion, in which the motion of the particle is only driven by the thermal fluctuations of the environment. These active particles emerge in the description of many natural phenomena at different scales. Examples include bacteria \cite{Berg_book, Cates_bacteria}, vibrated granular materials \cite{WW_2017}, active gels \cite{R_2010,NVG2019}, and the motion of larger animals \cite{R_2010,V_1995,HB_2004, VZ12}. One of the most paradigmatic and most studied models of active matter is indeed the RTP \cite{Berg_book,TC_2008}. This model, which was previously known as ``persistent random walk'' \cite{Weiss_2002,ML_2017}, has been introduced in the context of active matter to describe the motion of a class of bacteria, including \emph{E. Coli} \cite{Berg_book}. The motion of a single RTP in $d$ dimensions can be described as follows. The RTP alternates phases of straight ballistic motion with constant velocity $v_0$ (``runs''), during a ``time of flight'' $\tau$, with abrupt events in which the particle  ``tumbles'', changing its direction of motion uniformly at random. In the simplest version of the model, the duration $\tau$ of a running phase is an exponential random variable with rate $\gamma$ and the velocity $v_0$ of the particle is fixed.

Many studies have shown that the RTP model displays a rich and peculiar behavior. Some of these interesting features, for instance clustering at boundaries \cite{Bechinger16}, motility-induced phase 
separation \cite{CT_2015}, jamming \cite{SEB2016}, emerge from the interactions of many RTPs. However, relevant properties, such as non-Boltzmann stationary state in a confining potential \cite{MBE2019,Sevilla_19,Dhar_19,basu2020,DCR20}, can be observed even at the single-particle level. Moreover, many interesting quantities have been computed exactly in the one-dimensional case \cite{artuso14,DM_2018,Malakar_2018, LDM_2019,Dhar_19,singh19,LDM20,banerjee20}. Examples include the persistence properties \cite{ADP_2014,artuso14,Dhar_19,  Malakar_2018, LDM_2019, DM_2018} as well as the distribution of the time of the maximum \cite{singh19}. Variants of the RTP model in which the velocity of each flight is random \cite{GM_2019}, in which the tumbling rate $\gamma$ is space-dependent \cite{Singh2020,LDM20}, or in which the particle undergoes stochastic resetting to its starting position \cite{EM_2018,M2019} have also been investigated.
When $d=1$, the analytical description of the system is greatly simplified since the particle is either going to the left or to the right. On the other hand, already at $d=2$ the direction of the particle is continuously varying and performing exact computations becomes more difficult \cite{Santra2020}. Nevertheless, approximate methods have been used to compute the mean passage time in confined geometries in $d=2$ and $d=3$ \cite{RBV16}. However, to the best of our knowledge, our recent Letter \cite{mori20} provided the first exact results (at all time $t$) for the first-passage properties as well as for the record statistics for an RTP in $d$ dimensions, for any $d \geq 1$. In this paper, we extend these exact results to other observables as well as to other generalised models of self-propelled particles in $d$ dimensions.

The rest of the paper is organized as follows. In Section \ref{sec:model_results} we define the different RTP models under consideration and we provide a summary of our main results. In Section \ref{sec:IT} we focus on the RTP model with instantaneous tumblings. In particular, in Section \ref{sec:surv} we compute exactly the probability that the $x$-component of the particle does not become negative up to time $t$. In Section \ref{sec:tmax} we compute the distribution of the time of the maximum $t_{\max}$, while the record statistics is studied in Section \ref{sec:record}. A variant of the RTP model in which the particle takes instantaneous jumps is considered in Section \ref{sec:IR}, where we compute exactly the survival probability, the distribution of $t_{\max}$, and the record statistics for this model. In Section \ref{sec:mixed_model}, we introduce non-instantaneous tumblings in the model and we compute exactly the survival probability, the distribution of $t_{\max}$, and the record statistics also for this model. Finally, in Section \ref{sec:conclusion}, we conclude with a summary and few open questions. Some details of the computations are relegated to the appendices.

\section{Model and summary of the main results}\label{sec:model_results}

\begin{figure*}[t]
\includegraphics[width=1\textwidth]{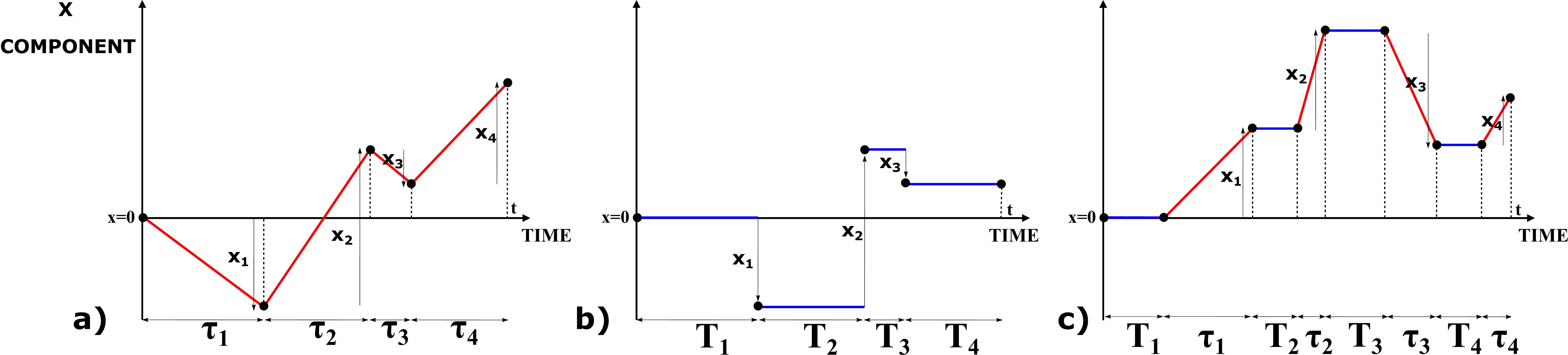} 
\caption{\label{fig:model} {\bf a)} Typical trajectory of the $x$ component of the RTP, moving according to the IT model. The total number of tumblings is $n$ (here $n=4$) and the displacements in the $x$ direction are $x_1,\ldots x_n$.
{\bf b)} Trajectory of the $x$ component in the case of the IR model. The durations of the waiting phases are denoted by $T_i$'s and the jumps by $x_i$'s, with $i=1,2, \cdots, n-1$ (here $n=4$). Note that no displacement is associated to the last time interval $T_n$. {\bf c)} Typical trajectory of the $x$ component in the mixed model. The full blue lines represent the waiting phases, while the red lines represent the running phases. The duration of the $i^{\rm th}$ waiting (running) phase is denoted by $T_i$ ($\tau_i$) and $x_i$ denotes the $x$-component displacement of the interval $\tau_i$. At the final time $t$, the particle can be in a running phase (as in figure) or in a waiting phase. The number of waiting phases is denoted by $n$, while $m$ is the number of running phases (in this case $m=n=4$).
}
\end{figure*}

In this section we first present the main RTP model considered in the paper and its generalizations. Since the paper is quite long, we also provide a summary of our main results.

\subsection{Description of the model}\label{sec:model}

We consider three different models to describe the motion of an RTP: the instantaneous-tumbling (IT) model, the instantaneous-run (IR) model, and the mixed model. The details of these models are presented below.

\vspace*{0.3cm}
\noindent{\bf Instantaneous-tumbling (IT) model:}
We first introduce the most common RTP model, which we will refer to as IT model. We consider a single run-and-tumble particle in $d$ dimensions, starting from the origin O and evolving up to time $t$. The motion is defined in continuous time as follows. The particle initially chooses a direction uniformly at random and  moves ballistically in that direction during a random time interval $\tau_1$, drawn from the running-time distribution $P_R(\tau)$, which is assumed to be exponential with rate $\gamma$, i.e. $P_R(\tau)=\gamma\, e^{-\gamma \tau}$. Calling $v_0$ the fixed velocity of the particle, the distance $l_1=v_0 \tau_1$ travelled during this flight is thus also a random variable. After that, the particle tumbles instantaneously, i.e. it chooses randomly a new direction. After the first tumbling, it moves ballistically in the new direction with the same velocity $v_0$ for an exponentially distributed time $\tau_2$
drawn independently from the same distribution $P_R(\tau)= \gamma\, e^{-\gamma \tau}$, and so on. More precisely, in a small time interval $dt$:

\begin{itemize}

\item With probability $\gamma\, dt$, the particle changes its direction of motion randomly. 

\item With the complementary probability $(1-\gamma dt)$, the particle retains its direction and moves forward in that direction by a distance $v_0\, dt$.

\end{itemize}
Note that the number $n$ of tumblings is also random. We consider the starting point $O$ as a tumbling. Thus, we always have $n\ge 1$. Throughout the paper, we focus on the statistical properties of the continuous-time process obtained by projecting the position of the particle in the $x$ direction. An example of such an $x$-component process is shown in Fig. \ref{fig:model}a. Note that, even if for the sake of simplicity we consider the $x$ component of the particle, the results of this paper are valid for the process obtained projecting the motion of the RTP in any fixed direction. This is a consequence of the isotropy of the RTP process. Moreover, we also consider a variant of this IT model in which the velocity of the particle, and not only the direction, is renewed afresh after each tumbling. To be precise, we study the case in which the velocity of each running-phase is drawn from a PDF $W(v)$, with $v\geq 0$. Note that the simple case with fixed velocity corresponds to choosing $W(v)=\delta(v-v_0)$.

\vspace*{0.3cm}
\noindent{\bf Instantaneous-run (IR) model:} The first variation of the IT model that we consider is a model in which, instead of moving with constant velocity during a running phase, the particle waits without moving for a random time and then it jumps instantaneously to its new position. We will refer to this model as the IR model. Specifically, we assume that the particle starts from the origin of the $d$-dimensional space and it first waits a random time $T_1$, drawn from the waiting-time distribution $P_W(T)$. After that, the particle picks a direction uniformly at random and takes an instantaneous jump of length $l_1=v_1 T_1$, where $v_1$ is drawn from the speed distribution $W(v)$. Note that here $v_1$ can not be interpreted as the velocity of the particle, since the jumps are instantaneous. After the first jump, the RTP stays still for a time $T_2$, drawn independently from $P_W(T)$ and it jumps again in a random direction to a distance $l_2=v_2 T_2$, where $v_2$ is drawn from $W(v)$. This process repeats up to time $t$. Even if we will mainly focus on the case of exponentially distributed waiting times with rate $a$, it turns out that, in the case of the IR model, it is possible to compute exactly many statistical properties for any probability distribution $P_W(T)$. Notably, this will allow us to infer the late time behavior of the IT model with a generic running-time distribution $P_R(\tau)$. A schematic representation of the evolution of the $x$ component of a particle moving according to this IR model is shown in Fig. \ref{fig:model}b.

\vspace*{0.3cm}
\noindent{\bf Mixed model:} One of the key assumptions behind the IT model is that the tumbling times are negligible with respect to the running times. However, in many cases this assumption is not realistic and it is worth to investigate the effect of finite-time tumblings. For this reason, we also consider a model in which the tumblings are not instantaneous.  In this model, which we will call mixed model, the particle alternates running phases, in which it moves ballistically in a random direction with random velocity $v$, drawn from $W(v)$, with waiting (or tumbling) phases, in which the particle does not move. These waiting phases are introduced to model the time required for the particle to change its direction when tumbling. To be precise, we assume that the RTP starts from the origin and initially waits up to time $T_1$, drawn from the distribution $P_W(T)$. We will first consider a generic waiting-time distribution $P_W(T)$ and then we will focus on the special case in which this distribution is exponential with fixed rate $a$, i.e. $P_W(T)=a e^{-aT}$. At the end of the first waiting phase, the particle chooses a random direction and moves in that direction with constant velocity $v_1$, drawn from $W(v)$, for a random time $\tau_1$, drawn from the exponential distribution $P_R(\tau)=\gamma e^{-\gamma \tau}$. Then, the particle tumbles, i.e. it waits in its position for a random time $T_2$, independently drawn from $P_W(T)$. After this time interval $T_2$, the particle starts moving again in a random direction for a time interval $\tau_2$, distributed according to $P_R(\tau)$, and with a random velocity $v_2$. The particle alternates waiting and running phases up to time $t$. A typical $x$-component trajectory of the mixed model is shown in Fig. \ref{fig:model}c.

Note that, focusing on the case $P_W(T)=a \,e^{-aT}$, from the mixed model one can obtain the IT model by taking the limit $a\to \infty$, while keeping $\gamma$ fixed. Indeed, since the typical duration of a tumbling phase is $1/a$, taking this limit the tumblings become instantaneous. On the other hand, if one takes the limit $\gamma\to \infty$ with $a$ fixed, one does not recover the IR model. It is true that, when $\gamma\to \infty$, the running-phases become instantaneous, however in this limit, at variance with the IR model, the distance travelled during each running phase goes to zero. Indeed, in the mixed model, the distance travelled during each running phase depends on the duration of the phase itself, while in the IR model the length of each instantaneous jump depends on the duration of the previous waiting phase.

\subsection{Summary of the main results}\label{sec:results}

Throughout the paper we will mainly focus on few statistical properties of the $x$ component of a single RTP, for each of the three models presented above. These quantities, that we define precisely below, are the survival probability $S(t)$, the probability distribution of the time $t_{\max}$ of the maximum and the record statistics. We will use the notation $S^{\rm IT}(t)$, $S^{\rm IR}(t)$, and $S^{\rm Mixed}(t)$ to denote the survival probability computed for the IT, IR and mixed models, respectively. For the other observables, i.e. for the distribution of $t_{\max}$ and for the statistics of records, we will use, for simplicity, the same notation for the three models. In the cases of the IR model and the mixed model, here we present the results obtained when the waiting-time distribution is exponential with waiting rate $a$, i.e.  $P_W(T)=ae^{-aT}$. However, many of our results extend to any arbitrary $P_W(T)$. Remarkably, all the results presented are universal, i.e. they do not depend on the dimension $d$ nor on the speed distribution $W(v)$.

\vspace*{0.3cm}
\noindent{\bf Survival probability:}
We consider a single RTP starting from the origin and evolving in $d$ dimensions over a time interval $[0,t]$. Let $S(t)$ be the probability that the $x$ component of the particle does not become negative up to time $t$. Using the SA theorem, we compute $S(t)$ exactly at all $t$ for the three RTP models presented above and we show that, for each model, it is independent of the dimension $d$ and of the speed distribution $W(v)$.

In particular, for the IT model with running-time distribution $P_R(\tau)=\gamma\,e^{-\gamma \tau}$, we find that the survival probability is given by
\begin{equation}\label{eq:S_IT_intro}
S^{\rm IT}(t)= \frac{1}{2}\, e^{-\gamma t/2}\, \left(I_0\left(\frac{\gamma}{2}t\right)+ I_1\left(\frac{\gamma}{2}t\right)\right)\,,
\end{equation}
where $I_0(t)$ and $I_1(t)$ are modified Bessel functions. This result in Eq. (\ref{eq:S_IT_intro}) is shown in Fig. \ref{fig:s} and was first derived in the one-dimensional case via Fokker-Plank approaches \cite{Weiss_2002,Malakar_2018,LDM_2019}. In our previous Letter \cite{mori20}, we showed that it is valid in any dimension $d$ and for any speed distribution $W(v)$. When $t\to 0$, $S^{\rm IT}(t)$ goes to the limit value $1/2$, while for large $t$ the survival probability decays as $S^{\rm IT}(t)\sim 1/\sqrt{\pi \gamma t}$.

Moreover, in the case of the IT model with a generic time distribution $P_R(\tau)$, we show that for late times
\begin{equation}
S^{\rm IT}(t)\sim t^{-\theta}\,,
\end{equation}
where $\theta=1/2$ if the first moment $\langle \tau \rangle=\int_{0}^{\infty}d\tau\,\tau P_R(\tau)$ is finite, while $\theta=\mu/2$ when $P_R(\tau)\sim 1/\tau^{\mu+1}$ for large $\tau$, with $0<\mu<1$ .

For the IR model, in the special case of the exponential waiting-time distribution $P_W(T)=a\,e^{-a T}$, we obtain that the survival probability is
\begin{equation}\label{eq:S_IR_intro}
S^{\rm IR}(t)= e^{-a\, t/2}\, I_0\left(\frac{a}{2}t\right)\, .
\end{equation}
When $t\to 0$ the survival probability $S^{\rm IR}(t)$ goes to the limit value $1$ (see Fig. \ref{fig:S_wait}), while for large $t$ it decays as $S^{\rm IR}(t)\sim 1/\sqrt{\pi a t}$.

Finally, for the mixed model, i.e. for the RTP with non-instantaneous tumblings, assuming that the running times and the waiting times are both exponentially distributed with rates $\gamma$ and $a$, respectively, we obtain
\begin{eqnarray}\label{eq:S_mixed_intro}
S^{\rm Mixed}(t)&=&\frac{\gamma}{4} e^{-\gamma t/2}\int_{0}^{t}dt'\,e^{-a t'}\left(I_0\left(\frac{\gamma}{2}t'\right)+I_1\left(\frac{\gamma}{2}t'\right)\right)\nonumber \\
&\times & \left(I_0\left(\frac{\gamma}{2} (t-t')\right)+I_1\left(\frac{\gamma}{2}( t-t')\right)\right) \\
&+&\frac12 \left(1+e^{-at}\right) e^{-\gamma t/2}\left(I_0\left(\frac{\gamma}{2} t\right)+I_1\left(\frac{\gamma}{2} t\right)\right)\,. \nonumber
\end{eqnarray}
When $t\to 0$, $S^{\rm Mixed}(t)$ goes to the limit value $1$ and for late times it decays as $S^{\rm Mixed }(t)\sim \sqrt{(1/a+1/\gamma)/(\pi t)}$. The survival probability $S^{\rm Mixed}$ is shown in Fig. \ref{fig:S_mixed_model} for $\gamma=1$ and different values of $a$.

\vspace*{0.3cm}
\noindent{\bf Time to reach the maximum:}
We consider again a single RTP starting from the origin and moving up to time $t$ in $d$ dimensions with speed distribution $W(v)$. Let $t_{\max}$ be the time at which the $x$ component reaches its maximal value for the first time. We compute exactly the PDF $P(t_{\max}|t)$ of $t_{\max}$ at fixed total time $t$ for the three RTP models, showing that for each model this PDF is independent of $d$ and $W(v)$. 

In the case of the IT model with tumbling rate $\gamma$, we find that for any $t_{\max}$ and $t$
\begin{eqnarray}\label{eq:P_tmax_IT_intro}
P(t_{\max}|t)&=&\gamma S^{\rm IT}(t_{\max})S^{\rm IT}(t-t_{\max})\\
&+& S^{\rm IT}(t)\left(\delta(t_{\max})+\delta(t-t_{\max})\right)\,,\nonumber
\end{eqnarray}
where $S^{\rm IT}(t)$ is given in Eq. (\ref{eq:S_IT_intro}). This result in Eq. (\ref{eq:P_tmax_IT_intro}) was derived in the one-dimensional case by solving the Fokker-Plank equation \cite{singh19}. Here we show that it is valid for any dimension $d$ and for any speed distribution $W(v)$. The cumulative probability $P(t_{\max}<t'|t)$, obtained by numerical integration of the exact PDF in Eq. (\ref{eq:P_tmax_IT_intro}), is plotted, as a function of $t'$, in Fig. \ref{fig:cumulative_tmax}.

For a single RTP evolving according to the IR model with waiting-time distribution $P_W(T)=a\, e^{-aT}$, we find that for any $t_{\max}$ and $t$
\begin{eqnarray}\label{eq:P_tmax_IR_intro}
&&P(t_{\max}|t)=\delta(t_{\max}) e^{-a\, t/2}\, I_0\left(\frac{a}{2}t\right)
\\&+&
\frac{a}{2}e^{-a t_{\max}/2}\left(I_0\left(\frac{a}{2}t_{\max}\right)+I_1\left(\frac{a}{2}t_{\max}\right)\right)\nonumber \\
&\times &  e^{-a\, (t-t_{\max})/2}\, I_0\left(\frac{a}{2}(t-t_{\max})\right)\,.
\end{eqnarray} 
By integrating this PDF in Eq. (\ref{eq:P_tmax_IR_intro}) numerically, we also obtain the cumulative probability $P(t_{\max}<t'|t)$, which is shown in Fig. \ref{fig:t_max_wait}.

In the case of the mixed model with waiting rate $a$ and running rate $\gamma$, we find
\begin{eqnarray}\label{eq:P_tmax_Mixed_intro}
P(t_{\max}|t)&=&P_{\rm I}(t_{\max})P_{\rm II}(t-t_{\max})\\
&+&\delta(t_{\max})P_{\rm II}(t)+\delta(t-t_{\max})\frac{1}{\gamma}P_{\rm I}(t)\,,\nonumber
\end{eqnarray}
the expressions for $P_{\rm I}(T)$ and $P_{\rm II}(t)$ are rather long and are given in Eqs. (\ref{eq:P_I_final}) and (\ref{eq:P_II_final}). The cumulative probability $P(t_{\max}<t'|t)$, obtained from the exact PDF in Eq. (\ref{eq:P_tmax_Mixed_intro}), is shown in Fig. \ref{fig:p_tmax_mixed_model}.

\vspace*{0.3cm}
\noindent{\bf Record statistics:}
For each of the RTP processes that we consider, we also show that the record statistics of the $x$ component is completely universal, i.e. it is independent of $d$ and $W(v)$. We will focus on lower records, but the results that we obtain are also valid for the upper records, due to the $x\to-x$ symmetry of the process. Let us first define a lower record. We consider the trajectory of an RTP moving in $d$ dimensions, with speed distribution $W(v)$. Let $m$ be the number of running phases and $x_1,x_2,\ldots x_m$ be the displacements in the $x$ component of the particle during each flight. We also define
\begin{equation}
X_k=x_1+x_2+\ldots +x_k\,,
\end{equation}
i.e. the $x$ component of the particle at the end of each flight. Then, we say that $X_k$ is a lower record if and only if $X_k<X_i$ for all $i<k$. We assume that the starting point $X_0=0$ is also a record. Then, the main quantities that we are interested in are the probability $S_N(t)$ that there are exactly $N$ records up to time $t$ and the average number of records $\langle N(t)\rangle$ at time $t$. It is clear that, since the starting point is counted as a record, $S_1(t)$ is the probability that the $x$ component of the particle has not become negative up to time $t$. Thus, one simply finds $S_1(t)=S(t)$.

Let us first consider the IT model with exponential running-phases with rate $\gamma$. Apart from the trivial case $S_1(t)$, we compute exactly $S_N(t)$ for $N=2$ and $N=3$ (see Fig. \ref{Fig_s2_s3})
\begin{equation}\label{eq:S2_IT_intro}
S_2(t) = S^{\rm IT}(t) = \frac{1}{2}e^{-\gamma t /2} \left( I_0 \left(\frac{\gamma}{2}t\right)+ I_1 \left(\frac{\gamma}{2}t\right) \right) \,, 
\end{equation}
\begin{equation}\label{eq:S3_IT_intro}
S_3(t) = e^{-{\gamma  t}/{2}} I_1\left(\frac{\gamma}{2}t\right) \,,
\end{equation}
where $S^{\rm IT}(t)$ is given in Eq. (\ref{eq:S_IT_intro}). The fact that $S_2(t)=S(t)$ a priori is unexpected. As we will see, this is not the case for the other two models. Moreover, we find that the average number of records is given by
\begin{eqnarray}\label{eq:N_IT_intro}
&&\langle N(t) \rangle\\  &=&  \frac{1}{2} e^{-\gamma t/2} \left((2 \gamma t+3) I_0\left(\frac{\gamma t}{2}\right)+(2 \gamma t+1)
   I_1\left(\frac{\gamma t}{2}\right)\right)  \;.\nonumber
\end{eqnarray}
The average number of records $\langle N(t)\rangle$ is plotted, as a function of $t$, in Fig. \ref{fig_av_record}.
The results in Eqs. (\ref{eq:S2_IT_intro}-\ref{eq:N_IT_intro}) were first derived in our previous Letter \cite{mori20}.

In the case of the IR model with $P_W(T)=a	 e^{-a T}$ we find
\begin{equation}
S_2(t)=e^{-a t/2}I_0\left(\frac{a}{2}t\right)-e^{-at}\,,
\end{equation}
\begin{eqnarray}
&&S_3(t)=e^{-a t/2}I_0\left(\frac{a}{2}t\right)-e^{-a t}\\
&+&\frac{a}{2}e^{-a t}\int_{0}^{t}dt'\,\left(I_1\left(\frac{a}{2}t'\right)-I_0\left(\frac{a}{2}t'\right)\right)\,.\nonumber
\end{eqnarray}
These probabilities $S_2(t)$ and $S_3(t)$ are shown in Fig. \ref{fig:s2_s3_wait}.
In this case, the average number of records is given by (see Fig. \ref{fig:avg_N_wait})
\begin{equation}
\langle N(t) \rangle=e^{-a t/2}\left(\left(1+a t\right)I_0\left(\frac{a}{2}t\right)+a t \,I_1\left(\frac{a}{2}t\right)\right)\,.
\end{equation}

Finally, we consider the mixed model with waiting rate $a$ and running rate $\gamma$ and we show that
\begin{eqnarray}
&& S_2(t)=S^{\rm Mixed}(t)-e^{-at}\,,\\
&&S_3(t)=2 S^{\rm Mixed}(t)-2e^{-at}-a\,e^{-\gamma t/2}\int_{0}^{t}dt'\,e^{-a(t-t')}\nonumber \\
&\times &I_0\left(\frac{\gamma}{2}(t-t')\right)\left(\left(1+\gamma t'\right)I_0\left(\frac{\gamma}{2}t'\right)+\gamma t' I_1\left(\frac{\gamma}{2}t'\right)\right)\,,\nonumber
\end{eqnarray}
where $S^{\rm Mixed}(t)$ is given in Eq. (\ref{eq:S_mixed_intro}). The probabilities $S_2(t)$ and $S_3(t)$ are plotted, as functions of $t$, in Fig. \ref{fig:s2_s3_mixed}.
The average number of records for the RTP with non-instantaneous jumps is given by (see Fig. \ref{fig_av_record_mixed})
\begin{eqnarray}
&&\langle N(t)\rangle=S^{\rm Mixed}(t)\\
&+&a e^{-\gamma t/2}\int_{0}^{t} dt'\,e^{-a(t-t')}I_0\left(\frac{\gamma}{2}(t-t')\right) \nonumber\\
&\times &\left((1+\gamma t')I_0\left(\frac{\gamma}{2}t'\right)+\gamma t'I_1\left(\frac{\gamma}{2}t'\right)\right)\,,\nonumber
\end{eqnarray}
where $S^{\rm Mixed}(t)$ is given in Eq. (\ref{eq:S_mixed_intro}).

\section{Instantaneous-tumbling model}\label{sec:IT}

In this section we consider one of the most common and studied models of RTPs: the instantaneous-tumbling model. This model is based on the assumption that the time in which the particle changes its direction is typically negligible with respect to the time of a flight, so that the tumblings can be assumed to be instantaneous. Below, we compute exactly the survival probability, the distribution of the time of the maximum and the statistics of records for this IT model, assuming that the tumblings happen with constant rate $\gamma$ and that the velocity of each flight is drawn from the probability distribution $W(v)$. The special case in which the velocity of the particle is fixed can be obtained by choosing $W(v)=\delta(v-v_0)$.

\subsection{Survival probability}\label{sec:surv}

We consider a single RTP in a $d-$dimensional space, starting at the origin $O$ and evolving according to the IT model for a total fixed time $t$. In this section we want to compute the probability $S^{\rm IT}(t)$ that the $x$ component of the particle does not change sign up to time $t$.

We denote by $\tau_i$ the time of the flight after the $i^{\rm th}$ tumbling, see Fig. \ref{fig:model}. As explained in the Section \ref{sec:model}, these times are i.i.d. drawn from the exponential distribution $P_R(\tau)=\gamma e^{-\gamma t}$. However, since we are fixing the total time $t$, the duration $\tau_n$ of the last time interval is not completed. Consequently, its distribution
is given by the probability $\int_{\tau_n}^{\infty}d\tau\,P_R(\tau)=e^{-\gamma\, \tau_n}$ that no tumbling happens during the interval $\tau_n$. Hence, the joint distribution of the time intervals $\{\tau_i\}=\{\tau_1,\tau_2,\ldots, \tau_n\}$ {\em and}
the number of tumblings $n$, for a fixed total duration $t$, is given by
\begin{equation}
P\left(\{\tau_i\},\, n|t\right)= 
\left[\prod_{i=1}^{n-1} \gamma\, e^{-\gamma\, \tau_i}\right]\, 
e^{-\gamma\, \tau_n}\, \delta\left( \sum_{i=1}^n \tau_i-t\right)\,,
\label{joint_tau_n.1}
\end{equation}
where the delta function enforces the constraint on the total time.

Let us now define $\{\vec l_i\}=\{\vec {l}_1,\vec{l}_2,\ldots \vec{l}_n\}$ as the $d$-dimensional displacement vectors of the particle up to time $t$. The direction of each of these random vectors is chosen uniformly at random at each tumbling and their norms $\{l_i\}=\{l_1,l_2,\ldots, l_n\}$, which are the straight distances travelled by the particle, are simply given by $l_i= v_i\, \tau_i$ for all $i$, where the random variables $v_i$ are drawn from the PDF $W(v)$. Thus, using Eq. (\ref{joint_tau_n.1}), we can compute the joint distribution of $\{l_i\}$, and the number of tumblings $n$ as
\begin{eqnarray}\label{joint_l_n.1}
&& P\left(\{l_i\},\, n|t\right)\\
&=& \frac{1}{\gamma}\,
\prod_{i=1}^{n}\int_{0}^{\infty}dv_i\,W(v_i)  \frac{\gamma}{v_i}\, e^{-\gamma\, l_i/v_i}\,
\delta\left(  \sum_{i=1}^n \frac{l_i}{v_i}-t\right)\,.\nonumber
\end{eqnarray}
Since we are interested in the survival probability $S^{\rm IT}(t)$ of the $x$ component of the process, we would like to obtain the joint distribution of $x$ components $\{x_i\}=\{x_1,x_2,\ldots x_n\}$ of these vectors $\{\vec{l_i}\}$. Thus, let us consider a random vector $\vec{l}$ in $d$ dimensions, with fixed norm $l$ and with uniformly distributed direction. Let $x$ be the $x$ component of this vector. Then, it is possible to show (see Appendix \ref{sec:f_d}) that the distribution of $x$, given the fixed norm $l$, is given by
\begin{equation}
P(x|l)=\frac{1}{l}f_d\left(\frac{x}{l}\right)\,,
\end{equation}
where
\begin{equation}
f_d(z)= \frac{\Gamma(d/2)}{\sqrt{\pi}\, \Gamma((d-1)/2)}\, (1-z^2)^{(d-3)/2}\, \theta(1-|z|)\,,
\label{fdz.1}
\end{equation}  
where $\Gamma(y)$ is the Gamma function and $\theta(y)$ is the Heaviside step function: $\theta(y)=1$ if $y\ge 0$ and $\theta(y)=0$ if $y<0$. Moreover, since the directions of the different flights are independent, the joint distribution of the $x$ components of the random vectors $\{{\vec l_i}\}$
with given norms $\{l_i\}$ factorises as:
\begin{equation}
P\left(\{x_i\}|\{l_i\}\right)= \prod_{i=1}^n \frac{1}{l_i} 
f_d\left(\frac{x_i}{l_i}\right)\, .
\label{x_comp_cond.1}
\end{equation}

We can then write down explicitly the joint distribution of the $x$ components $\{x_i\}$, the norms $\{l_i\}$ and the number of tumblings $n$ at fixed total time $t$ as 
\begin{eqnarray}
&& P\left(\{x_i\}, \{l_i\}, n|t\right) =  
P\left(\{x_i\}|\{l_i\}\right)\, P\left(\{l_i\},\, n|t\right)\nonumber
\\
&= &\frac{1}{\gamma}\,
\prod_{i=1}^{n}\int_{0}^{\infty}dv_i\,W(v_i) \frac{1}{l_i}f_d\left(\frac{x_i}{l_i}\right) \frac{\gamma}{v_i}\, e^{-\gamma\, l_i/v_i}\,\nonumber
\\ &\times &
\delta\left( \sum_{i=1}^n \frac{l_i}{v_i}-t\right)\,,
\label{full_joint.1}
\end{eqnarray}
where we used the results in Eqs. (\ref{joint_l_n.1}) and (\ref{x_comp_cond.1}). Having obtained this joint distribution, we can now integrate over the $\{l_i\}$ variables to
obtain the marginal joint distributions of $\{x_i\}$ and $n$, given $t$
\begin{eqnarray}\label{eq:joint_xn.1}
&& P\left(\{x_i\}\,, n|t\right)= \frac{1}{\gamma}\,
\prod_{i=1}^{n}\int_{0}^{\infty}dv_i\,W(v_i) \\ &\times &\int_{0}^{\infty}dl_i\,\frac{1}{l_i}f_d\left(\frac{x_i}{l_i}\right) \frac{\gamma}{v_i}\, e^{-\gamma\, l_i/v_i}\,
\delta\left(  \sum_{i=1}^n \frac{l_i}{v_i}-t\right)\,.\nonumber
\end{eqnarray}
The result in Eq. (\ref{eq:joint_xn.1})  then can be interpreted as an effective $x$-component process $\{x_i\}$ projected from the $d$-dimensional RTP of fixed duration $t$. To further simplify this $x$-component process, we take a Laplace transform
with respect to $t$ that decouples the integrals over the $\{l_i\}$ variables
\begin{eqnarray}\label{lt_joint.1}
&&\int_{0}^{\infty}\,dt\,e^{-st} P\left(\{x_i\}\,, n|t\right)\\&=& \frac{1}{\gamma}\,
\prod_{i=1}^{n}\int_{0}^{\infty}dv_i\,W(v_i)\int_{0}^{\infty}dl_i\,\frac{1}{l_i}f_d\left(\frac{x_i}{l_i}\right) \frac{\gamma}{v_i}\, e^{-(\gamma+s) l_i/v_i}\,\nonumber\\
&=&\frac{1}{\gamma}\left(\frac{\gamma}{\gamma+s}\right)^n\prod_{i=1}^{n}\tilde{p}_s(x_i)\,,\nonumber
\end{eqnarray}
where we have defined
\begin{equation}\label{eq:psv}
\tilde{p}_s(x)=\int_{0}^{\infty}dl\,\frac{1}{l}f_d\left(\frac{x}{l}\right)\int_{0}^{\infty}dv\,W(v)\,\frac{\gamma+s}{v}e^{(\gamma+s)l/v}\,.
\end{equation}
Note that in Eq. (\ref{lt_joint.1}) we 
have multiplied and divided by a factor $(\gamma+s)^n$ so that the function ${\tilde p}_s(x)$, which depends on $s$, $d$, $\gamma$ and on the speed distribution $W(v)$ can be interpreted as a PDF of a random variable $x$.
Manifestly ${\tilde p}_s(x)$ is non-negative and normalized to unity. Indeed, integrating over $x$ one gets
\begin{eqnarray}
&&\int_{-\infty}^{\infty} {\tilde p}_s(x)\, dx =(\gamma+s)\,\int_0^{\infty} dl\, \int_{-\infty}^{\infty} \frac{dx}{l}\, f_d\left(\frac{x}{l}\right)\,\\ &\times & \int_{0}^{\infty}\frac{dv}{v}W\left(v\right)
e^{-(\gamma+s)\, l/v}\, \nonumber = \left(\gamma+s\right)\,\int_0^{\infty} \frac{dv}{v}W\left(v\right)\\ &\times &\int_0^{\infty} dl\, e^{-(\gamma+s)\, l/v} \int_{-1}^1 dz\, f_d(z) =\int_0^{\infty} dv\,W\left(v\right) \,=1 ,\nonumber
\label{normal_psx_v}
\end{eqnarray}
where we have performed the change of variable $x\to z=x/l$ and we have used the fact that $f_d(z)$ and $W(v)$ are normalized to one. Moreover, due to the symmetry $f_d(z)=f_d(-z)$, $\tilde{p}_s(x)$ is also symmetric around $x=0$. Even if one could in principle compute $\tilde{p}_s(x)$, we will show that the precise expression for ${\tilde p}_s(x)$ is not relevant, as long as it is continuous and symmetric in $x$. Finally, inverting the Laplace transform in Eq. (\ref{lt_joint.1}) formally, we have the joint distribution of $\{x_i\}$ and $n$ for a fixed $t$
\begin{equation}
P\left(\{x_i\}, n|t\right)=  
\int \frac{ds}{2\pi\,i} e^{s\, t}\, 
\frac{1}{\gamma}\, \left(\frac{\gamma}{\gamma+s}\right)^n \prod_{i=1}^n {\tilde p}_s(x_i)\, ,
\label{Pxn.1}
\end{equation}
where the integral is over the Bromwich contour (imaginary axis in this case) in the complex $s$ plane. We observe that the projection of the motion in $d$ dimensions of the RTP in the $x$ direction can be interpreted as an effective one-dimensional RW with increments $\{x_i\}$. Note, however, that these increments are correlated in a complicated way (see Eq. (\ref{Pxn.1})).

\begin{figure}[t]
\includegraphics[width = \linewidth]{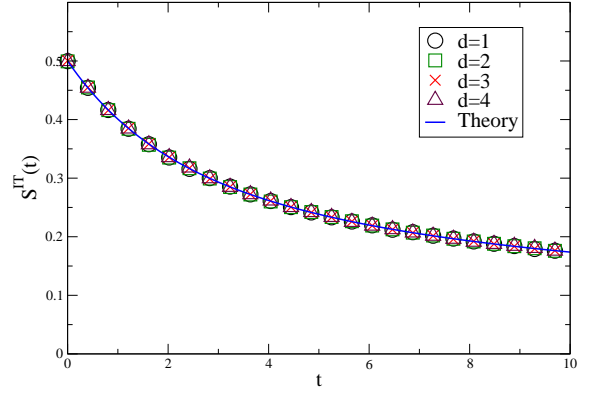} 
\caption{\label{fig:s} Survival probability $S^{\rm IT}(t)$ of a single RTP in the IT model as a function of time $t$, for $\gamma=1$. 
The continuous blue line corresponds to the exact result in Eq. (\ref{surv.2}). The symbols correspond to simulations with the choices $d=1,2,3,4$ and $P_R(\tau) = e^{-\tau}$. They all fall on the analytical blue line for all $t$. }
\end{figure}

The survival probability $S^{\rm IT}(t)$ of this $x$-component process up to time $t$ is the probability of the event that the successive sums
\begin{equation}\label{eq:Xk}
X_k=x_1+x_2+\ldots +x_k\,
\end{equation}
with $1\leq k\leq n$ are all positive. We recall that the number of tumblings $n$ is also a random variable. Thus, summing over $n$ one obtains
\begin{eqnarray}
&&S^{\rm IT}(t) = \sum_{n=1}^{\infty} \int_{-\infty}^{\infty} dx_1\ldots \int_{-\infty}^{\infty} dx_n\,
\theta(X_1)\ldots\theta(X_n)\, \nonumber\\ &\times & P\left(\{x_i\}, n|t\right)\, ,
\label{St.1}
\end{eqnarray}
where the product of theta functions constrains the $x$-component process to remain positive up to time $t$.  Plugging the expression for $P\left(\{x_i\}, n|t\right)$ given in Eq. (\ref{Pxn.1}) into Eq. (\ref{St.1}) gives
\begin{eqnarray}
&& S^{\rm IT}(t)  =\int \frac{ds\,e^{s\, t}}{2\pi\,i\gamma}  \sum_{n=1}^{\infty} \left(\frac{\gamma}{\gamma+s}\right)^n\,
\int_{-\infty}^{\infty} dx_1\ldots \int_{-\infty}^{\infty} dx_n\,\nonumber\\
&&\times \, \prod_{i=1}^n \theta (X_i){\tilde p}_s(x_i) =\int \frac{ds}{2\pi\,i} \,
\frac{e^{s\, t}}{\gamma}\, \sum_{n=1}^{\infty} \left(\frac{\gamma}{\gamma+s}\right)^n\, q_n\,,
\label{St.2}
\end{eqnarray}
where we have defined the multiple integral
\begin{equation}
q_n= \int_{-\infty}^{\infty} dx_1\ldots \int_{-\infty}^{\infty} dx_n\,
\, \prod_{i=1}^n\theta(X_i) {\tilde p}_s(x_i)\, .
\label{qn.1}
\end{equation}
Notably, this quantity $q_n$ can be interpreted as the probability that a one-dimensional RW does not visit the negative side of the $x$ axis. Indeed, let us consider the RW $X_k$ defined as 
\begin{equation}
X_k=X_{k-1}+x_k\,,
\end{equation}
with $X_0=0$. The increments $x_k$ are i.i.d. with distribution ${\tilde p}_s(x_k)$. As explained in the introduction, since ${\tilde p}_s(x)$ is continuous and symmetric, the SA theorem \cite{SA_54} states that $q_n$ is universal and its expression is given by:
\begin{equation}
q_n= {2n \choose n}\, 2^{-2n}\, \quad\quad\quad n=0,1,2,\ldots
\label{sa.1}
\end{equation}
Note that this formula is valid for any $n$. The generating function of $q_n$ is thus also universal
\begin{equation}
\sum_{n=0}^{\infty} q_n\, z^n = \sum_{n=0}^{\infty}  {2n \choose n}\, 
\left(\frac{z}{4}\right)^n= \frac{1}{\sqrt{1-z}}\, .
\label{sa.2}
\end{equation}
Using this result (\ref{sa.2}) in Eq. (\ref{St.2}) and noticing that the sum in Eq. (\ref{St.2}) does not include the $n=0$ term leads to the result  
\begin{equation}
S^{\rm IT}(t)= \int \frac{ds}{2\pi\,i} e^{s\, t}\,
\frac{1}{\gamma}\ \left[\sqrt{\frac{\gamma+s}{s}}-1\right]\, .
\label{sa.3}
\end{equation}
Remarkably, this result is universal in the sense that it does not depend on the dimension $d$ nor on the speed distribution $W(v)$. Indeed, $d$ and $W(v)$ appear only in Eq. (\ref{St.2}) through the PDF ${\tilde p}_s(x)$. However, we have seen that as a consequence of the SA theorem the result is completely independent of the particular expression of ${\tilde p}_s(x)$. The Laplace inversion in Eq. (\ref{sa.3}) can be computed explicitly using the inversion formula \cite{schiff_book}
\begin{equation}\label{eq:laplace_inversion_1}
\mathcal{L}^{-1}_{s\to t}\left(\sqrt{\frac{b+s}{s}}-1\right)\left(t\right)=\frac{b}{2} e^{-\frac{b}{2} t}\left(I_0\left(\frac{b}{2}t\right)+I_1\left(\frac{b}{2}t\right)\right)\,. 
\end{equation}
Thus, we obtain that
\begin{equation}
S^{\rm IT}(t)= \frac{1}{2}\, e^{-\gamma t/2}\, \left(I_0\left(\frac{\gamma}{2}t\right)+ I_1\left(\frac{\gamma}{2}t\right)\right)\,,
\label{surv.2}
\end{equation}
where $I_0(z)$ and $I_1(z)$ are modified Bessel functions. 

\begin{figure}
\includegraphics[width =\linewidth]{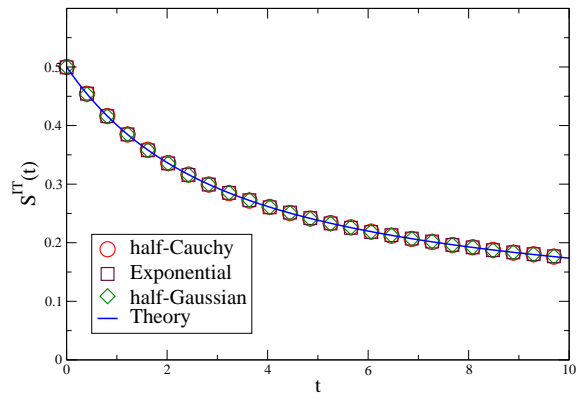}
\caption{Plot of $S^{\rm IT}(t)$ for the IT model evaluated numerically  
for different velocity distributions $W(v)$ and an exponential distribution $P_R(\tau)=e^{-\tau}$ in dimension $d=2$. The solid line corresponds to the
exact analytical result in Eq. (\ref{surv.2}).} \label{fig:S_velocity}
\end{figure}

The function $S^{\rm IT}(t)$ is shown in Fig. \ref{fig:s} and it is in good agreement with numerical simulations performed in dimensions $d=1,2,3,4$ with fixed velocity $v_0$, i.e. choosing $W(v)=\delta(v-v_0)$. Moreover, in Fig. \ref{fig:S_velocity}, we show that $S^{\rm IT}(t)$ is in perfect agreement also with numerical simulations performed in $d=2$ with different speed distributions $W(v)$. From Eq. (\ref{surv.2}) it is easy to derive the time asymptotics of $S^{\rm IT}(t)$. When $t\to 0$, the survival probability $S^{\rm IT}(t)$ can be approximated as
\begin{equation}
S^{\rm IT}(t)\simeq \frac12 -\frac{\gamma}{8}t\,.
\end{equation}
The limit value $1/2$ is the probability that the $x$ component of the initial direction is positive. On the other hand, when $t\to \infty$ 
\begin{equation}
S^{\rm IT}(t)\simeq \frac{1}{\sqrt{\pi \gamma t}}-\frac{1}{4\sqrt{\pi \gamma^3 t^3}}\,.
\end{equation}

\subsection{Time to reach the maximum} \label{sec:tmax}

\begin{figure}[t]
\includegraphics[width=0.45\textwidth]{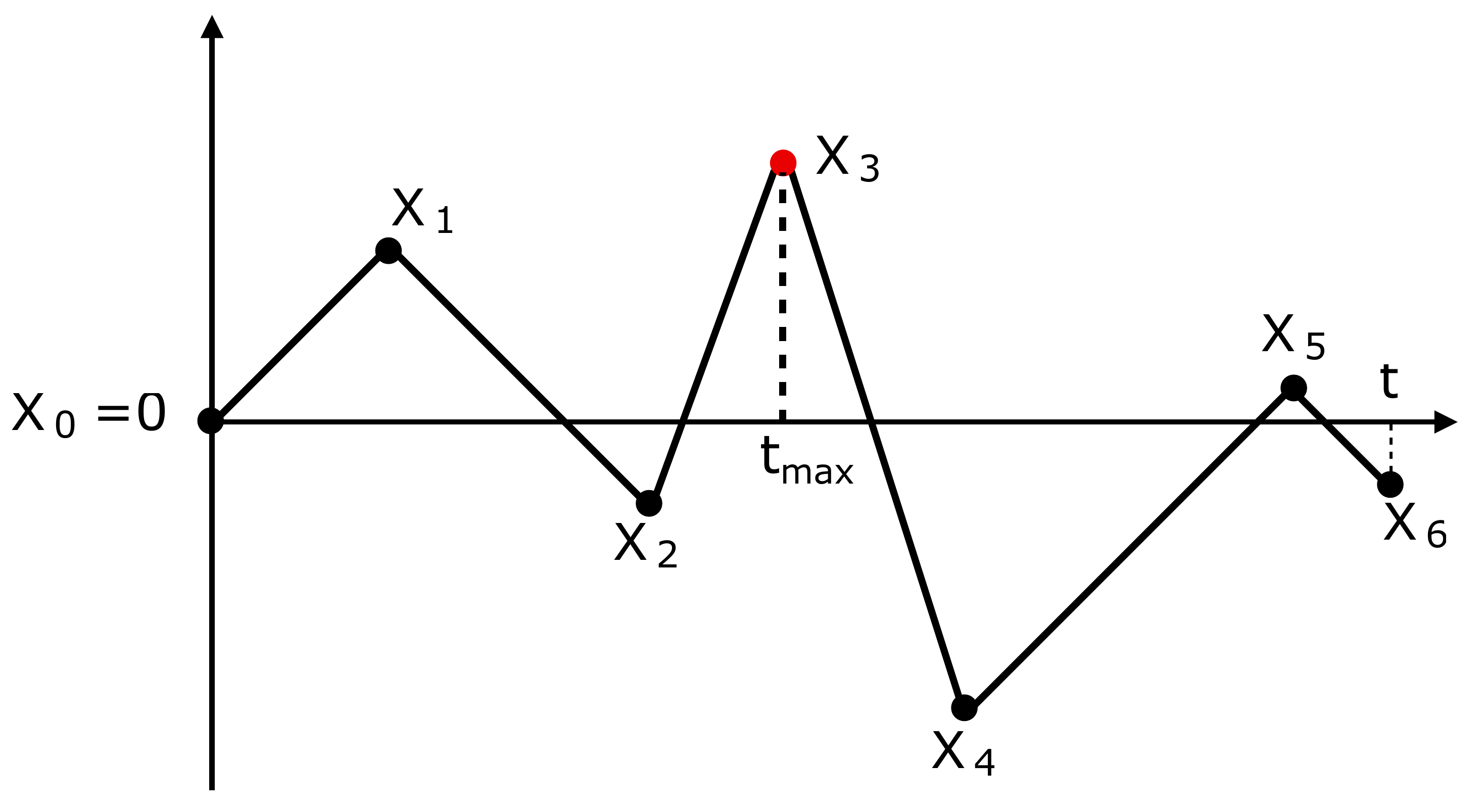} 
\caption{\label{fig:tmax} Typical trajectory of the $x$ component of a single RTP in the IT model (black solid line). The dots represent the position of the associated random walk $X_k$ for $0\leq k \leq n$ (here $n=6$). The global maximum (indicated by a red dot) is reached at time $t_{\max}$, corresponding to the step $n_1$ of the random walk (here $n_1=3$). Note that in principle also the starting position $X_0=0$ or the last position $X_n$ could be the global maximum.
}
\end{figure}

In this section we consider again an RTP starting at the origin and moving according to the IT model in $d$ dimensions up to time $t$. Focusing on the $x$ component of the particle, we want to compute the distribution of the time $t_{\max}$ at which the $x$ component reaches its maximal value.

First of all, we note that, since the motion of the particle is ballistic during each flight, the maximal value of the $x$ component can only be reached at the end of a running phase (except for the special case $t_{\max}=0$). It is useful to define the RW generated by the $x$ component of the RTP at the end of each run by
\begin{equation}\label{eq:rw}
X_k=x_1+x_2+\ldots +x_k\,,
\end{equation}
where $x_i$ is the displacement in the $x$ component of the $i^{\rm th}$ run (see Fig. \ref{fig:tmax}). We denote by $n\geq 1$ the number of running phases and we assume that the global maximum is reached at time $t_{\max}$, corresponding to the step $n_1$ of the random walker. As we will show, the events $t_{\max}=0$ and $t_{\max}=t$ happen with a finite probability. Thus, after calculating the distribution $P(t_{\max}|t)$ in the case $0<t_{\max}<t$, we will also need to compute the contributions corresponding to $t_{\max}=0$ and $t_{\max}=t$. When $0<t_{\max}<t$, it is useful to exploit the fact that, since a tumbling happens at time $t_{\max}$, the time intervals $[0,t_{\max}]$ (I) and $[t_{\max},t]$ (II) are independent. Note that $n_1$ is the number of running phases in the interval $[0,t_{\max}]$. We also define the additional variable $n_2=n-n_1$, denoting the number of running phases in the interval $[t_{\max},t]$. We observe that, when $0<t_{\max}<t$, one has $n_1\geq 1$ and $n_2\geq 1$. Thus, $P(t_{\max}|t)$ is given by the product of the probability weights $P_{\rm I}(t_{\max})$ and $P_{\rm II}(t-t_{\max})$ corresponding to the two intervals. In the interval (I), the $x$ component of the particle has to remain below its maximal value $X_{n_1}$ up to time $t_{\max}$ and it has to reach $X_{n_1}$ at time $t_{\max}$. In the second interval, the $x$ component of the RTP is $X_{n_1}$ at time $t_{\max}$ and it has to remain below $X_{n_1}$ up to time~$t$.

Let us now compute the probability weight $P_{\rm I}(t_{\max})$ of the first interval $[0,t_{\max}]$. Note that here we keep $t_{\max}$ as a random variable and not condition it to take a specific value. The joint distribution of the displacements $x_1,x_2,\ldots x_{n_1}$ and of the number $n_1$ of running phases up to time $t_{\max}$ was computed in Section \ref{sec:surv} and is given by (see Eq. (\ref{Pxn.1}))
\begin{eqnarray}\label{Pxn_tmax}
P\left(x_1,\ldots x_{n_1},n_1|t_{\max}\right)&=&  
\int \frac{ds_1}{2\pi\,i} e^{s_1\, t_{\max}}\, \left(\frac{\gamma}{\gamma+s_1}\right)^{n_1}\nonumber \\ &\times & \prod_{i=1}^{n_1} {\tilde p}_{s_1}(x_i)\, ,
\end{eqnarray}
where ${\tilde p}_{s_1}(x_i)$ is given in Eq. (\ref{eq:psv}), for the most general case in which the velocities of each run are random. Let us remind the reader that the notation $P\left(x_1,\ldots x_{n_1},n_1|t_{\max}\right)$ does not mean conditioning $t_{\max}$ to take a specific value, $t_{\max}$ is still a variable. Note also that, since in this case the final time $t_{\max}$ is also a tumbling time, Eq. (\ref{Pxn_tmax}) differs from Eq. (\ref{Pxn.1}) by a factor $\gamma$. The weight $P_{\rm I}(t_{\max})$ is the probability that, in the interval $[0,t_{\max}]$, the maximum is reached at the last step ${n_1}$, i.e. that $X_{n_1}>X_i$ for all $i<n_1$. This probability can be written as, summing over $n_1\geq 1$,
\begin{eqnarray}
P_{\rm I}(t_{\max})&=&\sum_{n_1=1}^{\infty}\int_{-\infty}^{\infty}dx_1 \,\ldots \int_{-\infty}^{\infty}dx_{n_1}\\
&\times & P\left(x_1,\ldots x_{n_1}, n_1|t_{\max}\right)\nonumber \\
&\times & \theta(X_{n_1}) \theta(X_{n_1}-X_1)\ldots \theta(X_{n_1}-X_{n_1-1})\nonumber\,,
\end{eqnarray}
where $X_i$ is defined in Eq. (\ref{eq:rw}). The term $\theta(X_{n_1}) \theta(X_{n_1}-X_1)\ldots \theta(X_{n_1}-X_{n_1-1})$ enforces the global maximum to be reached at step $n_1$. Using Eq. (\ref{Pxn_tmax}), we obtain
\begin{equation}\label{eq_Qm}
P_{\rm I}(t_{\max})=\int \frac{ds_1}{2\pi\,i} e^{s_1\, t_{\max}}\,\sum_{n_1=1}^{\infty} \left(\frac{\gamma}{\gamma+s_1}\right)^{n_1}  q_{n_1}\,,
\end{equation}
where 
\begin{equation}\label{eq:qm_tmax1}
q_{n_1}=\int_{-\infty}^{\infty}dx_1\,\ldots\int_{-\infty}^{\infty}dx_{n_1}\,\prod_{i=i}^{{n_1}}\tilde{p}_s(x_i)\theta(X_{n_1}-X_{n_1-i})\,.
\end{equation}

Clearly, this quantity $q_{n_1}$ is just the survival probability. This is best explained with the help of Fig. \ref{fig:tmax}: if we consider the interval $[0,t_{\max}]$, looking at the trajectory from position $X_{n_1}$ (with $n_1=3$ in this case) and inverting time, we can observe that the walker has to remain below its starting position up to step ${n_1}$. Thus, using the $x\to -x$ symmetry, we obtain that $q_{n_1}$ is a survival probability. More precisely, we perform the change of variables $z_k=x_{n_1-k}$ and we consider the RW $Z_k=X_{n_1}-X_{n_1-k}=z_1+z_2+\ldots z_{n_1}$. Then, $q_{n_1}$ can be rewritten as
\begin{equation}
q_{n_1}=\int_{-\infty}^{\infty}dz_1\,\ldots\int_{-\infty}^{\infty}dz_{n_1}\,\prod_{i=1}^{n_1}\tilde{p}_s(z_i)\theta(Z_i)\,,
\end{equation}
which is precisely the probability that the position $Z_k$ of the random walker remains positive up to step $n_1$. Since the probability distribution $\tilde{p}_s(z)$ is continuous and symmetric, as explained in Section \ref{sec:surv}, the survival probability $q_{n_1}$ is universal and its generating function is given by (see Eq. (\ref{sa.2}))
\begin{equation}\label{gen_fun_tmax}
\sum_{{n_1}=0}^{\infty}q_{n_1} \, z^{n_1}=\frac{1}{\sqrt{1-z}}\,.
\end{equation}
Thus, using this relation (\ref{gen_fun_tmax}) in Eq. (\ref{eq_Qm}) we obtain that the probability weight of the first interval is given by
\begin{equation}\label{eq:P_I}
P_{\rm I}(t_{\max})=\int \frac{ds_1}{2\pi\,i} e^{s_1\, t_{\max}}\left(\sqrt{\frac{\gamma+s_1}{s_1}}-1\right)\,.
\end{equation}
The Laplace inversion can be performed explicitly using Eq. (\ref{eq:laplace_inversion_1}) and one obtains
\begin{equation}\label{P_I_Laplace}
P_{\rm I}(t_{\max})=\frac{\gamma}{2}e^{-\frac{\gamma}{2}t_{\max}}\left(I_0\left(\frac{\gamma}{2}t_{\max}\right)+I_1\left(\frac{\gamma}{2}t_{\max}\right)\right)\,.
\end{equation}
Note that this expression is identical, apart from a factor $\gamma$, to the one obtained for the survival probability $S^{\rm IT}(t)$ computed in Section \ref{sec:surv}:
\begin{equation}\label{P_I_final}
P_{\rm I}(t_{\max})=\gamma S^{\rm IT}(t_{\max})\,,
\end{equation}
where $S^{\rm IT}(t)$ is given in Eq. (\ref{surv.2}).

Similarly, one can compute the probability weight $P_{\rm II}(t-t_{\max})$ of the second time interval $[t_{\max},t]$. The joint PDF of the displacements $x_{n_1+1},x_{n_1+2}\ldots x_n$ and of the number $n_2$ of tumbling phases in the interval $[t_{\max},t]$ can be written as
(see Eq. (\ref{Pxn.1}))
\begin{eqnarray}\label{Pxn_tmax2}
&&P\left(x_{n_1+1},\ldots x_{n_1+n_2}, n_2|t-t_{\max}\right)\\ &=& \frac{1}{\gamma} 
\int \frac{ds_2}{2\pi\,i} e^{s_2\, (t-t_{\max})}\, \left(\frac{\gamma}{\gamma+s_2}\right)^{n_2} \prod_{i=n_1+1}^{n_1+n_2} {\tilde p}_{s_2}(x_i)\, ,\nonumber
\end{eqnarray}
where recall that $n_2=n-n_1$. The weight $P_{\rm II}(t-t_{\max})$ of the second time interval is the probability that that the $x$ component of the RTP remains below position $X_{n_1}$ up to time $t$, starting from $X_{n_1}$ at time $t_{\max}$. This probability can be written as, summing over $n_2\geq 1$,
\begin{eqnarray}\label{eq:P_II_n2}
P_{\rm II}(t-t_{\max})&=&\sum_{n_2=1}^{\infty}\int_{-\infty}^{\infty}dx_{n_1+1}\ldots\int_{-\infty}^{\infty}dx_{n_1+n_2} \nonumber\\
&\times &P\left(x_{n_1+1},\ldots x_{n_1+n_2}, n_2|t-t_{\max}\right) \\
&\times & \theta(X_{n_1}-X_{n_1+1})\ldots\theta(X_{n_1}-X_{n_1+n_2})\,.\nonumber
\end{eqnarray}
Using the expression for $P\left(x_{n_1+1},\ldots x_n, n_2|t_1\right)$ in Eq. (\ref{Pxn_tmax2}), we can rewrite Eq. (\ref{eq:P_II_n2}) as
\begin{equation}\label{eq:Pn1}
P_{\rm II}(t-t_{\max})=\frac{1}{\gamma} 
\int \frac{ds_2}{2\pi\,i} e^{s_2 (t-t_{\max})}\sum_{n_2=1}^{\infty}\left(\frac{\gamma}{\gamma+s_2}\right)^{n_2}q_{n_2}\,,
\end{equation}
where 
\begin{eqnarray}
q_{n_2}&=&\int_{-\infty}^{\infty}dx_{n_1+1}\ldots\int_{-\infty}^{\infty}dx_{n_1+n_2}\\
&\times &\prod_{i=n_1+1}^{n_1+n_2}\tilde{p}_{s_2}(x_i)\theta(X_{n_1}-X_{i})\,.\nonumber
\end{eqnarray}
The probability $q_{n_2}$ can again be rewritten as a survival probability of a RW. Indeed, similarly to what we have done above, we perform the change of variables $z_k=x_{n_1+k}$ and we consider the RW $Z_k=X_{n_1}-X_{n_1+k}=z_1+z_2+\ldots z_{n_2}$. Looking at Fig. \ref{fig:tmax}, this transformation is equivalent to flip the figure and to look at the trajectory from position $X_{n_1}$. This transformation yields
\begin{equation}
q_{n_2}=\int_{-\infty}^{\infty}dz_1\ldots\int_{-\infty}^{\infty}dz_{n_2}\prod_{i=1}^{n_2}\tilde{p}_{s_2}(z_i)\theta(Z_i)\,,
\end{equation}
which is again the probability that a random walker starting from the origin remains in the positive side up to step $n_2$. As stated above, this probability is universal and its generating function is given by Eq. (\ref{gen_fun_tmax}). Thus, using Eq. (\ref{gen_fun_tmax}), Eq. (\ref{eq:Pn1}) can be rewritten as
\begin{equation}
P_{\rm II}(t-t_{\max})=\frac{1}{\gamma} 
\int \frac{ds_2}{2\pi\,i} e^{s_2\, (t-t_{\max})}\,\left(\sqrt{\frac{\gamma+s_2}{s_2}}-1\right)\,.
\end{equation}
Note that the probability weight $P_{\rm II}(t-t_{\max})$ of the interval $[t_{\max},t]$ turns out to be completely independent of the position $X_{n_1}$ at time $t_{\max}$.
Finally, using the formula in Eq. (\ref{eq:laplace_inversion_1}) to perform the Laplace inversion , we obtain that
\begin{eqnarray}\label{P_II_Laplace}
P_{\rm II}(t-t_{\max})&=&\frac{1}{2}e^{-\frac{\gamma}{2}(t-t_{\max})}\Bigg(I_0\left(\frac{\gamma}{2}(t-t_{\max})\right) \nonumber\\
&+&I_1\left(\frac{\gamma}{2}(t-t_{\max})\right)\Bigg)\,.
\end{eqnarray}
Note that this expression is identical to the one obtained for the survival probability $S^{\rm IT}(t)$ computed in Section \ref{sec:surv}:
\begin{equation}\label{P_II_final}
P_{\rm II}(t-t_{\max})=S^{\rm IT}(t-t_{\max})\,,
\end{equation}
where $S^{\rm IT}(t)$ is given in Eq. (\ref{surv.2}).  In principle, one could have guessed Eq. (\ref{P_II_final}). Indeed, after time $t_{\max}$ the particle has to remain below its starting  position up to time $t$ and, using the translation invariance and the $x\to -x$ symmetry of the process, it is clear that the weight of the second interval is given by the survival probability $S^{\rm IT}(t-t_{\max})$.  We can now compute the probability distribution of $t_{\max}$ as the product of the two factors $P_{\rm I}(t_{\max})$ and $P_{\rm II}(t-t_{\max})$. Using Eqs. (\ref{P_I_final}) and (\ref{P_II_final}), we obtain
\begin{equation}\label{eq:p_tmax_incomplete}
P(t_{\max}|t)=\gamma S^{\rm IT}(t_{\max}) S^{\rm IT}(t-t_{\max})\,,
\end{equation}
where $S^{\rm IT}(t)$ is given in Eq. (\ref{surv.2}).

Note, however, that Eq. (\ref{eq:p_tmax_incomplete}) is only valid when $0<t_{\max}<t$ and that we need to compute separately the contributions of the events $t_{\max}=0$ and $t_{\max}=t$. It is clear that the maximum will be reached at time $t_{\max}=0$ only if the $x$ component of the particle does not visit the positive side up to time $t$. Thus, using the $x\to -x$ symmetry of the process we find that 
\begin{equation}\label{eq:proba_tmax_0}
{\rm Prob.}(t_{\max}=0|t)=S^{\rm IT}(t)\,,
\end{equation}
where $S^{\rm IT}(t)$ is given in Eq. (\ref{surv.2}). Similarly, it is also easy to show that 
\begin{equation}\label{eq:proba_tmax_t}
{\rm Prob.}(t_{\max}=t|t)=S^{\rm IT}(t)\,.
\end{equation}
Thus, using Eqs. (\ref{eq:p_tmax_incomplete}), (\ref{eq:proba_tmax_0}), and (\ref{eq:proba_tmax_t}), we obtain that for $0\leq t_{\max}\leq t$
\begin{eqnarray}\label{eq:p_tmax_final}
P(t_{\max}|t)&=&\gamma S^{\rm IT}(t_{\max})S^{\rm IT}(t-t_{\max})\\
&+&S^{\rm IT}(t)\left(\delta(t_{\max})+\delta(t-t_{\max})\right)\,,\nonumber
\end{eqnarray}
where $S^{\rm IT}(t)$ is given in Eq. (\ref{surv.2}). Note that with a similar technique one can also derive the probability distribution of the number $n_1$ of running phases before the global maximum at fixed $t$ (see Appendix \ref{app:n1}).

\begin{figure}[t]
\includegraphics[width=0.5\textwidth]{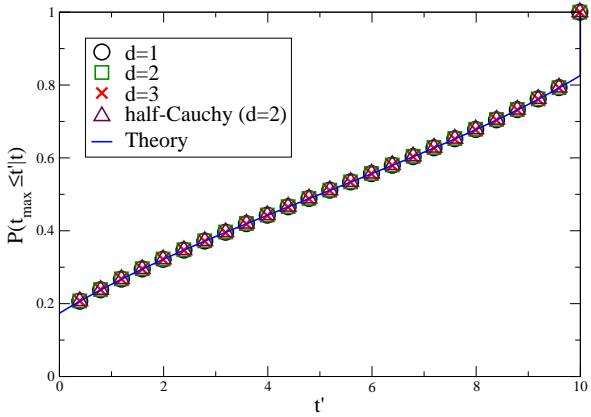} \caption{\label{fig:cumulative_tmax} 
Cumulative probability $P(t_{\max}\leq t'|t)$ for the IT model as a function of $t'$, with $\gamma=1$ and $t=10$ in this case. The continuous blue line corresponds to the exact result in Eq. (\ref{eq:cumulative_IT}). The symbols correspond to simulations with the choices $d = 1, 2, 3$ with $W(v)=\delta(v-1)$ and $d = 2$ with $W(v) =2/(\pi\left(1+v^2\right))$, for $v > 0$
(half-Cauchy). We observe that the numerical curves collapse on the corresponding analytical blue line. Note that the cumulative probability has a jump discontinuity at $t'=t$ (see Eq. (\ref{eq:cumulative_IT})).}
\end{figure}

This result in Eq. (\ref{eq:p_tmax_final}) was derived in the one-dimensional case by solving the Fokker-Plank equation associated to the system \cite{singh19}. Here we have proved that Eq. (\ref{eq:p_tmax_final}) is valid for any $t_{\max}$ and $t$, independently of the dimension $d$ of the system and of the speed distribution $W(v)$. Integrating Eq. (\ref{eq:p_tmax_final}) we obtain that the cumulative probability of $t_{\max}$ is given by, for $0\leq t'\leq t$,
\begin{eqnarray}\label{eq:cumulative_IT}
P(t_{\max}&\leq& t'|t)=\gamma\int_{0}^{t'}dt_{\max}\,S^{\rm IT}(t_{\max})S^{\rm IT}(t-t_{\max})\nonumber \\
&+& S^{\rm IT}(t)\left[1+\theta(t'-t)\right]\,,
\end{eqnarray}
where $\theta(t'-t)=0$ if $t'<t$ and $\theta(t-t')=1$ if $t'=t$, and $S^{\rm IT}(t)$ is given in Eq. (\ref{surv.2}). Consequently, one should observe a jump discontinuity at $t'=t$ in the cumulative distribution. Indeed, this discontinuity can be observed in Fig. \ref{fig:cumulative_tmax}, where we also show that the exact result in Eq. (\ref{eq:cumulative_IT}) is in excellent agreement with numerical simulations performed with different choices of $W(v)$ and $d$.

Finally, in order to check the expression for $P(t_{\max}|t)$ given in Eq. (\ref{eq:p_tmax_final}) is correctly normalized to one, it is useful to take a Laplace transform with respect to $t_{\max}$ and $t$ on both sides of Eq. (\ref{eq:p_tmax_final}). This yields
\begin{eqnarray}
&& \int_{0}^{\infty}dt\,\int_{0}^{t}dt_{\max}\,P(t_{\max}|t)e^{-st-s_1 t_{\max}}\\
&=& \gamma \tilde{S}^{\rm IT}(s_1+s)\tilde{S}^{\rm IT}(s)+\tilde{S}^{\rm IT}(s_1+s)+\tilde{S}^{\rm IT}(s)\,,\nonumber
\end{eqnarray} 
where $\tilde{S}^{\rm IT}(s)$ is the Laplace transform of $S^{\rm IT}(t)$. Using the expression for $\tilde{S}^{\rm IT}(s)$, given in Eq. (\ref{sa.3}), we obtain, after few steps of algebra
\begin{eqnarray}\label{eq:normalization_1}
&& \int_{0}^{\infty}dt\,\int_{0}^{t}dt_{\max}\,P(t_{\max}|t)e^{-st-s_1 t_{\max}}\\
&=& \frac{1}{\gamma}\left(\sqrt{\frac{\gamma+s}{s}}\sqrt{\frac{\gamma+s+s_1}{s+s_1}}-1\right)\,.\nonumber
\end{eqnarray}
Setting $s_1=0$ on both sides of Eq. (\ref{eq:normalization_1}), we get
\begin{equation}
 \int_{0}^{\infty}dt\,\int_{0}^{t}dt_{\max}\,P(t_{\max}|t)e^{-st}=\frac{1}{s}\,.
\end{equation}
Finally, inverting the Laplace transform we obtain 
\begin{equation}
\int_{0}^{t}dt_{\max}\,P(t_{\max}|t)=1\,.
\end{equation}
Thus, we have verified that $P(t_{\max}|t)$ is normalized to one.

\subsection{Record statistics}\label{sec:record}

\begin{figure}
\includegraphics[width =  \linewidth]{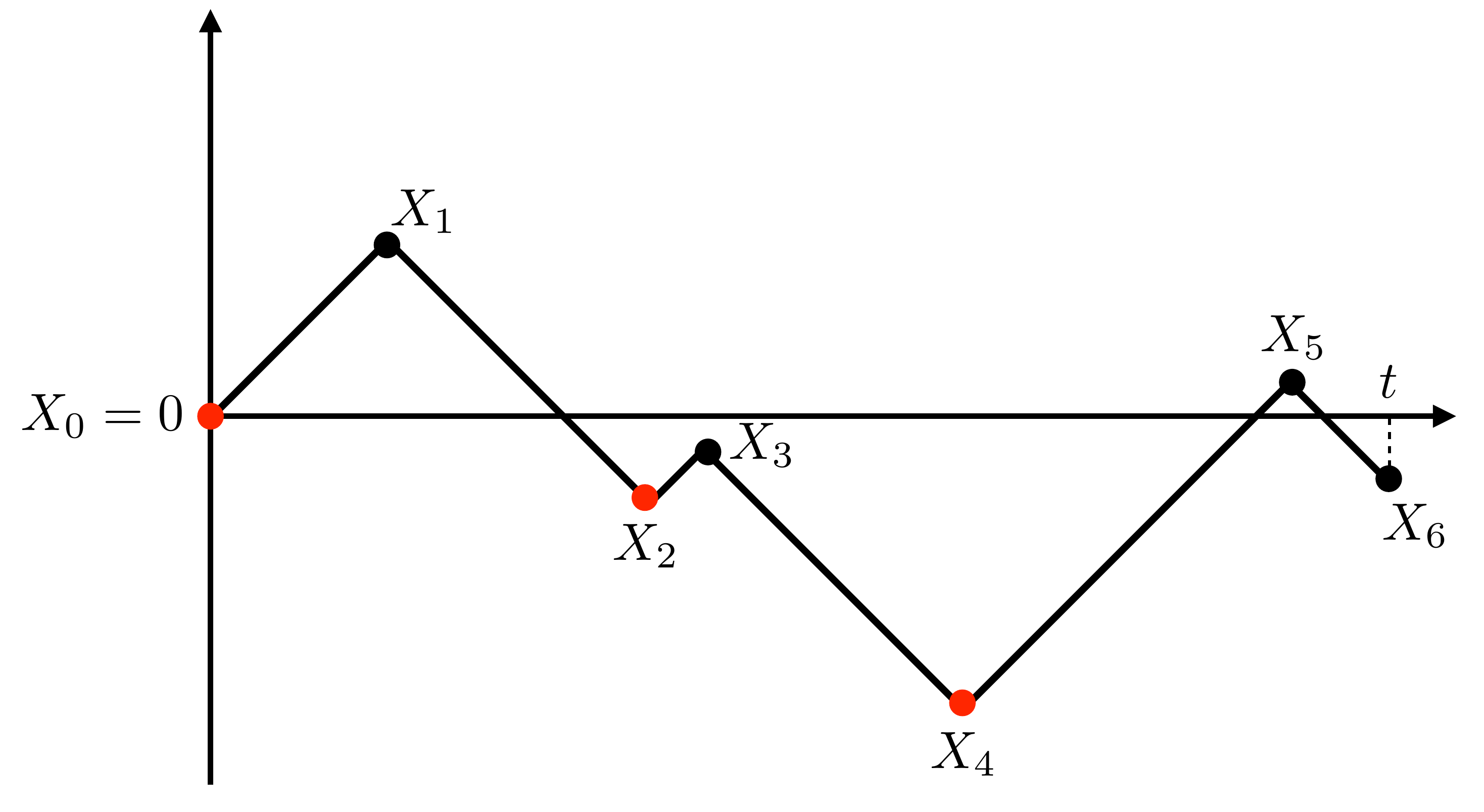}
\caption{Illustration of an $x$-component trajectory of the IT RTP (black solid line) together with the positions of the associated random walk $X_0 = 0, X_1, \ldots X_{n}$ with $n=6$, up to time $t$ (dots). The lower records are indicated in red, the first position $X_0$ being counted as a lower record. Note that the final position $X_n$ can in principle be also a record -- although not in the above figure.}\label{fig_record}
\end{figure}

In this section we show that our result for the survival probability $S^{\rm IT}(t)$ for a $d$-dimensional RTP with instantaneous tumblings can be used to compute the statistics of records for the $x$ component of the RTP process. Indeed, the universality of $S^{\rm IT}(t)$ for the RTP with an exponential distribution of the flight times (corresponding to a constant tumbling rate $\gamma$) also renders the statistics of the records for the $x$ component 
universal in this problem, i.e. independent of the dimension $d$ as well as the speed distribution $W(v)$. In general, it is 
quite hard to obtain exact results for the record statistics for a correlated sequence. Below, we see that, using the method presented in Section \ref{sec:surv}, we can compute the exact record statistics of the $x$ component of the RTP with instantaneous tumblings in $d$ dimensions and show that it is universal. This is one of the rare examples of an exact solution for the record statistics for a correlated sequence.

\begin{figure*}[t]
 \includegraphics[width = \linewidth]{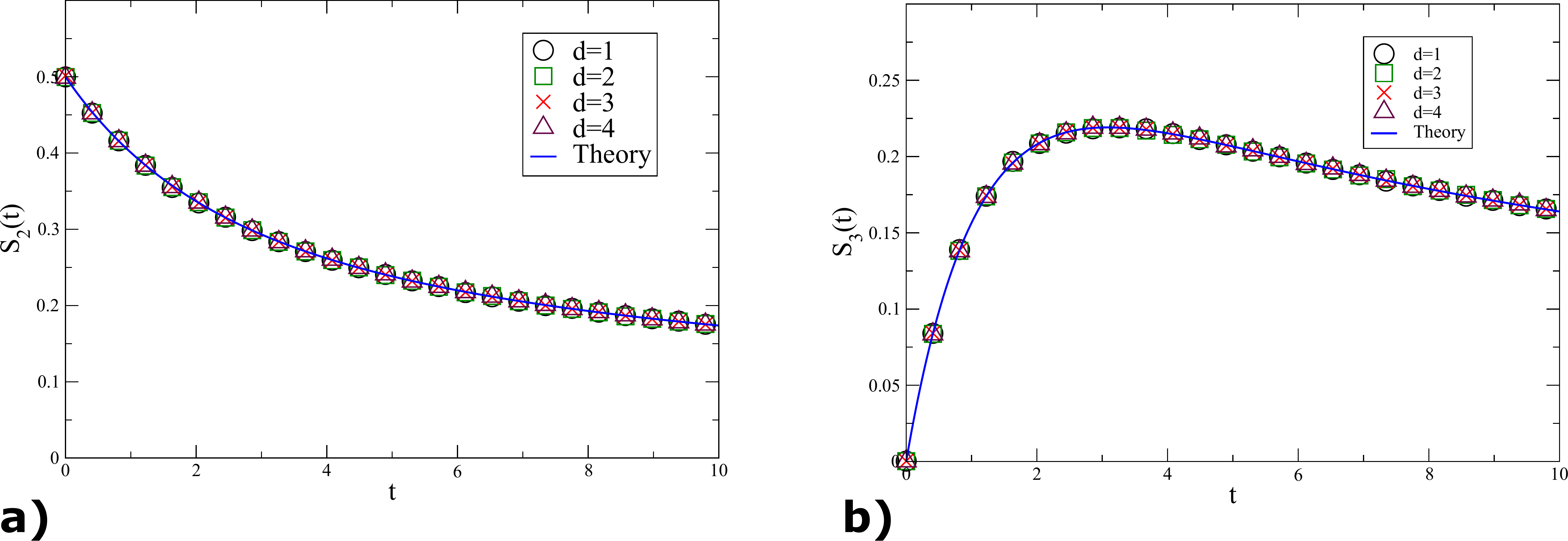}
\caption{Plot of $S_2(t)$ in (a) and $S_3(t)$ in (b) in the IT model for different dimensions $d=1,2,3$ and $d=4$ (symbols correspond to numerical simulations) and an exponential distribution $P_R(\tau) = \gamma e^{-\gamma \tau}$ with $\gamma = 1$. The solid line corresponds to the exact results for $S_2(t) = S^{\rm IT}(t)$ in (\ref{S2}) and $S_3(t)$ in (\ref{S3}).}\label{Fig_s2_s3}
\end{figure*}

Let us first recall the definition of a record. We consider a trajectory in $d$ dimensions of the RTP of duration $t$ starting at the origin. We recall that $n$ denotes the number of tumblings in this trajectory and that the $x$ components of the $n$ successive runs are denoted by $x_1, x_2, \cdots, x_n$. The $x$ components of the positions of the RTP at the end of each running-phase are denoted by (see Fig. \ref{fig_record})
\begin{equation}
X_k=x_1+x_2+\ldots +x_k\,,
\end{equation}
where $1\leq k\leq n$.
The joint distribution of the $x_i$'s and $n$ has been computed in Eq. (\ref{Pxn.1}) and is given by
\begin{equation}
P\left(\{x_i\}, n|t\right)=  
\int \frac{ds}{2\pi\,i} e^{s\, t}\, 
\frac{1}{\gamma}\, \left(\frac{\gamma}{\gamma+s}\right)^n \prod_{i=1}^n {\tilde p}_s(x_i) \;,
\label{Pxn.1_supp}
\end{equation} 
where $\tilde p_s(x_i)$ is given in Eq. (\ref{eq:psv}) for a generic speed distribution $W(v)$ and any dimension $d$. Therefore, the $X_i$'s can be viewed as the position of a one-dimensional discrete-time random walker with correlated steps given in Eq. (\ref{Pxn.1_supp}). A lower record happens at step $k$ if and only if the value $X_k$ is lower than all the previous values, i.e., $X_k < \min \{X_0=0, X_1, \cdots, X_{k-1} \} $ (see Fig. \ref{fig_record}). By convention, $X_0 = 0$ is a lower record. Note that the final position $X_n$ can also be a record. A natural question is then: how many records occur in time $t$? We denote by $S_N(t)$ the probability that there are exactly $N$ lower records up to time $t$. Clearly, when $N=1$ this corresponds to the event that the position has never gone below $0$ up to time $t$. But this precisely the survival probability $S^{\rm IT}(t)$ that we have computed in Section \ref{sec:surv}, thus $S_1(t) = S^{\rm IT}(t)$. We can then think of $S_N(t)$ as a natural generalization of the survival probability $S^{\rm IT}(t)$. One can similarly define upper records for the $x$ component of the RTP, whose statistics are exactly identical to the lower records, due to the $x \to -x$ symmetry of the RTP. {An alternative physical picture of this record process is as follows: whenever the particle achieves a new lower record, one can imagine that the absorbing barrier gets pushed to this new record value. For example, before the second record happens the absorbing barrier is at $X_0=0$. If the second lower record happens at step $k$ with value $X_k < 0$ (for example in Fig. \ref{fig_record} the second record happens at $k=2$), the absorbing barrier gets shifted to $X_k$, till the occurrence of the next lower record (see Fig. \ref{fig_record}).}

Thanks to our mapping to the one-dimensional discrete-time RW via Eq. (\ref{Pxn.1_supp}), we can use the known results for the record statistics of an $n$-step discrete-time RW, whose steps are i.i.d. variables, each drawn from $\tilde p_s(x_i)$ which is continuous and symmetric. We recall that the probability $q^N_n$ that a $n$-step RW has exactly $N$ lower records is universal, i.e. independent of the distribution $\tilde p_s(x_i )$~\cite{Ziff_Satya}. In particular, its generating function with respect to $n$ is given by~\cite{Ziff_Satya} 
\begin{eqnarray} \label{GF_record_nber}
\sum_{n = N-1}^\infty q^N_n z^n = \frac{(1-\sqrt{1-z})^{N-1}}{\sqrt{1-z}} \;.
\end{eqnarray}
The result in Eq. (\ref{Pxn.1_supp}) conveniently translates the results for any observable 
in the discrete-time $n$-step RW problem to the RTP in continuous time $t$. The statistics
of records is one such observable. Therefore, from Eq. (\ref{Pxn.1_supp}) one can show that (for $N \geq 2$)
\begin{eqnarray}\label{SN}
S_N(t) = \int \frac{ds}{2\pi\,i} e^{s\, t}\,
\frac{1}{\gamma}\, \sum_{n=N-1}^{\infty} \left(\frac{\gamma}{\gamma+s}\right)^n\, q^N_n\,,
\end{eqnarray}
where the integral is over the Bromwich contour (imaginary axis in this case). Recall that, for $N=1$,
\begin{equation}\label{S1}
S_1(t) = S^{\rm IT}(t)=\frac{1}{2}e^{-\gamma t /2} \left( I_0 \left(\gamma t /2 \right) + I_1 \left( \gamma t /2\right) \right)\,.
\end{equation}  
Using (\ref{GF_record_nber}) for $z = \gamma/(\gamma+s)$ in (\ref{SN}) one finds, for $N \geq 2$
\begin{eqnarray}\label{SN_2}
S_N(t) = \int \frac{ds}{2\pi\,i} e^{s\, t}\,
\frac{1}{\gamma}  \sqrt{\frac{\gamma + s}{s}}\left( 1 - \sqrt{\frac{s}{\gamma+s}}\right)^{N-1} \;.
\end{eqnarray}
The inverse Laplace transform on the right-hand side of Eq. (\ref{SN_2}) can be performed explicitly for the first few values of $N$. For $N=2$, using the formula in Eq. (\ref{eq:laplace_inversion_1}), we obtain
\begin{equation}
S_2(t) = S^{\rm IT}(t) = \frac{1}{2}e^{-\gamma t /2} \left( I_0 \left(\gamma t /2 \right) + I_1 \left( \gamma t /2\right) \right) \;.\label{S2} 
\end{equation}
Setting $N=3$ in Eq. (\ref{SN_2}), we obtain
\begin{eqnarray}\label{S3_LT}
S_3(t) = \int \frac{ds}{2\pi\,i} e^{s\, t}\,
\frac{1}{\gamma}  \sqrt{\frac{\gamma + s}{s}}\left( 1 - \sqrt{\frac{s}{\gamma+s}}\right)^{2} \;.
\end{eqnarray}
The Laplace transform in Eq. (\ref{S3_LT}) can be inverted using Eq. (\ref{eq:laplace_inversion_1}) and the following Laplace inversion formula \cite{schiff_book}
\begin{eqnarray}\label{eq:laplace_inversion_3}
&&\mathcal{L}^{-1}_{s\to t}\left(\sqrt{\frac{s}{(s+b)}}-1\right)\left(t\right)=
 \frac{b}{2}e^{-\frac{b}{2} t}\\ &\times &\left(I_1\left(\frac{b}{2}t\right)-I_0\left(\frac{b}{2}t\right)\right)\,,\nonumber
\end{eqnarray}
we obtain that
\begin{equation}
S_3(t) = e^{-{\gamma  t}/{2}} I_1\left({\gamma t}/{2}\right) \;. \label{S3}
\end{equation}

\begin{figure}
\includegraphics[width = \linewidth]{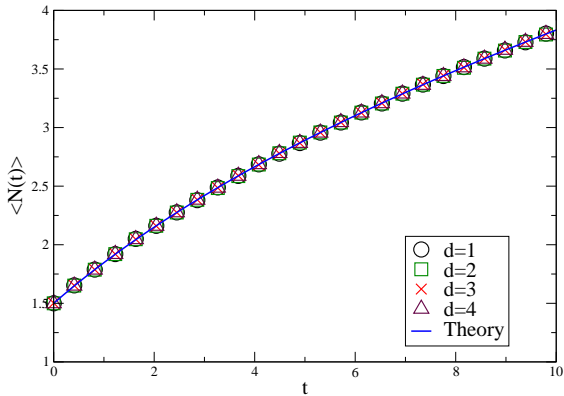}
\caption{Plot of the average number of records $\langle N(t) \rangle$ vs $t$ in the IT model. The solid line is given by the exact formula (\ref{av_record}) while the symbols represent numerical simulations in $d=1,2,3,4$ with $\gamma =1$ and $v_0 = 1$.}\label{fig_av_record}
\end{figure}

The fact that $S_2(t)=S_1(t) = S^{\rm IT}(t)$ at all $t$ is quite remarkable and is far from obvious. These results for $N=2$ and $N=3$ are plotted in Fig. \ref{Fig_s2_s3} and one sees that $S_3(t)$ exhibits a maximum at some characteristic time $t_3^*$ (actually for all $N \geq 3$, $S_N(t)$ exhibits a maximum at some characteristic time $t^*_N$ which can be shown to grow linearly with $N$ for large $N$). It seems hard to evaluate explicitly $S_N(t)$ for higher values of $N$. One can however compute the generating function $\tilde S(z,t)$ of $S^{\rm IT}_N(t)$, i.e.
\begin{eqnarray}\label{GF_S}
\tilde S(z,t) &=& \sum_{N=1}^\infty z^N S_N(t) \\
&=&  \int \frac{ds}{2\pi\,i} e^{s\, \gamma t}\, \left[ \frac{1+s}{s+ \frac{1-z}{z} \sqrt{s(1+s)}} - z\right] \;,\nonumber
\end{eqnarray}
where we have made the change of variable $s \to s/\gamma$. Clearly $S_N(t)$ is universal, i.e. independent of the dimension $d$ and the speed distribution $W(v)$. From this expression, we can compute the average number of records $\langle N(t) \rangle$ up to time $t$ and we get, for all $t$ (see also Fig. \ref{fig_av_record})
\begin{eqnarray}\label{av_record}
&&\langle N(t) \rangle\\  &=&  \frac{1}{2} e^{-\gamma t/2} \left((2 \gamma t+3) I_0\left(\frac{\gamma t}{2}\right)+(2 \gamma t+1)
   I_1\left(\frac{\gamma t}{2}\right)\right)  \;.\nonumber
\end{eqnarray}
For large $t$, it grows like $\langle N(t) \rangle \approx 2\sqrt{\gamma t}/\sqrt{\pi}$.

The Bromwich integral on the right-hand side of Eq. (\ref{GF_S}) can be computed explicitly. Skipping details, we get
\begin{eqnarray}\label{GF_S2}
\tilde S(z,t) &=& \frac{z(1-z)}{1-2z} S^{\rm IT}(t)  - \frac{z^3}{1-2z} e^{-\frac{(1-z)^2}{1-2z} \gamma t}\\
& - &\frac{z^3(1-z)}{(1-2z)^2} \gamma \int_0^{t} e^{-\frac{(1-z)^2}{1-2z}\gamma(t-t')} S^{\rm IT}(t') dt' \;,\nonumber
 \end{eqnarray}
 where $S^{\rm IT}(t)$ is given in Eq. (\ref{S1}). By setting $z=1$ in Eq. (\ref{GF_S2}), we can check the normalization condition, i.e. $\sum_{N=1}^\infty S_N(t) = \tilde S(z=1,t) = 1$, for $t>0$. 
 We can also check, by expanding the generating function in (\ref{GF_S2}) in powers of $z$ up to order $z^3$, that we recover the results for $S_N(t)$ for $N=1,2,3$ in Eqs. (\ref{S1}-\ref{S3}). For generic $N$, we can check by expanding in powers of $z$ and performing the integral over $t'$ in Eq. (\ref{GF_S2}) that, for all $N$, $S_N(t)$ has the following structure, 
\begin{eqnarray}\label{GF_S3}
S_N(t) &=&  e^{-\gamma t/2} \left(P_{0,N}(\gamma t) I_0(\gamma t/2) + P_{1,N}(\gamma t) I_1(\gamma t/2)  \right)\nonumber\\
& + & e^{-\gamma t} Q_{N}(\gamma t) \;,
\end{eqnarray}
where $P_{0,N}(x), P_{1,N}(x)$ and $Q_N(x)$ are some polynomials.

One can also extract the asymptotic behaviors of $S_N(t)$ at small and large time $t$. At small time, from Eq. (\ref{SN}), one sees that the large-$s$ behavior of the Laplace transform of $S_N(t)$ is $\sim \gamma^{N-2}\, q^N_{N-1}/s^{N-1}$, for $N \geq 2$. Using the known expression of $q^N_{N-1} = 2^{-N+1}$, from Ref. \cite{Ziff_Satya}, one obtains
\begin{equation}\label{SN_small_t}
S_N(t) \sim \frac{(\gamma t)^{N-2}}{(N-2)!} \, q^N_{N-1}= \frac{1}{2^{N-1}(N-2)!} \, (\gamma t)^{N-2} \;.
\end{equation} 
One sees explicitly that the small time behavior of $S_N(t)$ is dominated by trajectories where the RTP goes downwards at time $t$ and 
breaks a record at time $t$.  

The behavior of $S_N(t)$ for large time is easily obtained from the small-$s$ expansion of the Laplace transform in Eq. (\ref{SN_2}) and one finds, at leading order,
\begin{eqnarray} \label{SN_large_t}
S_N(t) \sim \frac{1}{\sqrt{\pi \gamma t}}  \;,
\end{eqnarray}
independently of $N$. This behavior indicates that $S_N(t)$ is dominated by the probability that, after breaking exactly $N$ lower records, the particle needs to stay above the value of the $N$ record, which, for large $t$, coincides with the survival probability $S_1(t)\sim 1/\sqrt{\pi \gamma t}$.

\section{Instantaneous-run model}\label{sec:IR}

In this section, we present a variant of the RTP model, which we will refer to as the instantaneous-run model. In this model, the particle waits a random time during a tumbling and then jumps instantaneously to its new position. We assume that the particle starts from the origin and evolves in $d$ dimensions up to time $t$. At the beginning the particle remains at the origin for a random time $T_1$, distributed according to the PDF $P_W(T)$, then it chooses a new direction uniformly at random and takes an instantaneous jump of length $v_1 T_1$ in that direction, where $v_1\geq 0$ is drawn from $W(v)$. Then, it waits a random time $T_2$, drawn from the distribution $P_W(T)$, then it jumps, and so on.

The $x$-component process obtained by projecting the motion of a particle moving according to this IR model is part of a more general class of RWs with spatio-temporal correlations, which we will call wait-then-jump walks, defined as follows. Let us consider a RW on the real line. Let
\begin{equation}
(x_{1},T_{1}),\,(x_{2},T_{2}),\ldots,(x_{j},T_{j}),\ldots
\end{equation}
be a sequence of i.i.d. pairs of random variables corresponding to the step length $x_{i}$ and the associated time $T_{i}$ (see Fig. \ref{fig:model}b)). We assume that each pair $(x_{i},T_{i})$ is distributed according to the joint PDF $p(x,T),$ which is continuous in $x$ and spatially symmetric: $p(x,T)=p(-x,T)$. After $n$ steps the RW will be in position $X_{n}$ at
time $t_{n}$, where
\begin{equation}
X_{n}=\sum_{k=1}^{n}\,x_{k},\quad\quad t_{n}=\sum_{k=1}^{n}\,T_{k}\,.\label{eq:S_n}
\end{equation}
To study the statistical properties of this class of RWs we also need to specify how the walker moves when taking a step. One possibility is that in order to take a step $x_i$ in a time $T_i$ the walker moves with constant velocity $v_i=x_i/T_i$, as in the case of the IT model. However, here we assume that the walker remains in its position for a time $T_i$ and then takes an instantaneous jump $x_i$. It turns out that for this latter wait-then-jump model many interesting properties can be computed exactly for any distribution $p(x,T)$.

Combining the SA theorem with additional combinatorial arguments, Artuso et al. recently computed exactly the survival probability of a wait-then-jump RW with arbitrary distribution $p(x,T)$ \cite{artuso14}. However, their clever technique cannot be used if the trajectory of the particle is continuous in time, as for the IT model. In this section we show that our method, presented in Section \ref{sec:IT}, turns out to be more general. Indeed, it not only provides the exact expression for the survival probability of the IT model, but it also recovers the result of \cite{artuso14} by a simpler non-combinatorial method. In addition, our technique allows us to compute exactly the distribution of the time of the maximum and the record statistics of a wait-then-jump walk with any $p(x,T)$. Note that when the RW describes the $x$-component process of an RTP with instantaneous runs, the joint distribution is given by, 
\begin{equation}\label{eq:pxt_IR}
p(x,T)=P_W(T)\int_{0}^{\infty}dv\,W(v)\,\frac{1}{v\,T}f_d\left(\frac{x}{v T}	\right)\,,
\end{equation}
where $P_W(T)$ is the waiting-time distribution, $W(v)$ is the speed distribution and $f_d(z)$ is given in Eq. (\ref{fdz.1}). Here, we first perform the computation with arbitrary $p(x,T)$ and then, using Eq. (\ref{eq:pxt_IR}), we will focus on the special case of a single RTP with instantaneous runs.

Notably, using the results obtained for the IR model, we will also be able to infer the late-time behavior of the survival probability of the IT model with non-exponential flight-time distribution $P_R(\tau)$. Indeed, the main limitation of the method presented in Section \ref{sec:surv} is that, in order to apply the SA theorem, the flight times have to be distributed exponentially. In fact, since the last running phase is not completed, our method amounts to compute the survival probability of an $n$-step RW, where the distribution of the last step differs from the others. In the special case of the running-time distribution $P_R(\tau)=\gamma e^{-\gamma \tau}$, the probability of the last running time $\tau_n$ is given by $e^{-\gamma \tau_n}$. Thus, the weight of the last flight differs from the other $(n-1)$ only by a constant prefactor $\gamma$ and the SA theorem, which requires the same distribution for each step, can still be applied. However, for a generic distribution $P_R(\tau)$ the weight of the last step is not proportional to the weight of the other steps and thus our method can not be used. Indeed, it is easy to observe from numerical simulations (see Fig. \ref{fig:late_times}) that, when $P_R(\tau)$ is not exponentially distributed, the survival probability $S^{\rm IT}(t)$ is no longer given by Eq. (\ref{surv.2}). However, choosing $P_R(t)=P_W(t)$ one can expect that at late times the IT model and the IR model behave in a qualitatively similar way. In particular, the survival probability should decay at late times as $S(t)\sim t^{-\theta}$, with the same exponent $\theta>0$ for the two models.

\subsection{Survival probability}\label{sec:surv_wait}

In this section we show that the probability $S(t)$ that a wait-then-jump walk has not visited the negative $x$ axis up to time $t$, can be computed exactly for any distribution $p(x,T)$.

Consider a trajectory of a wait-then-jump walk up to the total fixed time $t$, as in Fig \ref{fig:model}b). Let $n\geq 1$ be the number of waiting phases such that $\sum_{i=1}^n T_i=t$. The probability weight of the first $(n-1)$ intervals is $p(x_i,T_i)$, where $x_i$ is the length of the $i^{\rm th}$ jump. In contrast, the last time interval $T_n$ is not completed and no jump will be associated to this interval (see Fig. \ref{fig:model}b)). Thus, the weight of the last interval $T_n$ is given by
\begin{equation}
\int_{-\infty}^{\infty}dx\,\int_{T_n}^{\infty}dT\, p(x,T)\,,
\end{equation}
which is the probability that no jump happens in the last interval. We can now write the joint probability of the jumps $x_1,\ldots x_{n-1}$, the waiting times $T_1,\ldots T_n$, and of the number $n$ of waiting phases as
\begin{eqnarray}\label{eq:joint_xT_1}
&&P(x_1,\ldots x_{n-1},T_1,\ldots T_n,n|t)=\prod_{i=1}^{n-1}p(x_{i},T_{i})\,\\
&\times &\int_{-\infty}^{\infty}dx\,\int_{T_n}^{\infty}dT\, p(x,T)\,
\delta\left(\sum_{i=1}^{n}T_{i}-t\right),\nonumber
\end{eqnarray}
where the delta function enforces the constraint on the total time. Integrating both sides of Eq. (\ref{eq:joint_xT_1}) over the $T_i$ variables, we obtain the joint PDF of $x_1,\ldots x_{n-1}$ and of the number $n$ of waiting phases
\begin{eqnarray}
&&P(x_1,\ldots x_{n-1},n|t)=\int_{0}^{\infty}dT_1\,\ldots\int_{0}^{\infty}dT_n\,\prod_{i=1}^{n-1}p(x_{i},T_{i})\nonumber\,\\
&&\times \int_{-\infty}^{\infty}dx\,\int_{T_n}^{\infty}dT\, p(x,T)
\delta\left(\sum_{i=1}^{n}T_{i}-t\right).
\end{eqnarray}

Taking a Laplace transform with respect to $t$ we decouple the integrals over the $T_i$ variables
\begin{eqnarray}
&&\int_{0}^{\infty}dt \,e^{-st}P(x_1,\ldots x_{n-1},n|t)\\
&=&\,\prod_{i=1}^{n-1}\left(\int_{0}^{\infty}dT_i\,e^{-sT_i}p(x_{i},T_{i})\right)\,\,\nonumber\\
&\times &  \int_{0}^{\infty}dT_n\,e^{-sT_n}\int_{T_n}^{\infty}dT\, \int_{-\infty}^{\infty}dx\,p(x,T)\, .\nonumber
\end{eqnarray}
After an integration by parts, the integral over $T_n$ can be rewritten as
\begin{eqnarray}\label{eq:joint_xn1}
&&\int_{0}^{\infty}dt \,e^{-st}P(x_1,\ldots x_{n-1},n|t)\\
&=&\,\prod_{i=1}^{n-1}\left(\int_{0}^{\infty}dT_i\,p(x_{i},T_{i})\,e^{-sT_i}\right)\,\,\nonumber\\
&\times & \frac{1}{s}\left(1-\int_{0}^{\infty}dT_n\,\int_{-\infty}^{\infty}dx\,p(x,T_n)e^{-sT_n}\right)\,.\nonumber
\end{eqnarray}
It is useful to rewrite Eq. (\ref{eq:joint_xn1}) as
\begin{eqnarray}
\label{eq:joint_xn2}
&&\int_{0}^{\infty}dt \,e^{-st}P(x_1,\ldots x_{n-1},n|t)\\ &=&\frac{1-c(s)}{s}c(s)^{n-1}\prod_{i=1}^{n-1}\tilde{p}_s(x_i)\,,	\nonumber
\end{eqnarray}
where $c(s)$ is defined as
\begin{equation}\label{eq:c_s}
c(s)=\int_0^{\infty}dT\,\int_{-\infty}^{\infty}dx\,p(x,T)\,e^{-sT}
\end{equation}
 and 
\begin{equation}\label{eq:psx_wtj}
\tilde{p}_{s}(x)=\frac{1}{c(s)}\int_{0}^{\infty}dT\,p(x,T)\,e^{-sT}\,.
\end{equation}
Note that $\tilde{p}_{s}(x)$ can be interpreted as a PDF. Indeed, it is clearly non-negative and normalized to unity. Moreover, since we assume $p(x,T)$ to be continuous and symmetric
with respect to $x$, $\tilde{p}_{s}(x)$ will also be continuous and symmetric.
Finally, performing a formal Laplace inversion in Eq. (\ref{eq:joint_xn2}), we obtain the joint distribution of the jumps $x_1,\ldots x_{n-1}$ and of the number $n$ of waiting phases 
\begin{equation}\label{eq:joint_xn_final}
P(x_1,\ldots x_{n-1},n|t)=\int\,\frac{ds}{2\pi i}e^{st}\frac{1-c(s)}{s}c(s)^{n-1}\prod_{i=1}^{n-1}\tilde{p}_s(x_i)\,\,
\end{equation}
where the integral is over the Bromwich contour in the complex $s$ plane.

The exact result in Eq. (\ref{eq:joint_xn_final}) can be used to compute the survival probability $S(t)$, which is the probability that the positions $X_1,\ldots X_{n-1}$ of the walker after each jump are all positive. Thus, summing over $n\geq 1$, $S(t)$ can be written as
\begin{eqnarray}\label{eq:S_wait_1}
S(t)&=&\sum_{n=1}^{\infty}\int_{-\infty}^{\infty}dx_1\ldots\int_{-\infty}^{\infty}dx_{n-1}\,P(x_1,\ldots x_{n-1},n|t)\nonumber\\
&\times & \theta(X_1)\ldots\theta(X_{n-1})\,,
\end{eqnarray}
where the product of theta function enforces the walker to remain on the positive side. Plugging the expression for $P(x_1,\ldots x_{n-1},n|t)$, given in Eq. (\ref{eq:joint_xn_final}), into Eq. (\ref{eq:S_wait_1}), we obtain
\begin{eqnarray}\label{eq:S_wait_2}
S(t)&=&\int \frac{ds}{2\pi i}e^{st}\frac{1-c(s)}{s}\sum_{n=1}^{\infty}c(s)^{n-1}q_{n-1}\,,
\end{eqnarray}
where
\begin{equation}
q_{n}=\int_{-\infty}^{\infty}dx_1\ldots\int_{-\infty}^{\infty}dx_{n}\,\prod_{i=1}^{n}\tilde{p}_s(x_i) \theta(X_1)\ldots\theta(X_{n})\,.
\end{equation}
Similarly to what we have done in Section \ref{sec:surv}, since $\tilde{p}_s(x_i)$ is continuous and symmetric, $q_n$ can be interpreted as the survival probability of a discrete-time RW with continuous and symmetric jumps. Thus, as consequence of the SA theorem, the probability $q_n$ is completely independent of the particular form of $\tilde{p}_s(x_i)$ and its generating function is given by \cite{SA_54}
\begin{equation}\label{eq:SA_wait}
\sum_{n=0}^{\infty}z^n\,q_n=\frac{1}{\sqrt{1-z}}\,.
\end{equation}
Thus, using this relation (\ref{eq:SA_wait}), we can rewrite Eq. (\ref{eq:S_wait_2}) as
\begin{equation}\label{surv_final}
S(t)=\int \frac{ds}{2\pi i}e^{st}\frac{\sqrt{1-c(s)}}{s}\,.
\end{equation}
This is indeed the result of Artuso et. al.~\cite{artuso14}
obtained originally using a combinatorial method. Our derivation above is non-combinatorial and a bit simpler in our opinion.

\begin{figure}[t]
\includegraphics[width=0.5\textwidth]{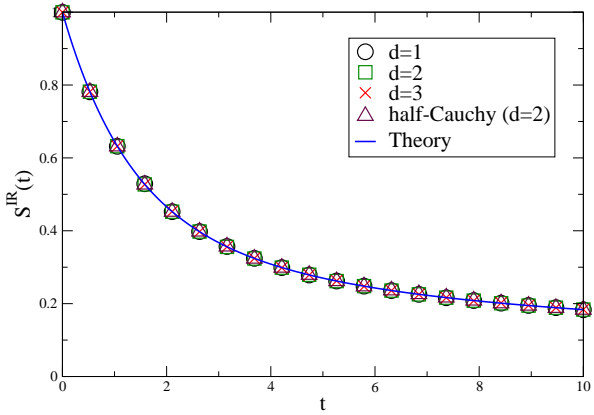} 
\caption{\label{fig:S_wait} Survival probability $S^{\rm IR}(t)$ of a single RTP in the IR model as a function of $t$ for $a=1$. The continuous blue line corresponds to the exact result in Eq. (\ref{surv_final.1}). The symbols correspond to simulations with the choices $d = 1, 2, 3$ with $W(v)=\delta(v-1)$ and $d = 2$ with $W(v) =2/(\pi\left(1+v^2\right))$, for $v > 0$ (half-Cauchy). The numerical curves collapse on the corresponding analytical blue line for all $t$.}
\end{figure}

We now want to study the IR model. Thus, we choose 
\begin{equation}\label{eq:pxt_vel}
p(x,T)= P_W(T)\,\int_{0}^{\infty}dv\,W(v)\frac{1}{vT}f_d\left(\frac{x}{vT}\right)\,,
\end{equation}
where $f_d(z)$ is given in Eq. (\ref{fdz.1}). Plugging this expression for $p(x,T)$ into Eq. (\ref{eq:c_s}), we get
\begin{eqnarray}
c(s)&=&\int_{0}^{\infty}dT\,e^{-sT}P_W(T)\\&\times &\int_{0}^{\infty}dv\,W(v)\int_{-\infty}^{\infty}dx\,\frac{1}{vT}f_d\left(\frac{x}{vT}\right)\,. \nonumber
\end{eqnarray}
Performing the change of variable $x\to z=x/(vT)$, we obtain
\begin{equation}
c(s)=\int_{0}^{\infty}dT\,e^{-sT}P_W(T)\int_{0}^{\infty}dv\,W(v)\int_{-\infty}^{\infty}dz\,f_d\left(z\right)\,.
\end{equation}
Using the fact that $W(v)$ and $f_d(z)$ are normalized to one, we find that
\begin{eqnarray}
c(s)= \int_{0}^{\infty}dT\,e^{-sT}P_W(T)\,=\tilde{P}_W(s)\,.
\end{eqnarray}
Note that $\tilde{P}_W(s)$ is simply defined as the Laplace transform of $P_W(T)$. Then, using Eq. (\ref{surv_final}), we obtain that the survival probability in the case of the IR model is given by
\begin{equation}
S^{\rm IR}(t)=\int \frac{ds}{2\pi i}e^{st}\frac{\sqrt{1-\tilde{P}_W(s)}}{s}\,.
\label{surv_final_rtm}
\end{equation}
Note that, for any waiting time distribution $P_W(T)$, the survival probability does not depend on the dimension $d$ nor on the distribution $W(v)$.

In the most relevant case of an exponential distribution $P_W(T)=a\,e^{-a T}$ one obtains that $\tilde{P}_W(s)= a/(a+s)$. Consequently, Eq. (\ref{surv_final_rtm}) gives
\begin{equation}\label{eq:S_IR_LT}
S^{\rm IR}(t)=\int \frac{ds}{2\pi i}e^{st}\frac{1}{\sqrt{s(a+s)}}\,.
\end{equation}
One can invert the Laplace transform using the inversion formula \cite{schiff_book}
\begin{eqnarray}\label{eq:laplace_inversion_2}
\mathcal{L}^{-1}_{s\to t}\left(\sqrt{\frac{1}{s(s+b)}}\right)\left(t\right)&=& e^{-\frac{b}{2} t}I_0\left(\frac{b}{2}t\right)\,.
\end{eqnarray}
Thus, we obtain that the exact survival probability at all $t$ for this specific IR model with exponential time distribution is given by
\begin{equation}
S^{\rm IR}(t)= e^{-a\, t/2}\, I_0\left(\frac{a\, t}{2}\right)\, ,
\label{surv_final.1}
\end{equation}
where $I_0(z)$ is again the modified Bessel functions. The result in (\ref{surv_final.1}) is manifestly different from the IT result in Eq.~(\ref{surv.2}). This clearly shows that the exact result in Eq. (\ref{surv_final}) for the wait-then-jump walks can not be used to derive our main result for the RTP in the IT setup. Note, however, that for late times the result in Eq. (\ref{surv_final.1}) has the same asymptotic behavior as the RTP result, namely $S^{\rm IR}(t)\sim 
1/\sqrt{\pi a t}$. In Fig. \ref{fig:S_wait} we observe that the exact result in Eq. (\ref{surv_final.1}) is in excellent agreement with numerical simulations for different choices of $d$ and $W(v)$.

\begin{figure}[t]
\includegraphics[width=0.45\textwidth]{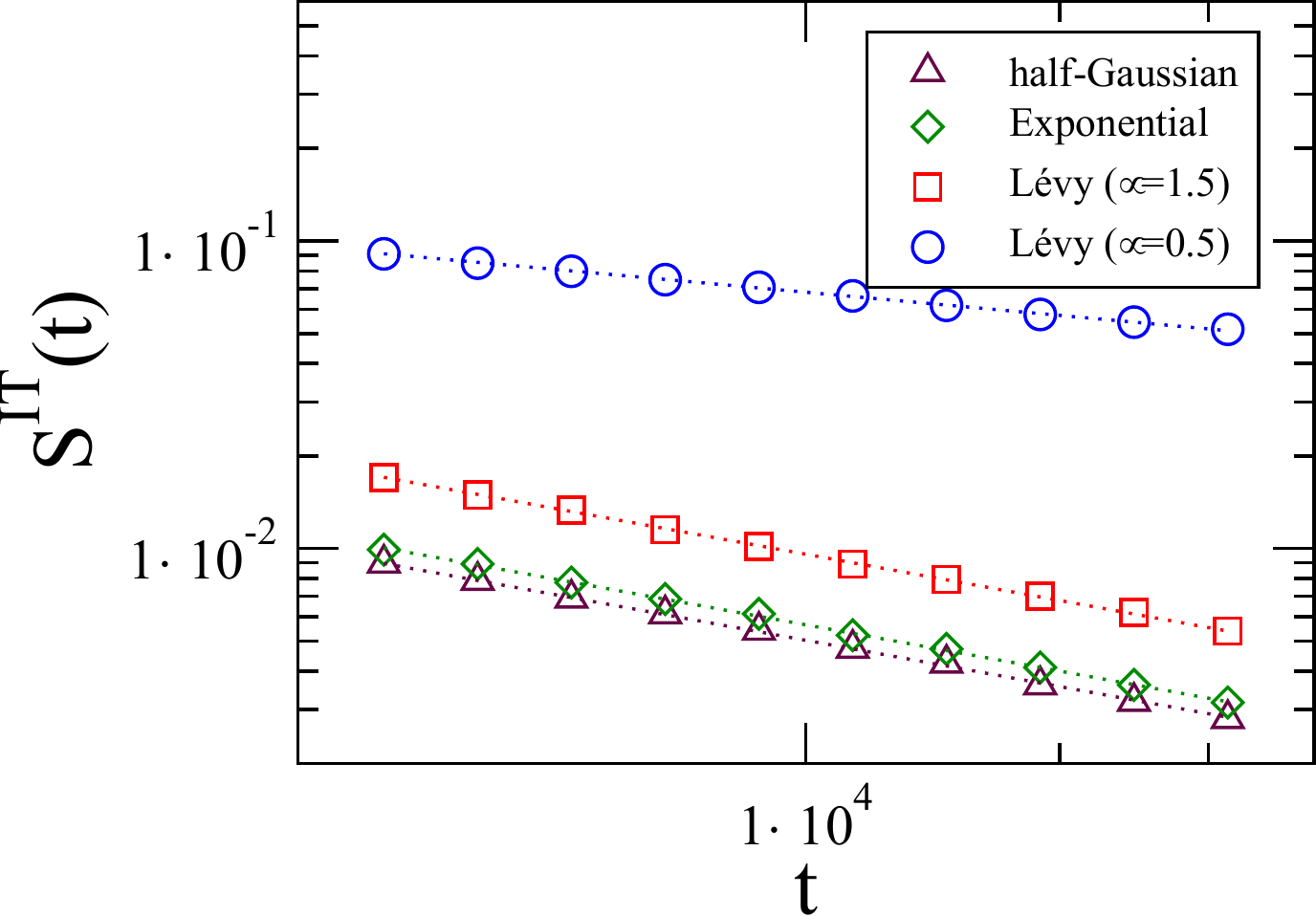} 
\caption{\label{fig:late_times}
Numerical computation of $S^{\rm IT}(t)$ in $d=2$ for the IT model with different distributions $P_R(\tau)$: (i) half-Gaussian, (ii) exponential, (iii)  asymmetric L\'evy distribution with L\'evy index $\mu = 3/2$ and (iv) asymmetric L\'evy with $\mu =1/2$. In all these cases, $S^{\rm IT}(t) \sim t^{-\theta}$ for large $t$ with $\theta = \frac{1}{2}$ in cases (i)-(iii) and $\theta = \mu/2 = 1/4$ for case (iv) corresponding to $\mu=1/2$.}
\end{figure}

Moreover, Eq. (\ref{surv_final_rtm}) can be useful to compute the late time behavior of $S^{\rm IT}(t)$ for the IT model with a generic time distribution $P_R(\tau)$. Indeed, one expects that $S(t)\sim t^{-\theta}$ when $t\to\infty$. Moreover, choosing $P_R(t)=P_W(t)$, for late times, it is natural to conjecture that the exponent $\theta$ is the same for the IT model and for the IR model. Here, we compute the exponent $\theta$ for different time distributions $P_W(T)$ in the IR setup. It is useful to distinguish two cases, depending on whether $P_W(T)$ has a well-defined first moment or not. 
\vspace*{0.3cm}

\noindent{\it The case where $P_W(T)$ has a well-defined first moment.} In this case, the Laplace transform $\tilde{P}_W(s)$ can be expanded, for small $s$, as
\begin{equation}\label{eq:laplace_p_finite}
\tilde{P}_W(s)\simeq 1- \langle T \rangle \,s+o(s)\,,
\end{equation}
where $\langle T \rangle=\int_{0}^\infty\,dT\,T\,P_W(T)$ is the first moment of $T$. Using Eq. (\ref{surv_final_rtm}) we obtain that, for small $s$
\begin{equation}
S^{\rm IR}(t)\sim \int \frac{ds}{2\pi i}e^{st}\sqrt{\frac{\langle T\rangle}{s}}\,.
\end{equation}
Inverting the Laplace transform gives, for late times,
\begin{equation}\label{S_av_T}
S^{\rm IR}(t)\sim \sqrt{\frac{\langle T \rangle}{\pi \, t}}\,.
\end{equation}
Hence, if $\langle T \rangle$ is finite we obtain that $\theta=1/2$. Note that for the exponential jump distribution with rate $a$, one has $\langle T \rangle = 1/a$ and this formula (\ref{S_av_T}) yields back $S^{\rm IR}(t) \sim 1/\sqrt{\pi a t}$, as it should.

\vspace*{0.3cm}

\noindent{\it The case where $P_W(T)$ has a diverging first moment.} If the average value of $T$ is diverging, i.e. if $P_W(T)\sim T^{-\mu-1}$ for $T\to\infty$ with $0<\mu<1$ (in the $1d$ case this corresponds to L\'evy walks, see e.g. \cite{metzler}), then $\tilde{P}_W(s)$ can be expanded for small $s$ as
\begin{equation}\label{eq:laplace_p_diverging}
\tilde{P}_W(s) =  1-(b\,s)^\mu+o(s^\mu)\;,
\end{equation}
where $b$ denotes a microscopic time scale. Using Eq. (\ref{surv_final_rtm}) we obtain that, when $s\to 0$,
\begin{equation}
S^{\rm IR}(t)\sim \int \frac{ds}{2\pi i}e^{st} s^{\mu/2-1}\,.
\end{equation}
Inverting the Laplace transform we get that when $t\to\infty$
\begin{equation}
S^{\rm IR}(t)\sim t^{-\mu/2}\,,
\end{equation}
and, hence, in this case $\theta=\mu/2$.

One can then conjecture that the late time behavior of the IR model is qualitatively similar to the one of the IT model, i.e. that $S^{\rm IT}(t)\sim S^{\rm IR}(t)$ for large $t$. Thus, we expect that if $\langle \tau \rangle =\int_{0}^{\infty}d\tau\,\tau P_R(\tau)$ is finite the decay exponent is $\theta=1/2$. On the other hand, when $P_R(\tau)\sim 1/\tau^{\mu+1}$ for $\tau\gg 1$ and $0<\mu<1$, the average running time $\langle \tau\rangle $ diverges and the decay exponent is $\theta=\mu/2$. In Fig. \ref{fig:late_times}, we show the results of numerical simulations of the IT model with different running-time distributions $P_R(\tau)$. We observe that the exponents computed in the IR setup describe well the late time behavior of $S^{\rm IT}(t)$ for the IT model. Note also that from Fig. \ref{fig:late_times} it is clear that the survival probability $S^{\rm IT}(t)$ for the IT model ceases to be universal if the distribution $P_R(\tau)$ is not exponential.

\subsection{Time to reach the maximum}\label{sec:t_max_wait}

\begin{figure}[t]
\includegraphics[width=0.5\textwidth]{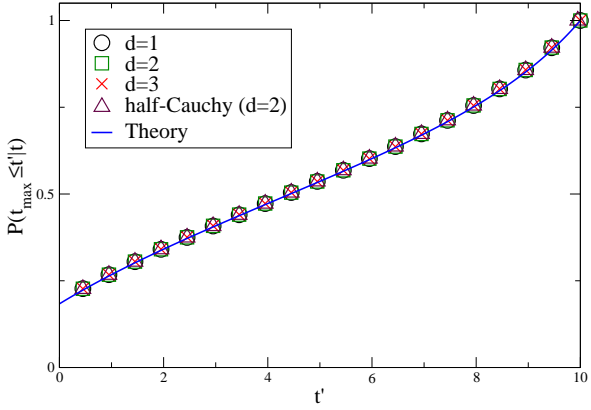} 
\caption{\label{fig:t_max_wait} 
Cumulative probability $P(t_{\max}\leq t'|t)$ under the IR model as a function of $t'$, with $a=1$ and $t=10$. The continuous blue line corresponds to the exact result in Eq. (\ref{eq:cumulative_wait}). The symbols correspond to simulations with the choices $d = 1, 2, 3$ with $W(v)=\delta(v-1)$ and $d = 2$ with $W(v) =2/(\pi\left(1+v^2\right))$, for $v > 0$
(half-Cauchy). We observe that the numerical curves collapse on the corresponding analytical blue line.}
\end{figure}

In this section we will first compute exactly the distribution of the time $t_{\max}$ of the global maximum of a wait-then-jump RW with a generic distribution $p(x,T)$, fixing the total time $t$. Then, we will focus on the particular case of the IR model for a single RTP. Since the walker does not move between two steps, to avoid degeneracies, it is useful to define $t_{\max}$ as the time at which the maximum is reached \emph{for the first time}. In order to compute the probability distribution $P(t_{\max}|t)$ of the time $t_{\max}$ given the total time $t$, we will follow the method presented in Section \ref{sec:tmax}. First of all, when $0<t_{\max}<t$, we can always split the interval $[0,t]$ into two subintervals: $[0,t_{\max}]$ (I) and $[t_{\max},t]$ (II). Note that, since $t_{\max}$ will always be the time of a jump, the two intervals (I) and (II) are independent. Thus, the distribution of $t_{\max}$ will be simply given by the product of the weights $P_{\rm I}(t_{\max})$ and $P_{\rm II}(t-t_{\max})$ of the two intervals. The cases $t_{\max}=0$ and $t_{\max}=t$ will be considered separately.

Let us first look at the interval (I) and define $n_1$ as the number of jumps, including the one at time $t_{\max}$, in the interval $[0,t_{\max}]$. Note that when $0<t_{\max}<t$ we have $n_1\geq 1$ and that $n_1$ is also the number of waiting phases in the first interval. In order to compute the probability weight of the first interval, we first need to compute the joint PDF of the jumps $x_1,\ldots x_{n_1}$ and of $n_1$. Note that in this case the time intervals $T_1,\ldots T_{n_1}$ are all completed, since the walker is jumping at time $t_{\max}$. Thus, the joint PDF of the pairs $\{(x_i,T_i)\}_{i=1}^{n_1}$ and of $n_1$, fixing the total time $t_{\max}$, is given by
\begin{equation}
P(\{(x_i,T_i)\},n_1|t_{\max})=\prod_{i=1}^{n_1}p(x_i,T_i)\delta(\sum_{i=1}^{n_1}T_i-t_{\max})\,.
\end{equation}
Integrating over the $T$ variables, we get
\begin{equation}
P(\{x_i\},n_1|t_{\max})=\prod_{i=1}^{n_1}\int_{0}^{\infty}dT_i\,p(x_i,T_i)\delta(\sum_{i=1}^{n_1}T_i-t_{\max})\,.
\end{equation}
We perform a Laplace transform with respect to $t_{\max}$ in order to decouple the integrals over the $T$ variables, and we obtain
\begin{equation}\label{eq:joint_x_n1_w}
\int_{0}^{\infty}dt_{\max}\,e^{-st_{\max}}P(\{x_i\},n_1|t_{\max})=c(s)^{n_1}\prod_{i=1}^{n_1}\tilde{p}_{s}(x_{i})\,,
\end{equation}
where $c(s)$ is given in Eq. (\ref{eq:c_s}) and $\tilde{p}_s(x)$ is given in Eq. (\ref{eq:psx_wtj}). Inverting the Laplace transform in Eq. (\ref{eq:joint_x_n1_w}) formally, we obtain
\begin{equation}\label{eq:joint_x_n1_wait}
P(x_1,\ldots x_{n_1},n_1|t_{\max})=\int \frac{ds}{2\pi i}e^{st_{\max}} c(s)^{n_1}\prod_{i=1}^{n_1}\tilde{p}_{s}(x_{i})\,.
\end{equation}

In the first segment, the walker has to reach the maximal value at time $t_{\max}$, thus the probability weight of the first interval can be written as, summing over $n_1\geq 1$
\begin{eqnarray}
&&P_{\rm I}(t_{\max})=\sum_{n_1=1}^{\infty}\int_{-\infty}^{\infty}dx_1\,\ldots\int_{-\infty}^{\infty} dx_{n_1}\\
&\times &P(x_1,\ldots x_{n_1},n_1|t_{\max})\theta(X_{n_1})\nonumber\\
&\times &\theta(X_{n_1}-X_{n_1-1})\ldots\theta(X_{n_1}-X_{1})\,,\nonumber
\end{eqnarray}
where $X_k=x_1+x_2+\ldots +x_k$. Using the expression for $P(x_1,\ldots x_{n_1},n_1|t_{\max})$, given in Eq. (\ref{eq:joint_x_n1_wait}), we obtain
\begin{equation}\label{eq:P_I_LT_wait}
P_{\rm I}(t_{\max})=\sum_{n_1=1}^{\infty}\int \frac{ds}{2\pi i}e^{st_{\max}} c(s)^{n_1}q_{n_1}\,,
\end{equation}
where 
\begin{equation}
q_{n_1}=\int_{-\infty}^{\infty}dx_1\,\ldots\int_{-\infty}^{\infty}dx_{n_1}\,\prod_{i=i}^{{n_1}}\tilde{p}_s(x_i)\theta(X_{n_1}-X_{n_1-i})\,.
\end{equation}
In Section \ref{sec:tmax} we have shown that when $\tilde{p}_s(x)$ is continuous and symmetric, $q_{n_1}$ is universal and that its generating function is given by Eq. (\ref{gen_fun_tmax}). Thus, using Eqs. (\ref{gen_fun_tmax}) and (\ref{eq:P_I_LT_wait}), we obtain
\begin{equation}\label{eq:P_I_wait}
P_{\rm I}(t_{\max})=\int \frac{ds}{2\pi i}e^{st_{\max}}\left(\frac{1}{\sqrt{1-c(s)}}-1\right)\,,
\end{equation}
where $c(s)$ is given in Eq. (\ref{eq:c_s}).

In the second segment $[t_{\max},t]$ the walker starts from position $X_{n_1}$ and has to remain below this position up to time $t$. Performing the translation $x\to x-X_{n_1}$, followed by the reflection $x\to -x$, it becomes clear that the weight of the second segment is given by
\begin{equation}
P_{\rm II}(t-t_{\max})=S(t-t_{\max})\,,
\end{equation}
where $S(t)$ is the survival probability of the wait-then-jump model, given in Eq. (\ref{surv_final}). Thus, using Eq. (\ref{surv_final}) we obtain
\begin{equation}\label{eq:P_II_wait}
P_{\rm II}(t-t_{\max})=\int \frac{ds}{2\pi i}e^{s(t-t_{\max})}\frac{\sqrt{1-c(s)}}{s}\,,
\end{equation}
where $c(s)$ is given in Eq. (\ref{eq:c_s}). Finally, the distribution of $t_{\max}$ is given by the product of the two factors
\begin{equation}\label{eq:p_tmax_final_wait}
P(t_{\max}|t)=P_{\rm I}(t_{\max})P_{\rm II}(t-t_{\max})\,,
\end{equation}
which is valid for $0<t_{\max}<t$.

Now, we need to consider the contributions of the events $t_{\max}=t$ and $t_{\max}=0$. First, it is clear that the event $t_{\max}=t$ can only happen if the time $t$ is the time of a jump, which happens with zero probability if $p(x,T)$ is continuous in $T$. On the other hand, $t_{\max}=0$ if the walker remains always in the negative side. Thus, using the $x\to -x$ symmetry, we get
\begin{equation}
{\rm Prob.}(t_{\max}=0)=S(t)\,,
\end{equation}
where $S(t)$ is the survival probability, given in Eq. (\ref{surv_final}).
Overall, we obtain that for any $t$ and for $0\leq t_{\max}\leq t$
\begin{equation}\label{eq:P_tmax_wait}
P(t_{\max}|t)=P_{\rm I}(t_{\max})P_{\rm II}(t-t_{\max})+P_{\rm II}(t)\delta(t_{\max})\,,
\end{equation}
where $P_{\rm I}(t)$ and $P_{\rm II}(t)$ are given in Eqs. (\ref{eq:P_I_wait}) and (\ref{eq:P_II_wait}).

We now want to check that the PDF $P(t_{\max}|t)$ is normalized to one. First of all, we perform a Laplace transform with respect to $t_{\max}$ and $t$ on both sides of Eq. (\ref{eq:P_tmax_wait}) and we obtain
\begin{eqnarray}
&&\int_{0}^{\infty}dt\,\int_{0}^{t}dt_{\max}\,P(t_{\max}|t)e^{-st-s_1 t_{\max}}\\
&=&\tilde{P}_{\rm I}(s+s_1)\tilde{P}_{\rm II}(s)+\tilde{P}_{\rm II}(s)\,,\nonumber
\end{eqnarray}
where $\tilde{P}_{\rm I}(s)$ and $\tilde{P}_{\rm II}(s)$ are the Laplace transforms of $P_{\rm I}(t)$ and $P_{\rm II}(t)$. Plugging the expressions for $P_{\rm I}(t)$ and $P_{\rm II}(t)$, given in Eqs. (\ref{eq:P_I_wait}) and (\ref{eq:P_II_wait}), we obtain, after few steps of algebra
\begin{eqnarray}\label{eq:normalization_wait}
&&\int_{0}^{\infty}dt\,\int_{0}^{t}dt_{\max}\,P(t_{\max}|t)e^{-st-s_1 t_{\max}}\\
&=&\frac{1}{s}\sqrt{\frac{1-c(s)}{1-c(s+s_1)}}\,.\nonumber
\end{eqnarray}
Setting $s_1=0$ on both sides of Eq. (\ref{eq:normalization_wait}), we obtain
\begin{equation}
\int_{0}^{\infty}dt\,\int_{0}^{t}dt_{\max}\,P(t_{\max}|t)e^{-st}=\frac{1}{s}\,.
\end{equation}
Inverting the Laplace transform with respect to $s$, we get that for all $t$
\begin{equation}
\int_{0}^{t}dt_{\max}\,P(t_{\max}|t)=1\,.
\end{equation}
Thus, the PDF $P(t_{\max}|t)$, given in Eq. (\ref{eq:P_tmax_wait}), is correctly normalized to one.

In the case of a wait-then-jump RTP with waiting-time distribution $P_W(T)$ and speed distribution $W(v)$, we just need to choose the joint distribution to be 
\begin{equation}
p(x,T)=P_W(T)\int_{0}^{\infty}dv\,W(v)\,\frac{1}{v T}f_d\left(\frac{x}{vT}\right)\,,
\end{equation}
where $f_d(z)$ is given in Eq. (\ref{fdz.1}). In the previous section, we have shown that, for this choice of $p(x,T)$ one obtains
\begin{equation}
c(s)=\tilde{P}_W(s)\,,
\end{equation}
where $\tilde{P}_W(s)$ is the Laplace transform of the time distribution $P_W(T)$. In the most relevant case of exponentially distributed waiting times, i.e. $P_W(T)=a e^{-a T}$, one obtains
\begin{equation}
c(s)=\frac{a}{a+s}\,.
\end{equation}
Plugging this expression into Eq. (\ref{eq:P_I_wait}) we obtain that 
\begin{equation}
P_{\rm I}(t)=\int \frac{ds}{2\pi i}e^{st_{\max}}\left(\sqrt{\frac{a+s}{s}}-1\right)\,.
\end{equation}
Using Eq. (\ref{eq:laplace_inversion_1}) to invert the Laplace transform, we get
\begin{equation}\label{eq:P_I_wait_final}
P_{\rm I}(t)=\frac{a}{2}e^{-a t/2}\left(I_0\left(\frac{a}{2}t\right)+I_1\left(\frac{a}{2}t\right)\right)\,.
\end{equation}
Similarly, one also finds
\begin{equation}\label{eq:P_II_wait_final}
P_{\rm II}(t)=e^{-a t/2}I_0\left(\frac{a}{2}t\right)\,.
\end{equation}
The cumulative distribution of $t_{\max}$ can be obtained from Eq. (\ref{eq:p_tmax_final_wait}) and is given by
\begin{equation}\label{eq:cumulative_wait}
P(t_{\max}\leq t'|t)=P_{\rm II}(t)+\int_{0}^{t'}dt_{\max}\,P_{\rm I}(t_{\max})P_{\rm II}(t-t_{\max})\,,
\end{equation}
where $P_{\rm I}(t)$ and $P_{\rm II}(t)$ are given in Eqs. (\ref{eq:P_I_wait_final}) and (\ref{eq:P_II_wait_final}). This exact result in Eq. (\ref{eq:cumulative_wait}) is shown in Fig. \ref{fig:t_max_wait}, where we observe that the agreement with numerical simulations is excellent.

\subsection{Record statistics}\label{sec:record_wait}

\begin{figure*}[t]
 \includegraphics[width = \linewidth]{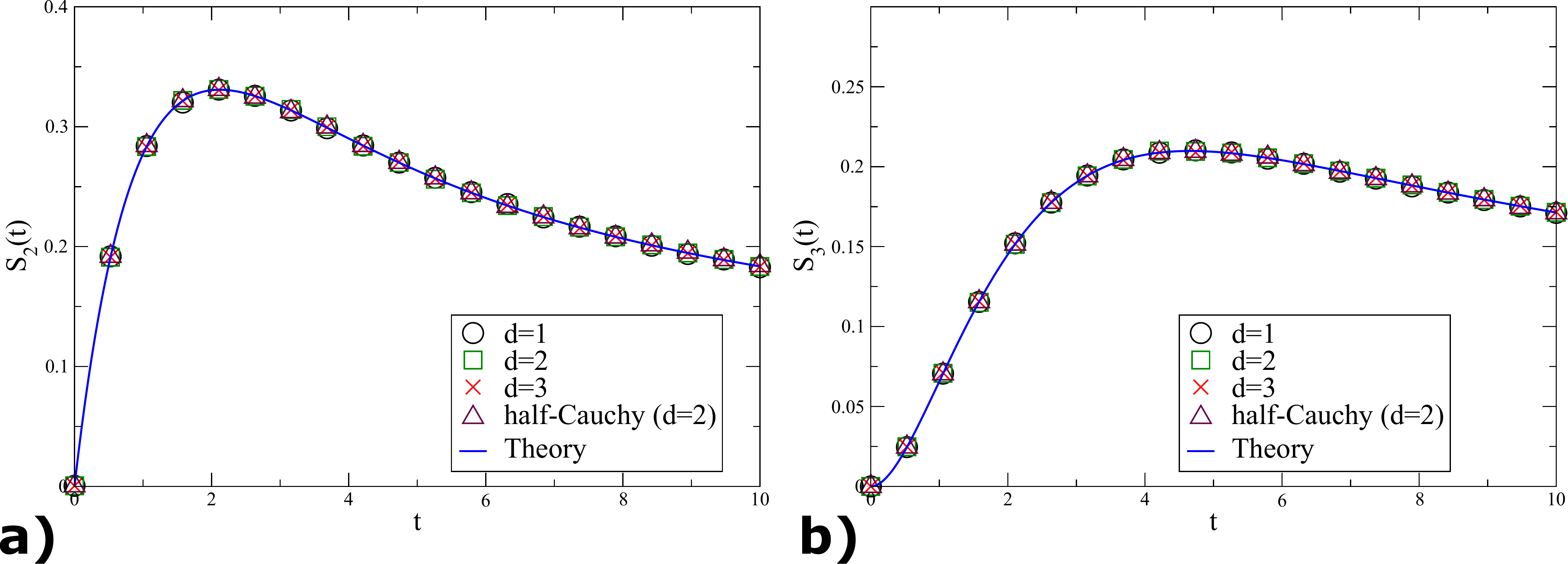}
\caption{Plot of $S_2(t)$ in (a) and $S_3(t)$ in (b) 
as functions of $t$ for the IR model, with waiting rate $a=1$. The continuous blue lines correspond to the exact result for $S_2(t)$ (a) and $S_3(t)$ (b), given in Eqs. (\ref{eq:S2t_wait}) and (\ref{eq:S3t_wait}). The symbols correspond to simulations with the choices $d = 1, 2, 3$ with $W(v)=\delta(v-1)$ and $d = 2$ with $W(v) =2/(\pi\left(1+v^2\right))$, for $v > 0$
(half-Cauchy). The numerical curves collapse on the corresponding analytical blue line for all $t$.}\label{fig:s2_s3_wait}
\end{figure*}

In this section, we investigate the record statistics of a wait-then-jump random walk with a generic distribution $p(x,T)$, following the steps presented in Section \ref{sec:record}. Then, we will use Eq. (\ref{eq:pxt_IR}) to study the case of the RTP under the IR model. We recall that the position $X_i$ at step $i$ is a lower record if it is lower than all the previous positions, i.e. if $X_j>X_i$ for all $0\leq j<i$. We adopt the convention that the starting point $X_0=0$ is also a record. As done in Section \ref{sec:record}, we compute the probability $S_N(t)$ that there are exactly $N$ lower records up to time $t$. 

Using the expression for the joint distribution of the jumps $x_1,\ldots x_n$ and of the number $n$ of waiting phases up to time $t$, given in Eq. (\ref{eq:joint_xn_final}), following the steps outlined in Section \ref{sec:record}, one can find that the probability $S_N(t)$ is given by
\begin{equation}\label{eq:SNt_LT}
S_N(t)=\int \frac{ds}{2\pi i}e^{st} \frac{1-c(s)}{s}\sum_{n=N-1}^{\infty}c(s)^{n} q_{n}^{N}\,,
\end{equation}
where  $c(s)$ is given in Eq. (\ref{eq:c_s}) and $q_{n}^{N}$ is the probability that, for a RW with continuous and symmetric increments, there are exactly $N$ lower record before step $n$. As stated in Section \ref{sec:record}, $q_n^N$ is completely universal and its generating function with respect to $n$ is given in Eq. (\ref{GF_record_nber}). Thus, using Eq. (\ref{GF_record_nber}) in Eq. (\ref{eq:SNt_LT}), we obtain
\begin{equation}\label{eq:SNt_wait}
S_N(t)=\int \frac{ds}{2\pi i}e^{st}\frac{1}{s}\left(1-\sqrt{1-c(s)}\right)^{N-1}\sqrt{1-c(s)}\,.
\end{equation}
Note that, as expected, that $S_1(t)=S(t)$, where $S(t)$ is the survival probability given in Eq. (\ref{surv_final}). Indeed, since the starting point is counted as a record, the number of records will be one if and only if the walker does not visit the negative side of the $x$ axis up to time $t$.

It is also useful to compute the generating function of $S_N(t)$ with respect to $N$, defined as
\begin{equation}
\tilde{S}(t,z)=\sum_{N=1}^{\infty}S_N(t)z^N\,.
\end{equation}
Using the expression for $S_N(t)$, given in Eq. (\ref{eq:SNt_wait}), we obtain
\begin{equation}\label{eq:Stz_wait}
\tilde{S}(t,z)=\int \frac{ds}{2\pi i}e^{st}\frac{1}{s}\frac{z\sqrt{1-c(s)}}{1-\left(1-\sqrt{1-c(s)}\right)z}\,.
\end{equation}
From $\tilde{S}(t,z)$ one can also obtain the average number of records up to time $t$. Indeed, differentiating Eq. (\ref{eq:Stz_wait}) with respect to $z$ and then setting $z=1$, we obtain
\begin{equation}\label{eq:avg_N_wait}
\langle N(t) \rangle=\int \frac{ds}{2\pi i}e^{st}\frac{1}{s\sqrt{1-c(s)}}\,,
\end{equation}
where $c(s)$ is given in Eq. (\ref{eq:c_s}).

In the case where the RW is the $x$-component process of an RTP in the IR model, we have already shown that $c(s)=\tilde{P}_W(s)$, where $\tilde{P}_W(s)$ is the Laplace transform of the distribution $P_W(T)$ of the waiting times. Note that this is true also when the velocity associated to each jump is drawn from a generic distribution $W(v)$. In the case of exponentially distributed waiting times with rate $a$ one has $c(s)=a /(a+s)$ and it is possible to find an explicit expression for some of the quantities computed above. Indeed, let us first consider the probability $S_N(t)$. Setting $c(s)=a/(a+s)$ in Eq. (\ref{eq:SNt_wait}), we obtain
\begin{equation}\label{eq:SNt_wait_exp}
S_N(t)=\int \frac{ds}{2\pi i}e^{st}\frac{1}{s}\left(1-\sqrt{\frac{s}{a+s}}\right)^{N-1}\sqrt{\frac{s}{a+s}}\,.
\end{equation}
For $N=2$, using Eq. (\ref{eq:laplace_inversion_2}), one can invert the Laplace transform and we obtain
\begin{equation}\label{eq:S2t_wait}
S_2(t)=e^{-a t/2}I_0\left(\frac{a}{2}t\right)-e^{-a t}\,.
\end{equation}
Similarly, for $N=3$, we obtain, using Eq. (\ref{eq:laplace_inversion_2}) and convolution theorem,
\begin{eqnarray}\label{eq:S3t_wait}
&&S_3(t)=e^{-a t/2}I_0\left(\frac{a}{2}t\right)-e^{\-a t}\\
&+&\frac{a}{2}e^{-a t}\int_{0}^{t}dt'\,\left(I_1\left(\frac{a}{2}t'\right)-I_0\left(\frac{a}{2}t'\right)\right)\,.\nonumber
\end{eqnarray}
The probabilities $S_2(t)$ and $S_3(t)$, shown in Fig. \ref{fig:s2_s3_wait}, are completely independent of the dimension $d$ and of the speed distribution $W(v)$. We observe that $S_2(t)$ and $S_3(t)$ assume their maximal value at the characteristic times $t^*_2$ and $t^*_3$. One can show that for any $N\geq 2$, $S_N(t)$ will reach its maximum at the characteristic time $t^*_N$, which can be shown to increase linearly with $N$. Computing explicitly $S_N(t)$ gets increasingly complicated for $N\geq 3$. However, one can compute the behavior of $S_N(t)$ for short and late times. Expanding Eq. (\ref{eq:SNt_wait_exp}) for large values of $s$ and inverting the Laplace transform, we obtain that for $t\to 0$
\begin{equation}
S_N(t)\simeq \frac{1}{(N-1)!}\left(\frac{a t}{2}\right)^{N-1}\,.
\end{equation}
On the other hand, expanding Eq. (\ref{eq:SNt_wait_exp}) for small $s$ and inverting the Laplace transform, we obtain that for late times
\begin{equation}
S_N(t)\simeq \frac{1}{\sqrt{\pi a t}}\,,
\end{equation}
independently of $N$.

\begin{figure}
\includegraphics[width = \linewidth]{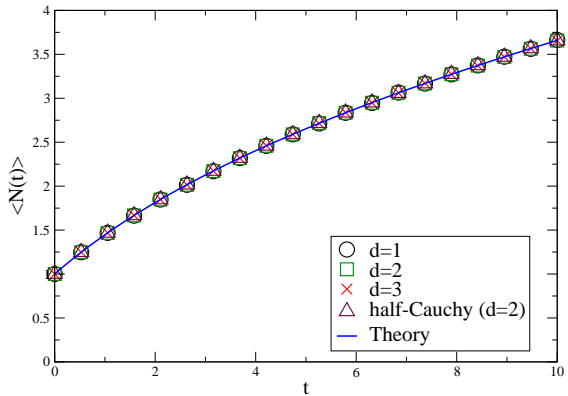}
\caption{Plot of the average number of records $\langle N(t) \rangle$ vs $t$ for the IR model, with waiting rate $a=1$. The solid blue lines are given by the exact formula in Eq. (\ref{eq:avg_N_wait_final}). The symbols correspond to simulations with the choices $d = 1, 2, 3$ with $W(v)=\delta(v-1)$ and $d = 2$ with $W(v) =2/(\pi\left(1+v^2\right))$, for $v > 0$
(half-Cauchy).}\label{fig:avg_N_wait}
\end{figure}

Finally, plugging $c(s)=a/(a+s)$ in the expression for $\langle N(t)\rangle$, given in Eq. (\ref{eq:avg_N_wait}), we get
\begin{equation}
\langle N(t) \rangle=\int \frac{ds}{2\pi i}e^{st}\sqrt{\frac{a+s}{s^3}}\,.
\end{equation}
Inverting the Laplace transform, we obtain
\begin{equation}\label{eq:avg_N_wait_final}
\langle N(t) \rangle=e^{-a t/2}\left(\left(1+a t\right)I_0\left(\frac{a}{2}t\right)+a t I_1\left(\frac{a}{2}t\right)\right)\,,
\end{equation}
which is again independent of $d$ and $W(v)$. For large $t$, the average number of records increases as $\langle N(t)\rangle\sim 2\sqrt{a t/\pi}$. The exact result in Eq. (\ref{eq:avg_N_wait_final}) is plotted in Fig. \ref{fig:avg_N_wait} and is in good agreement with numerical simulations.

\section{Mixed model: run-and-tumble particle with non-instantaneous tumblings}\label{sec:mixed_model}

One of the key assumptions of the IT model for a single RTP presented in the previous sections is that the tumblings can be considered instantaneous. In other words, we have assumed that the tumbling time is typically negligible with respect to the running time. However, in several situations this assumption may not be realistic. For example, from experiments on the dynamics of the bacterium \emph{E. Coli}, we know that the ratio of the average tumbling time to the average running time is typically as large as $0.1$ \cite{Berg_book}. Thus, it is relevant to investigate whether the universal properties described in the previous sections hold true when we include the effect of non-instantaneous tumblings in our model.

Let us consider a modified RTP model in $d$ dimensions, denoted as mixed model, in which the particle alternates the usual running phases to non-instantaneous tumbling phases in which it does not move (see Fig. \ref{fig:model}c). We denote by $n$ the number of waiting phases, or equivalently the number of tumblings, and by $m$ the number of running phases. Let $\{T_i\}=\{T_1,T_2,\ldots T_n\}$ be the  waiting times associated to each tumbling up to the total fixed time $t$.
These time intervals $\{T_i\}$ are assumed to be i.i.d. random variables from a generic probability distribution $P_W(T)$ with positive support. We also assume that the starting point is a tumbling. Thus, the particle initially remains at the origin for a random time $T_1$, drawn from the distribution $P_W(T)$. Then it runs in a random direction with random velocity $v_1$, drawn from $W(v)$, for a time $\tau_1$, exponentially distributed with rate $\gamma$, then waits a time $T_2$, and so on. Note that at the final time $t$ the particle could either be in its last running phase or in its last tumbling phase. In the first case the number $m$ of running phases will be equal to the number $n$ of tumblings, while in the second case we will have $m=n-1$.

In this section we will show that the universal properties described in Sections \ref{sec:surv}, \ref{sec:tmax}, and \ref{sec:record} remain valid also in the case of non-instantaneous tumblings. To show this, we will compute the survival probability, the distribution of the time of the maximum and the record statistics of the $x$ component of a single RTP with finite-time tumblings, showing that these quantities do not depend on the dimension of the system nor on the distribution of the velocities of each running phase. The key-ingredient to compute these three quantities is the joint distribution of the displacements $\{x_1,\ldots x_m\}$ in the $x$ component and the total number $m$ of running phases up to time $t$. Thus, we will first compute this joint probability and then we will use it to calculate the quantities listed above. We will perform the computation in the special case of fixed velocities, i.e. $W(v)=\delta(v-v_0)$, and then we will show how one can generalize the results to the case of arbitrary speed distribution $W(v)$.

First of all, note that, when the time $T_1$ of the first waiting phase exceeds the total time $t$, there will be no running phase, i.e. $m=0$. For the moment, we will focus on the case $m\geq 1$, considering the special case $m=0$ separately. When $m\geq 1$, at the final time $t$, the particle could be either running or waiting. In the first case, the running times are $\tau_1,\tau_2,\ldots ,\tau_m$ (where the last time $\tau_m$ is yet to be completed) and the waiting times are $T_1,T_2,\ldots T_m$ (in this case $m=n$). In the second case the running times are $\tau_1,\ldots \tau_m$ and the waiting times are $T_1,\ldots T_{m+1}$ (where the last waiting time is not completed and $m=n-1$).
The joint probability of the times $\{\tau_i\}=\{\tau_1,\ldots \tau_m\}$ and of the number $m$ of running phases is given by the sum of the terms corresponding to these two cases
\begin{eqnarray}\label{eq:P_tau_m}
&& P(\{\tau_i\},m|t)\\&=& \frac{1}{\gamma}\prod_{i=1}^{m}\int_{0}^{\infty}dT_i\,P_W(T_i)\gamma e^{-\gamma\tau_i}\delta\left(\sum_{i=1}^{m}(\tau_i+T_i)-t\right)\nonumber \\
&+&\int_{0}^{\infty}dT_{m+1}Q_W(T_{m+1})\prod_{i=1}^{m}\int_{0}^{\infty}dT_i\,P_W(T_i)\gamma e^{-\gamma\tau_i}\nonumber \\ &\times & \delta\left(\sum_{i=1}^{m}(\tau_i+T_i)+T_{m+1}-t\right)\,,\nonumber
\end{eqnarray}
where $Q_W(T)$ is the defined as
\begin{equation}\label{eq:Q_W}
Q_W(T)=\int_T^{\infty} dT' P_W(T')\,.
\end{equation}
Let us now explain the meaning of Eq. (\ref{eq:P_tau_m}). The first term of Eq. (\ref{eq:P_tau_m}) corresponds to the case in which the particle is in its last running phase at time $t$. Thus, the probability weight of each tumbling time $T_i$ is given by $P_W(T_i)$ and the probability each run time $\tau_i$ is $P_R(\tau_i)=\gamma e^{\gamma \tau_i}$, except for the last running interval $\tau_m$. Indeed, the last run is not completed and hence its probability weight is $e^{-\gamma \tau_m}$, i.e. the probability that no tumbling happens in the time interval $\tau_m$. On the other hand, the second term of Eq. (\ref{eq:P_tau_m}) corresponds to the complementary case, in which at time $t$ the particle is in a waiting phase. Thus, the weights of the times $T_i$ and $\tau_i$ for all $i\leq m$ are simply given by $P_W(T_i)$ and $P_R(\tau_i)$, respectively. The weight of the last tumbling interval $T_{m+1}$ is $Q_W(T_{m+1})$, i.e. the probability that the last waiting time lasts longer than $T_{m+1}$. For both terms, we integrate over the $T_i$ variables, keeping the total time $t$ fixed.

Let $\{l_1,\ldots l_m\}$ be the straight distances travelled by the particle during each running phase up to time $t$. Since $l_i=v_0 \tau_i$, we get
\begin{eqnarray}
&& P(\{\l_i\},m|t)\\
&=&\frac{1}{\gamma}\prod_{i=1}^{m}\int_{0}^{\infty}dT_i\,P_W(T_i)\frac{\gamma}{v_0} e^{-\gamma l_i/v_0}\delta\left(\sum_{i=1}^{m}(\frac{l_i}{v_0}+T_i)-t\right) \nonumber \\
&+&\int_{0}^{\infty}dT_{m+1}Q_W(T_{m+1})\prod_{i=1}^{m}\int_{0}^{\infty}dT_i\,P_W(T_i)\frac{\gamma}{v_0} e^{-\gamma l_i/v_0} \nonumber \\ &\times &\delta\left(\sum_{i=1}^{m}(\frac{l_i}{v_0}+T_i)+T_{m+1}-t\right)\,.\nonumber
\end{eqnarray}
We denote by ${x_i}$ the displacement of the $x$ component of the particle during the $i^{\rm th}$ running phase. We recall that (see Eq. (\ref{x_comp_cond.1}))
\begin{equation}\label{eq:Px|l}
P(\{x_i\}|\{l_i\})=\prod_{i=1}^{m}\frac{1}{l_i}f_d\left(\frac{x_i}{l_i}\right)\,,
\end{equation}
where $f_d(z)$ is given in Eq. (\ref{fdz.1}).
Thus, using Eq. (\ref{eq:Px|l}), we obtain
\begin{eqnarray}
&&P(\{x_i\},\{\l_i\},m|t)=P(\{x_i\}|\{l_i\})P(\{l_i\},m|t) \nonumber\\
&=&\frac{1}{\gamma}\prod_{i=1}^{m}\int_{0}^{\infty}dT_i\,P_W(T_i)\frac{\gamma}{v_0} e^{-\gamma\frac{l_i}{v_0}}  \frac{1}{l_i}f_d\left(\frac{x_i}{l_i}\right)\\ &\times &
\delta\left(\sum_{i=1}^{m}(\frac{l_i}{v_0}+T_i)-t\right)+\int_{0}^{\infty}dT_{m+1}Q_W(T_{m+1}) \nonumber \\ &\times &\prod_{i=1}^{m}\int_{0}^{\infty}dT_i\,P_W(T_i)\frac{\gamma}{v_0} e^{-\gamma \frac{l_i}{v_0}}
\frac{1}{l_i}f_d\left(\frac{x_i}{l_i}\right)\nonumber\\
& \times &\delta\left(\sum_{i=1}^{m}(\frac{l_i}{v_0}+T_i)+T_{m+1}-t\right)\,.\nonumber
\end{eqnarray}
Integrating over the $l_i$ variables, we obtain
\begin{eqnarray}
&& P(\{x_i\},m|t) \\
&=&\frac{1}{\gamma}\prod_{i=1}^{m}\int_{0}^{\infty}dT_i\,P_W(T_i)\int_{0}^{\infty}dl_i\,\frac{\gamma}{v_0} e^{-\gamma\frac{l_i}{v_0}}  \frac{1}{l_i}f_d\left(\frac{x_i}{l_i}\right) \nonumber \\ 
&\times &\delta\left(\sum_{i=1}^{m}(\frac{l_i}{v_0}+T_i)-t\right)+\int_{0}^{\infty}dT_{m+1}Q_W(T_{m+1})\nonumber  \\ &\times &\prod_{i=1}^{m}\int_{0}^{\infty}dT_i\,P_W(T_i)\int_{0}^{\infty}dl_i\,\frac{\gamma}{v_0} e^{-\gamma \frac{l_i}{v_0}}
\frac{1}{l_i}f_d\left(\frac{x_i}{l_i}\right) \nonumber \\
&\times &
\delta\left(\sum_{i=1}^{m}(\frac{l_i}{v_0}+T_i)+T_{m+1}-t\right)\,.\nonumber
\end{eqnarray}
Taking a Laplace transform with respect to $t$ and using Eq. (\ref{eq:Q_W}), we get
\begin{eqnarray}\label{eq:LT_1}
&& \int_{0}^{\infty}dt\,e^{-st}\,P(\{x_i\},m|t)=\frac{1}{\gamma}\prod_{i=1}^{m}\,\tilde{P}_W(s)\\ &\times &\int_{0}^{\infty}dl_i\,\frac{\gamma}{v_0} e^{-(\gamma+s)\frac{l_i}{v_0}}  \frac{1}{l_i}f_d\left(\frac{x_i}{l_i}\right) \nonumber +\frac{1}{s}\left(1-\tilde{P}_W(s)\right)\\ &\times &\prod_{i=1}^{m}\tilde{P}_W(s)\int_{0}^{\infty}dl_i\,\frac{\gamma}{v_0} e^{-(\gamma+s) \frac{l_i}{v_0}}
\frac{1}{l_i}f_d\left(\frac{x_i}{l_i}\right)
\,,\nonumber
\end{eqnarray}
where 
\begin{equation}\label{eq:tildeP_w}
\tilde{P}_W (s)=\int_{0}^{\infty}dT\,e^{-sT}P_W(T)\,.
\end{equation}
Eq. (\ref{eq:LT_1}) can be rewritten as
\begin{eqnarray}\label{eq:LT_2}
&&\int_{0}^{\infty}dt\,e^{-st}\,P(\{x_i\},m|t)=\left(\frac{1}{\gamma}+\frac{1}{s}\left(1-\tilde{P}_W(s)\right)\right)\nonumber\\&\times &\left(\tilde{P}_W(s)\frac{\gamma}{\gamma+s}\right)^m\prod_{i=1}^{m}\,\tilde{p}_s(x_i)
\,,
\end{eqnarray}
where
\begin{equation}\label{eq:tildep_s}
\tilde{p}_s(x)=\int_{0}^{\infty}dl\,\frac{\gamma+s}{v_0} e^{-(\gamma+s) \frac{l}{v_0}}
\frac{1}{l}f_d\left(\frac{x}{l}\right)\,.
\end{equation}
Finally, performing a formal inversion of the Laplace transform in Eq. (\ref{eq:LT_2}), we get 
\begin{eqnarray}\label{eq:P_1}
&&P(\{x_i\},m|t)=\int \frac{ds}{2\pi i}e^{st}\left(\frac{1}{\gamma}+\frac{1}{s}\left(1-\tilde{P}_W(s)\right)\right)\nonumber\\
&\times &\left(\tilde{P}_W(s)\frac{\gamma}{\gamma+s}\right)^m\prod_{i=1}^{m}\,\tilde{p}_s(x_i)
\,,
\end{eqnarray}
where the integral is over the Bromwich contour in the complex $s$ plane. As explained in Section \ref{sec:surv}, since $\tilde{p}_s(x)$ is positive and normalized to one, it can be interpreted as a probability distribution. Moreover, due to the symmetry $f_d(z)=f_d(-z)$ the probability $\tilde{p}_s(x)$ is also symmetric around $x=0$. Notably, Eq. (\ref{eq:P_1}) is also valid when the velocity of each running phase is a random variable. Indeed, following the steps outlined in Sec. \ref{sec:surv}, it is easy to show that Eq. (\ref{eq:P_1}) remains valid when the velocity $v$ of each running phase is drawn from a generic speed distribution $W(v)$. Even if in this case the distribution $\tilde{p}_s(x)$ will depend on $W(v)$ (the precise expression of $\tilde{p}_s(x)$ is given in Eq. (\ref{eq:psv})), it will still be continuous and symmetric, allowing us to apply the SA theorem. Thus, as we will see the results of this section will not depend on the precise form of $\tilde{p}_s(x)$, and will be valid even in the case of random velocities.

\subsection{Survival probability}

In order to compute $S^{\rm Mixed}(t)$, i.e. the probability that the $x$ component of the position of the particle has never become negative up to time $t$, we will use the method presented in Section \ref{sec:surv}. In this case it turns out to be easier to compute $S^{\rm Mixed}(t)$ fixing the total number $m$ of running phases. We denote by $S_{m}(t)$ the probability that the $x$ component of the particle does not become negative up to time $t$ and that there are exactly $m$ running phases. Then, the survival probability $S^{\rm Mixed}(t)$ can be computed as
\begin{equation}\label{eq:S_sum}
S^{\rm Mixed}(t)=\sum_{m=0}^{\infty}S_m(t)\,.
\end{equation}
The event $m=0$ can only happen when the initial waiting time $T_1$ is larger than $t$. Hence the survival probability is
\begin{equation}\label{eq:S1}
S_{m=0}(t)=Q_W(t)\,,
\end{equation}
where $Q_W(t)$, given in Eq. (\ref{eq:Q_W}) denotes the probability that the first waiting time is larger than $t$.

Looking at the case $m\geq 1$, the probability $S_m(t)$ that the $x$ component of the particle has not become negative up to time $t$ and that the particle has undergone exactly $m$ running phases can be written as
\begin{eqnarray}\label{eq:Sm}
S_m(t)&=&\int_{-\infty}^{\infty}dx_1\ldots\int_{-\infty}^{\infty}dx_m\,\theta(X_1)\ldots \theta(X_m)\,\nonumber\\ 
&\times &P(\{x_i\},m|t)\,,
\end{eqnarray}
where  $P(\{x_i\},m|t)$ is given in Eq. (\ref{eq:P_1}) and 
\begin{equation}
X_k=x_1+x_2+\ldots +x_k\,.
\end{equation}
The term $\theta(X_1)\ldots\theta(X_m)$ constraints the $x$ component of the particle to remain positive up to time $t$.
Using the expression in Eq. (\ref{eq:P_1}), we obtain
\begin{eqnarray}\label{eq:Sm2}
S_m(t)&=&\int \frac{ds}{2\pi i}e^{st}\left(\frac{1}{\gamma}+\frac{1}{s}\left(1-\tilde{P}_W(s)\right)\right)\\&\times &\left(\tilde{P}_W(s)\frac{\gamma}{\gamma+s}\right)^m
q_m\,,\nonumber
\end{eqnarray}
where 
\begin{equation}
q_m=\int_{-\infty}^{\infty}dx_1\,\ldots\int_{-\infty}^{\infty}dx_m\,\theta(X_1)\ldots \theta(X_m)\,\prod_{i=1}^{m}\tilde{p}_s(x_i)\,.
\end{equation}
The probability $\tilde{p}_s(x)$, given in Eq. (\ref{eq:psv}) for the case of a generic speed distribution $W(v)$, is continuous and symmetric.
Notably, as explained in Section \ref{sec:surv}, $q_m$ can be interpreted as the probability that the discrete-time RW $X_k$ has not visited the negative side up to step $m$. We recall that, since the increments of the walk are continuous and symmetric, $q_m$ is universal for all $m$ and that its generating function is given by 
\begin{equation}\label{eq:gen_fun}
\sum_{m=0}^{\infty}q_m\,z^m=\frac{1}{\sqrt{1-z}}\,.
\end{equation}

Plugging the expressions for $S_m(t)$, given in Eqs. (\ref{eq:S1}) (for $m=0$) and (\ref{eq:Sm2}) (for $m\geq 1$), into Eq. (\ref{eq:S_sum}), we obain
\begin{eqnarray}\label{eq:S2}
S^{\rm Mixed}(t)&=& Q_W(t)+\int\frac{ds}{2\pi i}e^{st}\left(\frac{1}{\gamma}+\frac{1}{s}\left(1-\tilde{P}_W(s)\right)\right)\nonumber\\
&\times &\sum_{m=1}^{\infty}\left(\tilde{P}_W(s)\frac{\gamma}{\gamma+s}\right)^m
q_m\,,
\end{eqnarray}
where $Q_W(t)$ is given in Eq. (\ref{eq:Q_W}). Using Eq. (\ref{eq:gen_fun}) and taking care of the fact that the sum in Eq. (\ref{eq:S2}) starts from $m=1$, we get that
\begin{eqnarray}
S^{\rm Mixed}(t)&=&Q_W(t)+\int\frac{ds}{2\pi i}e^{st}\left(\frac{1}{\gamma}+\frac{1}{s}\left(1-\tilde{P}_W(s)\right)\right) \nonumber \\ &\times &\left(\frac{1}{\sqrt{1-\tilde{P}_W(s)\frac{\gamma}{\gamma+s}}}-1\right)\,.
\end{eqnarray}
Using Eq. (\ref{eq:Q_W}), the term $Q_W(t)$ can be rewritten as
\begin{eqnarray}
&& S^{\rm Mixed}(t) = \int\frac{ds\,e^{st}}{2\pi i}\frac{1}{s}\left(1-\tilde{P}_W(s)\right)+\int\frac{ds\,e^{st}}{2\pi i} \\ &&\times \left(\frac{1}{\gamma}+\frac{1}{s} 
\left(1-\tilde{P}_W(s)\right)\right)\left(\frac{1}{\sqrt{1-\tilde{P}_W(s)\frac{\gamma}{\gamma+s}}}-1\right)\,.\nonumber
\end{eqnarray}
Finally, we obtain
\begin{equation}\label{eq:S_fin}
S^{\rm Mixed}(t)=\int\frac{ds}{2\pi i}e^{st}\left(\frac{\left(1-\tilde{P}_W(s)\right)}{s\, h(s)}+\frac{1}{\gamma}\left(\frac{1}{h(s)}-1\right)\right)\,,
\end{equation}
where
\begin{equation}\label{eq:h}
h(s)=\sqrt{1-\tilde{P}_W(s)\frac{\gamma}{\gamma+s}}\,.
\end{equation}
Remarkably, we observe once again that the survival probability $S^{\rm Mixed}(t)$ is completely independent of the dimension $d$ of the system and of the speed distribution $W(v)$. However, we observe that $S^{\rm Mixed}(t)$ will in general depend on the waiting-time distribution $P_W(T)$.

From Eq. (\ref{eq:S_fin}) one can obtain the long time behavior of $S^{\rm Mixed}(t)$. We recall that in general one expects $S^{\rm Mixed}(t)\sim t^{-\theta}$ for large $t$, where $\theta$ is some positive exponent. Let us consider first the case in which $P_W(T)$ has a well-defined first moment $\langle T\rangle$. In this case, one can show that for small $s$
\begin{equation}\label{Pw_expansion1}
\tilde{P}_W(s)\simeq 1-\langle T\rangle s\,.
\end{equation}
Thus, using Eq. (\ref{Pw_expansion1}) and expanding the right-hand side of Eq. (\ref{eq:S_fin}) for small $s$, we obtain
\begin{equation}\label{eq:S_mixed_large_t}
S^{\rm Mixed}(t)\simeq\int\frac{ds}{2\pi i}e^{st}
\frac{\sqrt{<T>+1/\gamma}}{\sqrt{s}}\,.
\end{equation}
Performing the Laplace inversion, we get that, for late times,
\begin{equation}
S^{\rm Mixed}(t)\sim t^{-1/2}\,.
\end{equation}
Thus, when the distribution $P_W(T)$ of the waiting times has a finite first moment we get $\theta=1/2$.

On the other hand, when the first moment is not well-defined, i.e. when $P_W(T)\sim 1/T^{\mu+1}$ with $0<\mu<1$, one can show that for small values of $s$
\begin{equation}\label{Pw_expansion2}
\tilde{P}_W(s)\sim 1-(b s)^\mu\,,
\end{equation}
where $b$ denotes a microscopic time scale. Using Eq. (\ref{Pw_expansion2}) to expand the right-hand side of Eq. (\ref{eq:S_fin}) for small values of $s$, we obtain
\begin{equation}
S^{\rm Mixed}(t)\sim\int\frac{ds}{2\pi i}e^{st}
s^{\mu/2-1}\,.
\end{equation}
Performing the Laplace inversion, we get that, for late times,
\begin{equation}
S^{\rm Mixed}(t)\sim t^{-\mu/2}\,.
\end{equation}
Thus, in this case we obtain $\theta=\mu/2$.

\vspace*{0.5cm}

\noindent{\bf Exponential waiting times:}\\

\begin{figure}[t]
\includegraphics[width=0.45\textwidth]{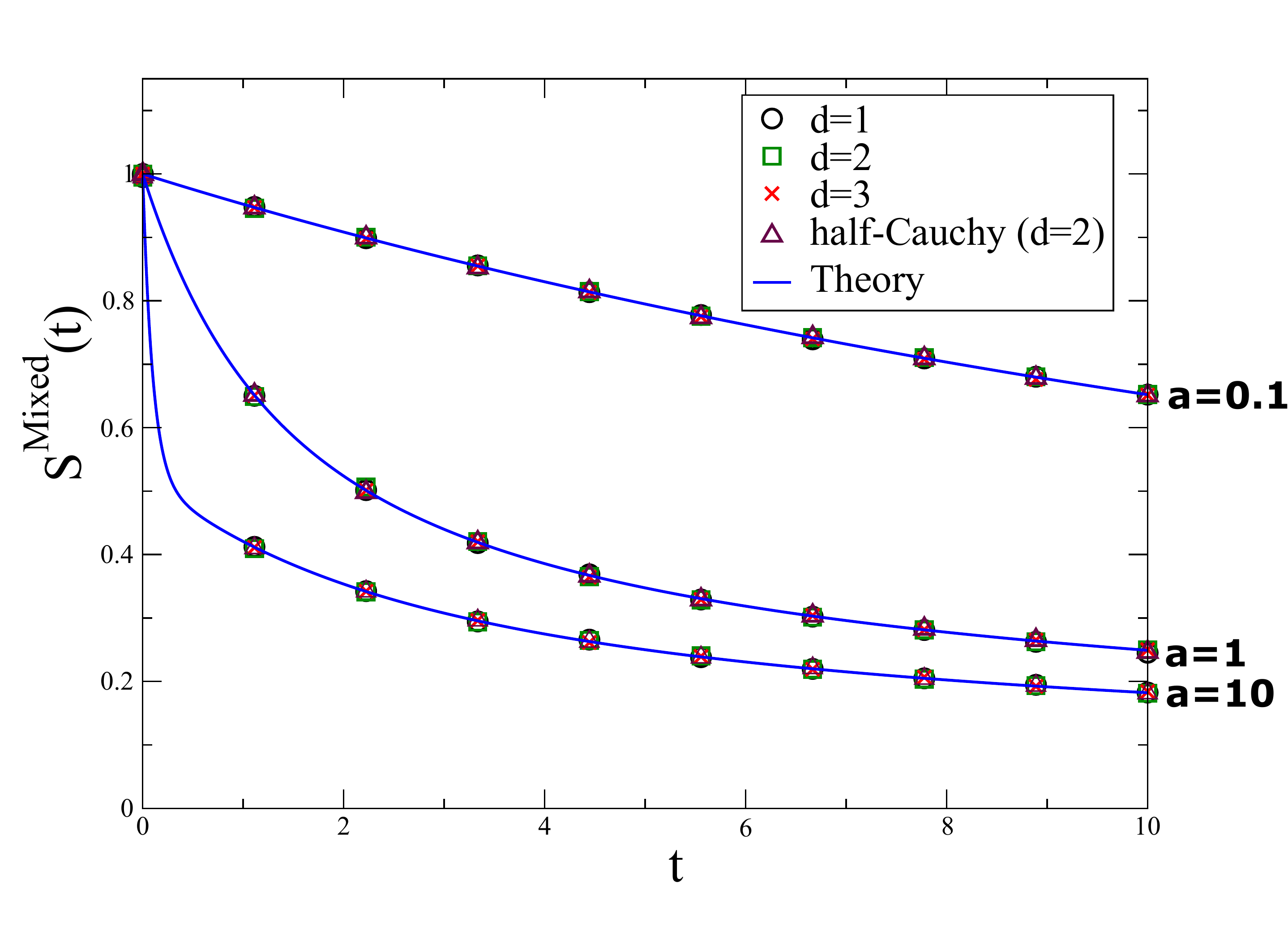} 
\caption{\label{fig:S_mixed_model} Survival probability $S^{\rm Mixed}(t)$ for a RTP with non-instantaneous tumblings as a function of $t$ for $\gamma=1$ and for different values of $a$ (from top to bottom $a=0.1,1,10$). The continuous blue lines correspond to the exact result in Eq. (\ref{eq:S_mixed}). The symbols correspond to simulations with the choices $d = 1, 2, 3$ with $W(v)=\delta(v-1)$ and $d = 2$ with $W(v) =2/(\pi\left(1+v^2\right))$, for $v > 0$
(half-Cauchy). For each value of $a$, the numerical curves collapse on the corresponding analytical blue line for all $t$.}
\end{figure}

Let us now assume that the waiting times are exponentially distributed with fixed rate $a$, i.e. that $P_W(t)=a e^{-at}$. This choice for the distribution $P_W(t)$ is relevant to describe the motion of \emph{E. Coli}. Indeed, for these bacteria the waiting times are known from the experiments to be exponentially distributed \cite{Berg_book}. In this case, the Laplace transform of $P_W(t)$ is $\tilde{P}_W(s)=a/(a+s)$ and Eq. (\ref{eq:S_fin}) becomes
\begin{equation}\label{eq:S_mixed_LT}
S^{\rm Mixed}(t)=\int\frac{ds}{2\pi i}e^{s t}\frac{1}{\gamma}\left(\sqrt{\frac{(\gamma+s)(a+\gamma+s)}{s(a+s)}}-1\right)\,.
\end{equation}

It is interesting to observe that taking the limit $\gamma\to \infty$, with $a$ fixed, in Eq. (\ref{eq:S_mixed_LT}), one obtains
\begin{equation}\label{eq:S_mixed_LT_limit1}
S^{\rm Mixed}(t)=\int\frac{ds}{2\pi i}e^{s t}\left(\sqrt{\frac{1}{s(a+s)}}-1\right)\,.
\end{equation}
Even if in Section \ref{sec:model} we have explained that in this limit one does not obtain the IR model, we observe that, the survival probability $S^{\rm Mixed}(t)$ is equal to $S^{\rm IR}(t)$, i.e. the survival probability of the IR model (see Eq. (\ref{eq:S_IR_LT})). This fact is rather unexpected and is a further indication of the universality of the survival probability. 

Taking the opposite limit, i.e. $a\to \infty$ with $\gamma$ fixed, in Eq. (\ref{eq:S_mixed_LT}) one obtains
\begin{equation}\label{eq:S_mixed_LT_limit2}
S^{\rm Mixed}(t)=\int\frac{ds}{2\pi i}e^{s t}\frac{1}{\gamma}\left(\sqrt{\frac{\gamma+s}{s}}-1\right)\,.
\end{equation}
Note that in this limit we find that $S^{\rm Mixed}(t)$ is identical to the survival probability $S^{\rm IT}(t)$ of the IT model (see Eq. (\ref{sa.3})). Indeed, as explained in Section \ref{sec:model}, in this limit the waiting times become instantaneous.

To perform the Laplace inversion in Eq. (\ref{eq:S_mixed_LT}), it is useful to rewrite Eq. (\ref{eq:S_mixed_LT}) as
\begin{eqnarray}\label{eq:Sr2}
&& S^{\rm Mixed}(t)\\
&=&\int\frac{ds}{2\pi i}e^{st}\frac{1}{\gamma}\Bigg[\left(\sqrt{\frac{(\gamma+s)}{s}}-1\right)\left(\sqrt{\frac{(\gamma+a+s)}{s+a}}-1\right) \nonumber\\
&+&\left(\sqrt{\frac{(\gamma+s)}{s}}-1\right)+\left(\sqrt{\frac{\gamma+a+s}{s+a}}-1\right)\Bigg]\,.\nonumber
\end{eqnarray}
Using the Laplace inversion formula in Eq. (\ref{eq:laplace_inversion_1}) and convolution theorem, we obtain
\begin{eqnarray}\label{eq:S_mixed}
S^{\rm Mixed}(t)&=&\frac{\gamma}{4} e^{-\gamma t/2}\int_{0}^{t}dt'\,e^{-a t'}\left(I_0\left(\frac{\gamma}{2}t'\right)+I_1\left(\frac{\gamma}{2}t'\right)\right)\nonumber \\
&\times & \left(I_0\left(\frac{\gamma}{2} (t-t')\right)+I_1\left(\frac{\gamma}{2}( t-t')\right)\right) \\
&+&\frac12 \left(1+e^{-at}\right) e^{-\gamma t/2}\left(I_0\left(\frac{\gamma}{2} t\right)+I_1\left(\frac{\gamma}{2} t\right)\right)\,. \nonumber
\end{eqnarray}
Computing the integral in Eq. (\ref{eq:S_mixed}) appears to be challenging, however one can easily perform the integration numerically for given values of $a$ and $\gamma$. In Fig. \ref{fig:S_mixed_model} we compare this theoretical result in Eq. (\ref{eq:S_mixed}) with numerical simulations, finding an excellent agreement. One can also easily derive the behavior of $S^{\rm Mixed}(t)$ for short and late times. To study the limit $t\to 0$ we need to expand the integrand on the right-hand side of Eq. (\ref{eq:S_mixed_LT}) for large $s$, yielding
\begin{equation}
S^{\rm Mixed}(t)\simeq \int \frac{ds}{2\pi i}e^{st}\left(\frac{1}{s}-\frac{a}{2s^2}\right)\,.
\end{equation}
Thus, when $t\to 0$
\begin{equation}
S^{\rm Mixed}(t)\simeq 1-\frac{a}{2}t\,.
\end{equation}
We observe that the survival probability goes to the limit value $1$ when $t\to 0$. It is easy to explain this limit if we think that the starting point is assumed to be at the beginning of a waiting phase. Thus, at small enough $t$, the initial waiting time will be larger than the total time with high probability. On the other hand, since the mean waiting time $\langle T\rangle$ is finite, we have already computed the late time behavior in Eq. (\ref{eq:S_mixed_large_t}). Thus, using $\langle T\rangle=1/a$ we find that for $t\to \infty$
\begin{equation}
S^{\rm Mixed}(t)\simeq \sqrt{\frac{1/a+1/\gamma}{\pi t}}\,.
\end{equation}

Notably, the expression for $S^{\rm Mixed}(t)$ becomes much simpler when the waiting rate is equal to the tumbling rate. Indeed, setting $a=\gamma$ in Eq. (\ref{eq:S_mixed_LT}), one obtains
\begin{equation}\label{eq:S_mixed_1}
S^{\rm Mixed}(t)=\int\frac{ds}{2\pi i}e^{s  t}\left(\sqrt{\frac{(2\gamma+s)}{s}}-1\right)\,.
\end{equation}
Using the inversion formula in Eq. (\ref{eq:laplace_inversion_1}), we invert the Laplace transform in Eq. (\ref{eq:S_mixed_1}) and we obtain that
\begin{equation}
S^{\rm Mixed}(t)=e^{-\gamma t}\left(I_0\left(\gamma t\right)+I_1\left(\gamma t\right)\right)\,,
\end{equation}
where $I_0(z)$ and $I_1(z)$ are modified Bessel functions.

As we have shown above, in the limits $\gamma\to \infty$ one obtains that $S^{\rm Mixed}(t)$ goes to the survival probability of the IR model. Similarly, when $a\to \infty$ one obtains the survival probability of the IT model. It is also relevant to compute the lowest order corrections to these two limits. Let us first consider the limit of instantaneous runs, i.e. $\gamma \gg a$. Expanding Eq. (\ref{eq:S_mixed_LT}) in powers of $a/\gamma$, one obtains 
\begin{eqnarray}
&&S^{\rm Mixed}(t)=\int\frac{ds}{2\pi i}e^{st}\frac{1}{\gamma}\left(\frac{\gamma+s}{\sqrt{s(s+a)}}\sqrt{1+\frac{a}{\gamma+s}}-1\right)\nonumber\\
&&\simeq \int\frac{ds}{2\pi i}e^{st}\frac{1}{\gamma}\left(\frac{\gamma+s}{\sqrt{s(s+a)}}\left(1+\frac{a}{2(\gamma+s)}\right)-1\right).
\end{eqnarray}
Using Eqs. (\ref{eq:laplace_inversion_3}) and  (\ref{eq:laplace_inversion_2}), we obtain that for $\gamma\gg a$
\begin{equation}\label{eq:S_mixed_limit1}
S^{\rm Mixed}(t)\simeq e^{-at/2}I_0\left(\frac{a}{2}t\right)+\frac{1}{2}\frac{a}{\gamma}e^{-at/2}I_1\left(\frac{a}{2}t\right)\,.
\end{equation}
Let us now consider the IT limit $a\gg \gamma$. Expanding Eq. (\ref{eq:S_mixed_LT}) in powers of $\gamma/a$ we obtain
\begin{eqnarray}
&&S^{\rm Mixed}(t)=\int\frac{ds}{2\pi i}e^{st}\frac{1}{\gamma}\left(\sqrt{\frac{(\gamma+s)}{s}}\sqrt{1+\frac{\gamma}{a+s}}-1\right)\nonumber\\
&&\simeq \int\frac{ds}{2\pi i}e^{st}\frac{1}{\gamma}\left(\sqrt{\frac{(\gamma+s)}{s}}\left(1+\frac{\gamma}{2(a+s)}\right)-1\right). 
\end{eqnarray}
Using Eq. (\ref{eq:laplace_inversion_1}) and convolution theorem, we obtain that when $a\gg \gamma$
\begin{eqnarray}\label{eq:S_mixed_limit2}
&&S^{\rm Mixed}(t)\\
&&\simeq \frac{1}{2}e^{-\gamma t/2}\left(I_0\left(\frac{\gamma}{2}t\right)+I_1\left(\frac{\gamma}{2}t\right)\right)+\frac12 e^{-at}\nonumber \\
&&+\frac{1}{2}\int_{0}^{t}dt'e^{-a(t-t')-\gamma t'/2}\left(I_0\left(\frac{\gamma}{2}t'\right)+I_1\left(\frac{\gamma}{2}t'\right)\right)\,.\nonumber
\end{eqnarray}

\subsection{Time to reach the maximum}

In this section we investigate the time $t_{\max}$ of the maximum of the $x$ component of a single RTP with non-instantaneous tumblings. We assume again that the particle starts from the origin and it evolves in $d$ dimensions up to time $t$. We recall that we are assuming that the starting point is at the beginning of a tumbling phase. Note that, since in this model the particle does not move during a tumbling, the time $t_{\max}$ of the maximum will be in general ill-defined since the $x$ component of the particle may remain at its maximal value for a finite amount of time. To avoid this issue, we define $t_{\max}$ as the time at which the $x$ component is maximal for the first time.

The strategy that we will adopt to compute the probability density $P(t_{\max}|t)$ of the first time $t_{\max}$ of the maximum up to time $t$ is similar to the one presented in Section \ref{sec:tmax}: in the case $0< t_{\max}< t$ we will decompose this probability in two factors, corresponding to the independent intervals $[0,t_{\max}]$ (I) and $[t_{\max},t]$ (II), and we will show that each of these factors can be rewritten as a survival probability. The two intervals are independent because the time $t_{\max}$ will always be at the end of a running phase. This implies that, denoting by $m_1\geq 1$ the number of waiting phases in the first interval $[0,t_{\max}]$, the particle will also go through exactly $m_1$ running phases up to time $t_{\max}$. On the other hand, denoting by $m_2\geq 1$ the number of waiting phases in the second interval, the number of running phases in the second interval will be either $m_2$, when the particle is in a running phase at time $t$, or $m_2-1$, when the particle is waiting at the final time. Finally, we will also include the contributions of the events $t_{\max}=0$ and $t_{\max}=t$.

Let us now consider the case $0<t_{\max}<t$. We denote by $P_{\rm I}(t_{\max})$ and $P_{\rm II}(t-t_{\max})$ the probabilities corresponding to the intervals (I) and (II). Let us start by computing the probability weight $P_{\rm I}(t_{\max})$ of the first interval $[0,t_{\max}]$. We consider again the RW $X_k=x_1+x_2+\ldots +x_k$ associated to the displacements $\{x_i\}$ in the $x$ component of the particle. In the first interval, the RW $X_k$ will start from zero and will reach the maximal value $X_{m_1}$ for the first time after $m_1$ steps. The joint probability of the displacements $x_1,x_2,\ldots x_{m_1}$ and of the number $m_1$ of waiting phases can be easily written following the steps that led to the result in Eq. (\ref{eq:LT_2}) but keeping only the term corresponding to the case in which the final time is happening during a running phase. This yields
\begin{eqnarray}\label{eq:joint_x_m_1}
&&P(x_1,\ldots x_{m_1},m_1|t_{\max})=\int \frac{ds_1}{2 \pi i}e^{s_1 t_{\max}}\\
&\times &\left(\tilde{P}_W(s_1)\frac{\gamma}{\gamma+s_1}\right)^{m_1} \prod_{i=1}^{m_1}\tilde{p}_{s_1}(x_i)\,,\nonumber
\end{eqnarray}
where  $\tilde{P}_W(s_1)$ is the Laplace transform of $P_W(T)$ and $\tilde{p}_{s_1}(x)$, given in Eq. (\ref{eq:psv}) for the case of random velocities, is the usual continuous and symmetric probability distribution. We can now compute the probability $P_{\rm I}(t_{\max})$ that in the first interval the RW $X_k$ reaches its maximal value at the final step $m_1$, i.e. that $X_{m_1}>X_i$ for all $i<m_1$. This probability can be written as, summing over $m_1$,
\begin{eqnarray}
P_{\rm I}(t_{\max})&=&\sum_{m_1=1}^{\infty}\int_{-\infty}^{\infty}dx_1\,\ldots\int_{-\infty}^{\infty}dx_{m_1}\,\theta(X_{m_1})\nonumber\\ &\times & \theta(X_{m_1}-X_1)\ldots\theta(X_{m_1}-X_{m_1-1})\nonumber\\ &\times & P(x_1,\ldots x_{m_1},{m_1}|t_{\max}) \,.
\end{eqnarray}
Plugging the expression for $P(x_1,\ldots x_{m_1},m_1|t_{\max})$, given in Eq. (\ref{eq:joint_x_m_1}), we obtain
\begin{equation}\label{eq:P_I_m_mixed}
P_{\rm I}(t_{\max})=\int \frac{ds_1}{2\pi i}e^{s_1 t_{\max}}\sum_{m_1=1}^{\infty}\left(\tilde{P}_W(s_1)\frac{\gamma}{\gamma+s_1}\right)^{m_1} q_{m_1}\,,
\end{equation}
where 
\begin{equation}\label{eq:q_m_mixed}
q_{m_1}=\int_{-\infty}^{\infty}dx_1\ldots\int_{-\infty}^{\infty}dx_{m_1}\prod_{i=1}^{m_1} \tilde{p}_s(x_i)\theta(X_{m_1}-X_{m_1-i})\,.
\end{equation}
In Section \ref{sec:tmax}, we have shown that the probability $q_{m_1}$ in Eq. (\ref{eq:q_m_mixed}) can be rewritten as the survival probability of a RW with symmetric jumps. Thus, as a consequence of the SA theorem, $q_{m_1}$ is universal and its generating function is given in Eq. (\ref{eq:gen_fun}). Using Eq. (\ref{eq:gen_fun}), we obtain that the probability $P_{\rm I}(t_{\max})$ of the first segment is given by
\begin{equation}\label{eq:P_I_mixed}
P_{\rm I}(t_{\max})=\int \frac{ds_1}{2\pi i}e^{s_1 t_{\max}}\left(\frac{1}{h(s_1)}-1\right)\,,
\end{equation}
where $h(s)$ is given in Eq. (\ref{eq:h}).

Let us now consider the second interval $[t_{\max},t]$. In this interval the $x$ component of the particle starts from position $X_m$ at time $t_{\max}$ and has to remain below this position up to time $t$. Thus, it is easy to show that the probability of this segment is exactly given by the survival probability in Eq. (\ref{eq:S_fin}). Indeed, applying first the translation $x\to x-X_m$ and then the reflection $x\to -x$, it is clear that the probability of the segment is identical to the probability that the $x$ component of an RTP starting from the origin remains positive for a time $t-t_{\max}$. Thus, the probability of the second interval is given by
\begin{eqnarray}\label{eq:P_II_mixed}
P_{\rm II}(t-t_{\max})&=&\int\frac{ds_2}{2\pi i}e^{s_2 (t-t_{\max})}\Bigg(\frac{\left(1-\tilde{P}_W(s_2)\right)}{s_2 h(s_2)}\nonumber\\ &+&\frac{1}{\gamma}\left(\frac{1}{h(s_2)}-1\right)\Bigg)\,,
\end{eqnarray}
where $h(s)$ is given in Eq. (\ref{eq:h}). When $0<t_{\max}<t$, the probability $P(t_{\max}|t)$ will be simply given by the product of the two factors $P_{\rm I}(t_{\max})$ and $P_{\rm II}(t-t_{\max})$. Thus, we obtain
\begin{eqnarray}\label{eq:P_tmax_incomplete}
P(t_{\max}|t)=P_{\rm I}(t_{\max})P_{\rm II}(t-t_{\max})\,,
\end{eqnarray}
where $P_{\rm I}(t_{\max})$ and $P_{\rm II}(t-t_{\max})$ are given in Eqs. (\ref{eq:P_I_mixed}) and (\ref{eq:P_II_mixed}).

Note, however, that the distribution in Eq. (\ref{eq:P_tmax_incomplete}) is not normalized to one, since we still need to include the contributions corresponding to the cases $t_{\max}=0$ and $t_{\max}=t$. Let us first consider the case $t_{\max}=0$. The global maximum will be reached at the starting point only if the $x$ component of the particle does not become positive up to time $t$. Thus, the probability ${\rm Proba.}(t_{\max}=0|t)$ is simply given by the $P_{\rm II}(t)$, given in Eq. (\ref{eq:P_II_mixed}). Similarly, the probability that the maximum is reached at the final time $t$ can be written in terms of the probability $P_{\rm I}(t)$. However, at variance to the case $t_{\max}<t$, in the case $t_{\max}=t$ the last running phase before the maximum is not completed. It is easy to show that this difference leads to an extra factor $1/\gamma$. Thus, including also these additional contributions, we obtain that the probability distribution of the time $t_{\max}$ of the maximum at fixed time $t$ is given by
\begin{eqnarray}\label{eq:P_tmax_mixed}
P(t_{\max}|t)&=&P_{\rm I}(t_{\max})P_{\rm II}(t-t_{\max})\\
&+&\delta(t_{\max})P_{\rm II}(t)+\delta(t-t_{\max})\frac{1}{\gamma}P_{\rm I}(t)\,,\nonumber
\end{eqnarray}
where $P_{\rm I}(t_{\max})$ and $P_{\rm II}(t-t_{\max})$ are given in Eqs. (\ref{eq:P_I_mixed}) and (\ref{eq:P_II_mixed}). The distribution of $t_{\max}$ is completely independent of the dimension $d$ of the system and of the speed distribution $W(v)$. Note, however, that $P(t_{\max}|t)$ will depend in general on the distribution $P_W(T)$ of the waiting times and that performing the Laplace inversions in Eqs. (\ref{eq:P_I_mixed}) and (\ref{eq:P_II_mixed}) is in general hard.

We now want to check that the PDF of $t_{\max}$, given in Eq. (\ref{eq:P_tmax_mixed}) is normalized to one for any $t$. First of all, we perform a double Laplace transform with respect to $t_{\max}$ and $t$ in Eq. (\ref{eq:P_tmax_mixed}) and we obtain
\begin{eqnarray}\label{eq:normaliz_mixed_1}
&&\int_{0}^{\infty}dt\,\int_{0}^{t}dt_{\max}\,P(t_{\max}|t)e^{-st-s_1t_{\max}}\\
&=&\tilde{P}_{\rm I}(s+s_1)\tilde{P}_{\rm II}(s)+\frac{1}{\gamma}\tilde{P}_{\rm I}(s+s_1)+\tilde{P}_{\rm II}(s)\,,\nonumber
\end{eqnarray}
where $\tilde{P}_{\rm I}(s)$ and $\tilde{P}_{\rm II}(s)$ are the Laplace transforms of $P_{\rm I}(t)$ and $P_{\rm II}(t)$. Setting $s_1=0$ on both sides of Eq. (\ref{eq:normaliz_mixed_1}), we obtain
\begin{eqnarray}\label{eq:normaliz_mixed_2}
&&\int_{0}^{\infty}dt\,\int_{0}^{t}dt_{\max}\,P(t_{\max}|t)e^{-st}\\
&=&\tilde{P}_{\rm I}(s)\tilde{P}_{\rm II}(s)+\frac{1}{\gamma}\tilde{P}_{\rm I}(s)+\tilde{P}_{\rm II}(s)\,.\nonumber
\end{eqnarray}
Plugging the expressions for $\tilde{P}_{\rm I}(s)$ and $\tilde{P}_{\rm II}(s)$, given in Eqs. (\ref{eq:P_I_mixed}) and (\ref{eq:P_II_mixed}), in Eq. (\ref{eq:normaliz_mixed_2}), we obtain 
\begin{eqnarray}\label{eq:normaliz_mixed_3}
&&\int_{0}^{\infty}dt\,\int_{0}^{t}dt_{\max}\,P(t_{\max}|t)e^{-st}\\
&=&\left(\frac{1}{h(s)}-1\right)\Bigg(\frac{\left(1-\tilde{P}_W(s)\right)}{s h(s)}+\frac{1}{\gamma}\left(\frac{1}{h(s)}-1\right)\Bigg)\nonumber\\
&+&\frac{1}{\gamma}\left(\frac{1}{h(s)}-1\right)+\Bigg(\frac{\left(1-\tilde{P}_W(s)\right)}{s h(s)}+\frac{1}{\gamma}\left(\frac{1}{h(s)}-1\right)\Bigg)\,,\nonumber
\end{eqnarray}
where $\tilde{P}_W(s)$ is the Laplace transform of the waiting-time distribution $P_W(T)$ and $h(s)$ is given in Eq. (\ref{eq:h}). Using the expression of $h(s)$, given in Eq. (\ref{eq:h}), in Eq. (\ref{eq:normaliz_mixed_2}), we obtain, after few steps of algebra, that
\begin{equation}
\int_{0}^{\infty}dt\,\int_{0}^{t}dt_{\max}\,P(t_{\max}|t)e^{-st}=\frac{1}{s}\,.
\end{equation}
Inverting the Laplace transform, we get
\begin{equation}
\int_{0}^{t}dt_{\max}\,P(t_{\max}|t)=1\,.
\end{equation}
Thus, we have shown that $P(t_{\max}|t)$, given in Eq. (\ref{eq:P_tmax_mixed}), is normalized to one for any waiting-time distribution $P_W(T)$ and for any $t$.

\vspace*{0.3cm}

\noindent{\bf Exponential waiting times}\\

\begin{figure}[t]
\includegraphics[width=0.5\textwidth]{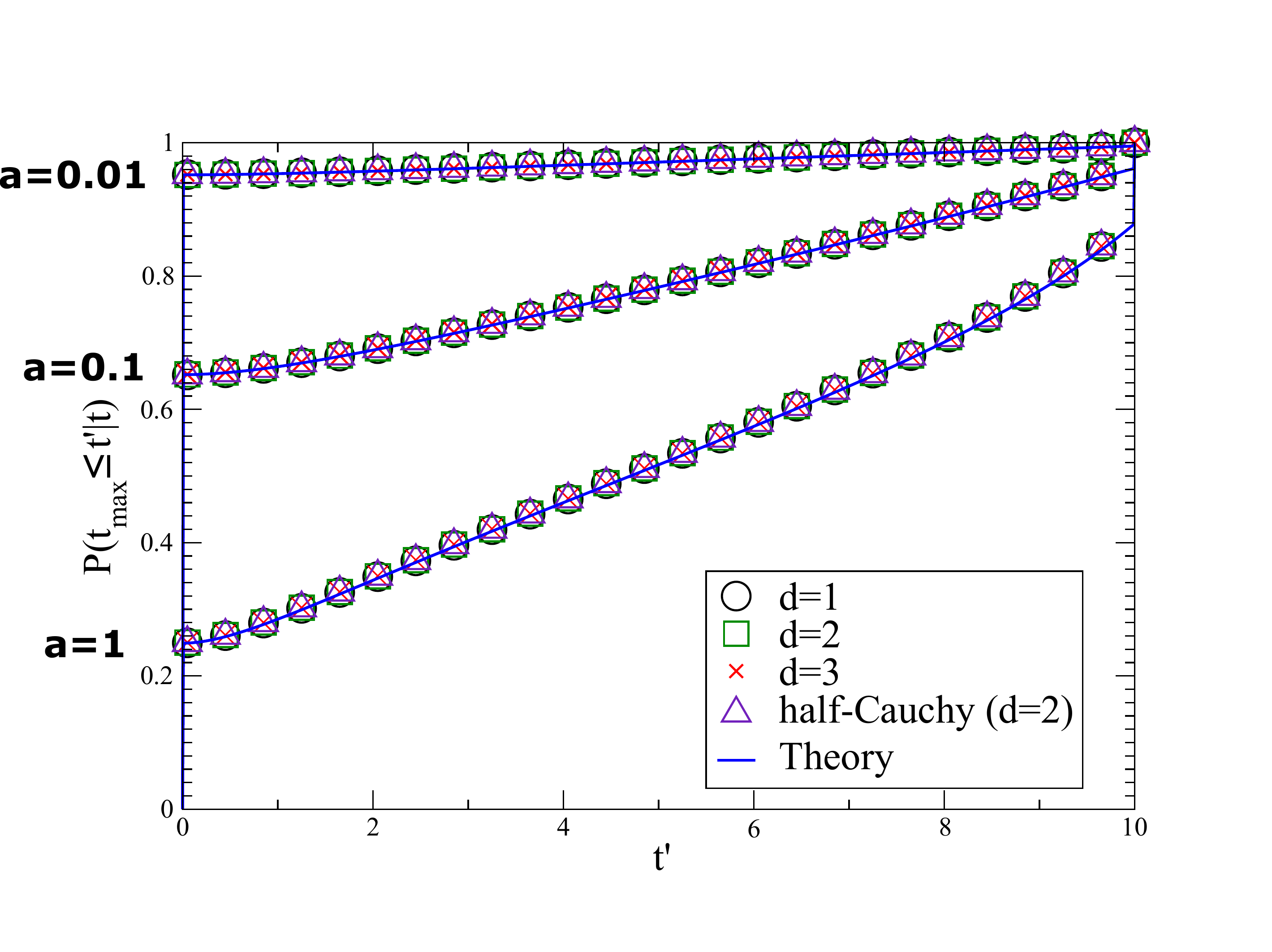} 
\caption{\label{fig:p_tmax_mixed_model} Cumulative probability $P(t_{\max}\leq t'|t)$ for the mixed model as a function of $t'$ for different values of the waiting rate $a$ (from top to bottom $a=0.01,0.1,1$), for $\gamma=1$ and $t=10$. The continuous blue lines correspond to the exact result in Eq. (\ref{eq:cumulative_mixed}). The symbols correspond to simulations with the choices $d = 1, 2, 3$ with $W(v)=\delta(v-1)$ and $d = 2$ with $W(v) =2/(\pi\left(1+v^2\right))$, for $v > 0$
(half-Cauchy). For each value of $a$, the numerical curves collapse on the corresponding analytical blue line for all $t$. We observe that the cumulative probability of $t_{\max}$ has a jump discontinuity at $t'=t$, see Eq. (\ref{eq:cumulative_mixed}).}
\end{figure}

Let us now assume that the waiting times $T_1\,\ldots T_m$ are exponentially distributed with rate $a$, i.e. that 
\begin{equation}\label{eq:P_w_exp}
P_W(T)=a e^{-aT}\,.
\end{equation} 
As we will show, for this particular choice of the waiting-time distribution, one can find an explicit expression for $P_{\rm I}(t)$ and $P_{\rm II}(t)$. Plugging the expression for $P_W(T)$, given in Eq. (\ref{eq:P_w_exp}) into the expression for $P_{\rm I}(t_{\max})$ given in Eq. (\ref{eq:P_I_mixed}) and using Eq. (\ref{eq:h}), we obtain
\begin{equation}\label{eq:P_I_exp}
P_{\rm I}(t)=\int \frac{ds_1}{2\pi i}e^{s_1 t}\left(\sqrt{\frac{(a+s_1)(\gamma+s_1)}{s_1(a+\gamma+s_1)}}-1\right)\,.
\end{equation}
Similarly, from Eqs. (\ref{eq:P_II_mixed}) and (\ref{eq:h}), we obtain
\begin{eqnarray}\label{eq:P_II_exp}
P_{\rm II}(t)&=&\int \frac{ds_1}{2\pi i}e^{s_2 t}\frac{1}{\gamma}\\ &\times &\left(\sqrt{\frac{(a+\gamma+s_2)(\gamma+s_2)}{s_2(a+s_2)}}-1\right)\,.\nonumber
\end{eqnarray}
The Laplace inversions in Eqs. (\ref{eq:P_I_exp}) and (\ref{eq:P_II_exp}) can be performed using the inversion formulae in Eqs. (\ref{eq:laplace_inversion_1}) and (\ref{eq:laplace_inversion_3}). Indeed, applying convolution theorem, we obtain
\begin{widetext}
\begin{eqnarray}\label{eq:P_I_final}
 P_{\rm I}(t)&=&\frac{a^2}{4}e^{- a t/2}\int_{0}^{t}dt'\left[ e^{-\gamma t'} \left(I_1\left(\frac{a}{2}t'\right)-I_0\left(\frac{a}{2}t'\right)\right) \left(I_0\left(\frac{a}{2}(t-t')\right)+I_1\left(\frac{a}{2}(t-t')\right)\right) \right]\\
 &+& \frac{a}{2}e^{-at/2}\left(I_0\left(\frac{a}{2}t\right)+I_1\left(\frac{a}{2}t\right)\right)+\frac{a}{2}e^{-(a/2+\gamma)t}\left(I_1\left(\frac{a}{2}t\right)-I_0\left(\frac{a}{2}t\right)\right)\,,\nonumber 
 \end{eqnarray}
 \begin{eqnarray}\label{eq:P_II_final}
 P_{\rm II}(t)&=&\frac{\gamma}{4}e^{- \gamma t/2}\int_{0}^{t}dt'\left[ e^{-a t'} \left(I_0\left(\frac{\gamma}{2}t'\right)+I_1\left(\frac{\gamma}{2}t'\right)\right) \left(I_0\left(\frac{\gamma}{2}(t-t')\right)+I_1\left(\frac{\gamma}{2}(t-t')\right)\right) \right]\\
 &+& \frac{1}{2}e^{-\gamma t/2}\left(I_0\left(\frac{a}{2}t\right)+I_1\left(\frac{a}{2}t\right)\right)(1+e^{-a t})\,.\nonumber
\end{eqnarray}
\end{widetext}
Computing the integrals over $t'$ in Eqs. (\ref{eq:P_I_final}) and (\ref{eq:P_II_final}) is challenging. However, one can evaluate these integrals numerically for a given set of values for $a$, $\gamma$ and $t$. Then, the cumulative distribution of $t_{\max}$ can be obtained from Eq. (\ref{eq:P_tmax_mixed}) and is given by
\begin{eqnarray}\label{eq:cumulative_mixed}
P(t_{\max}\leq t'|t)&=& \int_{0}^{t'}dt_{\max}P_{\rm I}(t_{\max})P_{\rm II}(t-t_{\max}) \nonumber \\
&+& P_{\rm II}(t)+P_{\rm I}(t)\theta(t'-t)\,,
\end{eqnarray}
where $P_{\rm I}(t)$ and $P_{\rm II}(t)$ are given in Eqs. (\ref{eq:P_I_final}) and (\ref{eq:P_II_final}). We recall that $\theta(t'-t)=0$ for $t'<t$ and $\theta(t'-t)=1$ at $t'=t$. In Fig. \ref{fig:p_tmax_mixed_model}, we compare the exact result in Eq. (\ref{eq:cumulative_mixed}) with numerical simulations, finding excellent agreement. We observe that, due to the term $\theta(t'-t)$ in Eq. (\ref{eq:cumulative_mixed}), the cumulative probability $P(t_{\max}\leq t'|t)$ has a jump discontinuity at $t'=t$.

The expressions for $P_{\rm I}(t)$ and $P_{\rm II}$ can be greatly simplified in the special case $a=\gamma$. Indeed, setting $a=\gamma$ in Eq. (\ref{eq:P_I_exp}), we obtain
\begin{equation}
P_{\rm I}(t)=\int \frac{ds_1}{2\pi i}e^{s_1 t}\left(\frac{\gamma+s_1}{\sqrt{s_1(2\gamma+s_1)}}-1\right)\,.
\end{equation}
This Laplace inversion can be easily performed using Eqs. (\ref{eq:laplace_inversion_2}) and (\ref{eq:laplace_inversion_3}), yielding
\begin{equation}
P_{\rm I}(t)=\gamma e^{-\gamma t}I_1(\gamma t)\,.
\end{equation}
Similarly, setting $a=\gamma$ in Eq. (\ref{eq:P_II_exp}) gives
\begin{equation}
P_{\rm II}(t)=\int \frac{ds_1}{2\pi i}e^{s_2 t}\frac{1}{\gamma}\left(\sqrt{\frac{2\gamma+s_2}{s_2}}-1\right)\,.
\end{equation}
Using the inversion formula in Eq. (\ref{eq:laplace_inversion_1}), we obtain
\begin{equation}
P_{\rm II}(t)= e^{-\gamma t}\left(I_0\left(\gamma t\right)+I_1\left(\gamma t\right)\right)\,.
\end{equation}

Finally, it is also is also relevant to investigate the behavior of the distribution $P(t_{\max}|t)$ in the limits $a\ll \gamma$ and $a\gg \gamma$. First, we consider the factor $P_{\rm I}(t)$. Expanding the expression for $P_{\rm I}(t)$ given in Eq. (\ref{eq:P_I_exp}) for $a\ll \gamma$, we obtain
\begin{equation}
P_{\rm I}(t)\simeq \int \frac{ds}{2\pi i}e^{st}\left(\sqrt{\frac{a+s}{s}}\left(1-\frac{a}{2(\gamma+s)}\right)-1\right)\,.
\end{equation}
Performing the Laplace inversion, we get that when $a\ll \gamma$
\begin{eqnarray}
&&P_{\rm I}(t) \simeq \frac{a}{2} e^{-at/2}\left(I_0\left(\frac{a}{2}t\right)+I_1\left(\frac{a}{2}t\right)\right)+\frac{a}{2}e^{-\gamma t}\\
&&+\frac{a^2}{4}\int_{0}^{t}dt'\,e^{-\gamma(t-t')}e^{-at'/2}\left(I_0\left(\frac{a}{2}t'\right)+I_1\left(\frac{a}{2}t'\right)\right)\nonumber\,.
\end{eqnarray}
On the other hand, when $a \gg \gamma$, expanding Eq. (\ref{eq:P_I_exp}), we obtain
\begin{equation}
P_{\rm I}(t)\simeq \int \frac{ds}{2\pi i}e^{st}\left(\sqrt{\frac{\gamma+s}{s}}\left(1-\frac{\gamma}{2(a+s)}\right)-1\right)\,.
\end{equation}
Inverting the Laplace transform, we obtain
\begin{eqnarray}
&&P_{\rm I}(t) \simeq \frac{\gamma}{2} e^{-\gamma t/2}\left(I_0\left(\frac{\gamma}{2}t\right)+I_1\left(\frac{\gamma}{2}t\right)\right)+\frac{\gamma}{2}e^{-a t}\\
&&+\frac{\gamma^2}{4}\int_{0}^{t}dt'\,e^{-a(t-t')}e^{-\gamma t'/2}\left(I_0\left(\frac{\gamma}{2}t'\right)+I_1\left(\frac{\gamma}{2}t'\right)\right)\nonumber\,.
\end{eqnarray}

To study the behavior of $P_{\rm II}(t)$ in the limits $a\ll \gamma$ and $a\gg \gamma$, we recall that $P_{\rm II}(t)=S^{\rm Mixed}(t)$, where $S^{\rm Mixed}(t)$ is given in Eq. (\ref{eq:S_mixed}). Thus, using Eq. (\ref{eq:S_mixed_limit1}), we obtain that for $a\ll \gamma$
\begin{equation}
P_{\rm II}(t)\simeq e^{-at/2}I_0\left(\frac{a}{2}t\right)+\frac{1}{2}\frac{a}{\gamma}e^{-at/2}I_1\left(\frac{a}{2}t\right)\,.
\end{equation}
From Eq. (\ref{eq:S_mixed_limit2}) we obtain that when $a\gg\gamma$
\begin{eqnarray}
P_{\rm II}(t)&\simeq &\frac{1}{2}e^{-\gamma t/2}\left(I_0\left(\frac{\gamma}{2}t\right)+I_1\left(\frac{\gamma}{2}t\right)\right)+\frac12 e^{-at}\\
&+&\frac{1}{2}\int_{0}^{t}dt'\,e^{-a(t-t')-\gamma t'/2}\left(I_0\left(\frac{\gamma}{2}t'\right)+I_1\left(\frac{\gamma}{2}t'\right)\right)\,.\nonumber 
\end{eqnarray}

We observe that, as expected, taking the limit $a\to\infty$ with $\gamma$ fixed, we find the expression for $P(t_{\max}|t)$ computed for the IT model and given in Eq. (\ref{eq:p_tmax_final}). Notably, in the opposite limit $\gamma\to \infty$ with $a$ fixed one finds the expression for $P(t_{\max}|t)$ computed in the IR setup (see Eq. (\ref{eq:p_tmax_final_wait})). We recall that this last result is unexpected, since the model obtained from the mixed model in the limit $\gamma\to \infty$ is different from the IR model, as explained in Section \ref{sec:model}.

\subsection{Record statistics}

\begin{figure*}[t]
 \includegraphics[width = \linewidth]{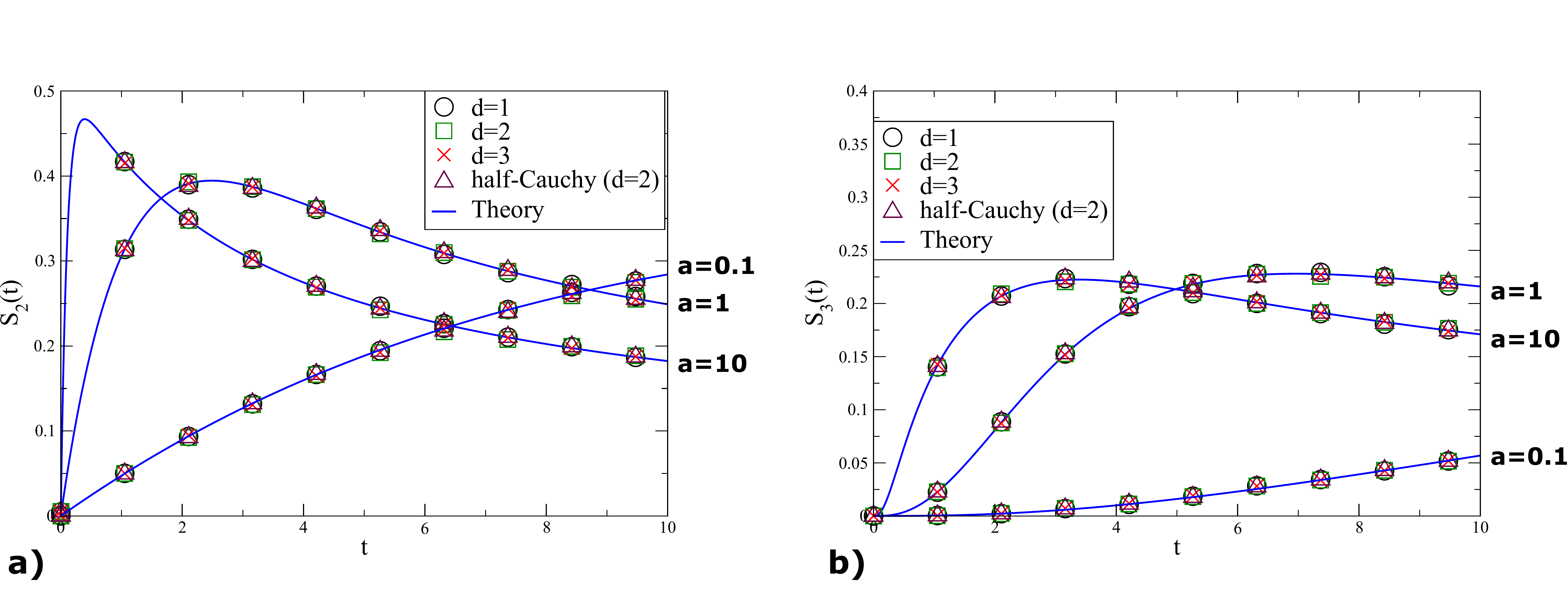}
\caption{Plot of $S_2(t)$ in (a) and $S_3(t)$ in (b) in the case of the mixed model as functions of $t$ for $\gamma=1$ and for different values of $a=0.1,1,10$. The continuous blue lines correspond to numerical integration of the exact result for $S_2(t)$ (a) and $S_3(t)$ (b), given in Eqs. (\ref{eq:S2_mixed}) and (\ref{eq:S3_mixed}). The symbols correspond to simulations with the choices $d = 1, 2, 3$ with $W(v)=\delta(v-1)$ and $d = 2$ with $W(v) =2/(\pi\left(1+v^2\right))$, for $v > 0$
(half-Cauchy). For each value of $a$, the numerical curves collapse on the corresponding analytical blue line for all $t$.}\label{fig:s2_s3_mixed}
\end{figure*}

In this section, we study the record statistics of an RTP with non-instantaneous tumblings, extending the results of Section \ref{sec:record} to the mixed model. In particular, we want to study the statistical properties of lower records of the $x$ component. The joint distribution of the displacements $x_1,\ldots x_m$ in the $x$ direction and of the number $m$ of running phases has been computed in Eq. (\ref{eq:LT_2}) is given by 
\begin{eqnarray}\label{eq:joint_x_m_record}
&&P(\{x_i\},m|t)=\int \frac{ds}{2\pi i}\,e^{st}\,\left(\frac{1}{\gamma}+\frac{1}{s}\left(1-\tilde{P}_W(s)\right)\right)\nonumber\\&\times &\left(\tilde{P}_W(s)\frac{\gamma}{\gamma+s}\right)^m\prod_{i=1}^{m}\,\tilde{p}_s(x_i)
\,,
\end{eqnarray}
where $\tilde{p}_s(x_i)$ is given in Eq. (\ref{eq:psv}), in the case of random velocities, and $\tilde{P}_W(s)$ is the Laplace transform of the distribution of the waiting times. Similarly to what we have done in Section \ref{sec:record}, we can apply the well-known results on the record statistics of discrete-time random walks to the one-dimensional RW 
\begin{equation}
X_k=x_1+x_2+\ldots +x_k\,
\end{equation}
generated by the $x$ component of the particle at the end of each running phase. We recall that the starting point is by definition a record and that $X_k$ is a lower record if $X_i<X_k$ for any $0\leq i <k$. As in Section \ref{sec:record}, we denote by $S_N (t)$ the probability that there are exactly $N$ lower records up to time $t$. Note that when $N=1$ the particle has never visited the negative side of the $x$ axis, thus $S_1(t)=S^{\rm Mixed}(t)$, where $S^{\rm Mixed}(t)$ is given in Eq.~(\ref{eq:S_fin}).

To investigate the case $N\geq 2$, it is useful to recall that the probability $q^N_m$ that an $m$-step RW with continuous and symmetric jumps has exactly $N$ lower records is universal. The generating function of $q^N_m$ with respect to $m$ is thus also universal and it is given in Eq. (\ref{GF_record_nber}) \cite{Ziff_Satya}. Since the distribution $\tilde{p}_s(x)$ is continuous and symmetric, using Eq. (\ref{eq:joint_x_m_record}) one can show that the probability that there are exactly $N$ records up to time $t$ is given by (for $N\geq 2$)
\begin{eqnarray}\label{eq:SNt_mixed_LT}
&&S_N(t)=\int \frac{ds}{2\pi i}e^{st} \left(\frac{1}{\gamma}+\frac{1}{s}\left(1-\tilde{P}_W(s)\right)\right)\nonumber\\&\times &\sum_{m=N-1}^{\infty} \left(\tilde{P}_W(s)\frac{\gamma}{\gamma+s}\right)^m\,q^N_m
\,.
\end{eqnarray}
Using the universal expression for the generating function of $q^N_m$, given in Eq. (\ref{GF_record_nber}), we obtain that, when $N\geq 2$
\begin{eqnarray}\label{eq:S_N}
&&S_N(t)=\int \frac{ds}{2\pi i}e^{st} \left(\frac{1}{\gamma}+\frac{1}{s}\left(1-\tilde{P}_W(s)\right)\right)\nonumber\\&\times &\frac{\left(1-h(s)\right)^{N-1}}{h(s)}
\,,
\end{eqnarray}
where $h(s)$ is given in Eq. (\ref{eq:h}). Note that Eq. (\ref{eq:S_N}) is independent of the dimension $d$ of the system and of the speed distribution $W(v)$. However, Eq. (\ref{eq:S_N}) does depend on the distribution $P_W(T)$ of the waiting times through its Laplace transform $\tilde{P}_W(s)$. Thus, the Laplace transform in Eq. (\ref{eq:S_N}) is hard to invert for a generic distribution $P_W(T)$.

Notably one can also compute the generating function $\tilde{S}(z,t)$ of $S_N(t)$, defined as
\begin{equation}
\tilde{S}(z,t)=\sum_{N=1}^{\infty}S_N(t) z^N\,.
\end{equation}
Using the fact that $S_1(t)=S^{\rm Mixed}(t)$ and using Eq. (\ref{eq:S_N}) when $N\geq 2$, we get
\begin{eqnarray}\label{eq:Szt}
&& \tilde{S}(z,t)=z S^{\rm Mixed}(t)\\
&+&\int \frac{ds}{2\pi i}e^{st} \frac{\left(\frac{1}{\gamma}+\frac{1}{s}\left(1-\tilde{P}_W(s)\right)\right)}{h(s)}\nonumber
 \frac{\left(1 -h(s)\right)z^2}{\left(1-\left(1-h(s)\right)z\right)}\,,
\end{eqnarray}
where $S^{\rm Mixed}(t)$ and $h(s)$ are given in Eqs. (\ref{eq:S_fin}) and (\ref{eq:h}), respectively. From Eq. (\ref{eq:Szt}) one can obtain the average number $\langle N(t) \rangle$ of lower records up to time $t$. Indeed, differentiating Eq. (\ref{eq:Szt}) with respect to $z$ and then setting $z=1$ one obtains
\begin{eqnarray}\label{eq:Nt}
&& \langle N(t)\rangle = S^{\rm Mixed}(t)\\ &&+\int \frac{ds}{2\pi i}e^{st} \left(\frac{1}{\gamma}+\frac{1}{s}\left(1-\tilde{P}_W(s)\right)\right)\nonumber \frac{\tilde{P}_W(s)\frac{\gamma}{\gamma+s}}{h(s)^3}\,,
\end{eqnarray}
where $S^{\rm Mixed}(t)$ and $h(s)$ are given in Eqs. (\ref{eq:S_fin}) and (\ref{eq:h}), respectively.
It is in hard to perform the Laplace inversion in Eq. (\ref{eq:Nt}) for a generic waiting-time distribution $P_W(T)$. Before considering the case of exponential waiting times, it is relevant to ask how the average number of records behaves at late times. It turns out that it depends on the first moment of the waiting-time distribution $P_W(T)$, i.e. on $\langle T \rangle=\int_{0}^{\infty}dT \, T P_W(T)$. Indeed, when $\langle T\rangle$ is finite, one can expand $\tilde{P}_W(s)$ for small $s$ as
\begin{equation}
\tilde{P}_W(s)\simeq 1-\langle T\rangle s\,,
\end{equation}
and from Eq. (\ref{eq:Nt}) it is easy to get that when $t\to \infty$
\begin{equation}
\langle N(t)\rangle \sim \sqrt{t}\,.
\end{equation}
On the other hand, when $\langle T \rangle$ is diverging, i.e. when $P_W(T)\sim 1/T^{\mu +1}$ for $0<\mu<1$, it is possible to show that for $s\to 0$
\begin{equation}
\tilde{P}_W(s)\simeq 1-(bs)^{\mu}\,,
\end{equation}
where $b$ denotes a microscopic time scale. Thus, from Eq. (\ref{eq:Nt}) one can obtain that for late times
\begin{equation}
\langle N(t)\rangle\sim t^{\mu/2}\,.
\end{equation}

\vspace*{0.3cm}

\noindent{\bf Exponential waiting times}\\

We now assume that the waiting times are distributed according to the exponential distribution $P_W(T)=a \, e^{-a T}$ with waiting rate $a$. For this choice of the waiting-time distribution one can find an exact expression for $\langle N(t)\rangle$ and for $S_N(t)$ with $N=2,3$. Indeed, using $\tilde{P}_W(s)=a/(a+s)$, Eq. (\ref{eq:S_N}) becomes
\begin{eqnarray}\label{eq:S_N_exp}
S_N(t)&=&\frac{1}{\gamma}\int \frac{ds}{2\pi i}e^{st} \sqrt{\frac{(a+\gamma+s)(\gamma+s)}{s(a+s)}}\\
&\times &\left(1-\sqrt{\frac{s(a+\gamma+s)}{(a+s)(\gamma+s)}}\right)^{N-1}\nonumber
\,.
\end{eqnarray} 
For $N=2$, we obtain
\begin{eqnarray}\label{eq:S_2_exp}
S_2(t)&=&\frac{1}{\gamma}\int \frac{ds}{2\pi i}e^{st} \sqrt{\frac{(a+\gamma+s)(\gamma+s)}{s(a+s)}}\\
&\times &\left(1-\sqrt{\frac{s(a+\gamma+s)}{(a+s)(\gamma+s)}}\right)\,.\nonumber
\end{eqnarray}
Comparing this expression in Eq. (\ref{eq:S_2_exp}) with the expression for $S^{\rm Mixed}(t)$ given in Eq. (\ref{eq:S_mixed_LT}) it is easy to show that
\begin{eqnarray}\label{eq:S2_mixed}
 S_2(t)=S^{\rm Mixed}(t)-e^{-at}\,,
\end{eqnarray}
where the explicit expression of $S^{\rm Mixed}(t)$ is given in Eq. (\ref{eq:S_mixed}).
In the case $N=3$, from Eq. (\ref{eq:S_N_exp}) we obtain that
\begin{eqnarray}
S_3(t)&=&\frac{1}{\gamma}\int \frac{ds}{2\pi i}e^{st} \sqrt{\frac{(a+\gamma+s)(\gamma+s)}{s(a+s)}}\\
&\times &\left(1-\sqrt{\frac{s(a+\gamma+s)}{(a+s)(\gamma+s)}}\right)^{2}\nonumber
\,.
\end{eqnarray}
It is easy to show that $S_3(t)$ can be rewritten as
\begin{equation}
S_3(t)=2 S_2(t)-a\int \frac{ds}{2\pi i}e^{st}\sqrt{\frac{a+\gamma+s}{s(\gamma+s)(a+s)^3}}\,.
\end{equation}
Performing the Laplace inversion and using Eq. (\ref{eq:S2_mixed}), we obtain
\begin{eqnarray}\label{eq:S3_mixed}
&&S_3(t)=2 S^{\rm Mixed}(t)\\
&&-a\,e^{-\gamma t/2}\int_{0}^{t}dt'\,e^{-a(t-t')}I_0\left(\frac{\gamma}{2}(t-t')\right)\nonumber \\
&\times &\left(\left(1+\gamma t'\right)I_0\left(\frac{\gamma}{2}t'\right)+\gamma t' I_1\left(\frac{\gamma}{2}t'\right)\right)-2e^{-at}\,,\nonumber
\end{eqnarray}
where $S^{\rm Mixed}(t)$ is given in Eq. (\ref{eq:S_mixed}). Computing $S_N(t)$ for $N\geq 3$ appears to be challenging. However, one can easily compute the behavior of $S_N(t)$ for short and late times. Indeed, from Eq. (\ref{eq:SNt_mixed_LT}), it is easy to see that when $s$ is large
\begin{equation}
S_N(t)\simeq\int \frac{ds}{2\pi i}e^{st}\frac{(a\gamma)^{N-1}}{s^{2(N-1)}}q^N_{N-1}\,.
\end{equation}
Inverting the Laplace transform and using the fact that $q^N_{N-1} = 2^{-N+1}$, we obtain
\begin{equation}
S_N(t)\simeq \left(\frac{a\gamma}{2}\right)^{N-1} \frac{1}{(2N-3)!}t^{2N-3}\,.
\end{equation}
On the other hand, expanding Eq. (\ref{eq:S_N_exp}) for small $s$ and inverting the Laplace transform, we obtain that, for $t\to \infty$
\begin{equation}
S_N(t)\simeq \sqrt{\frac{1}{a}+\frac{1}{\gamma}}\frac{1}{\sqrt{\pi t}}\,,
\end{equation}
independently of $N$. The functions $S_2(t)$ and $S_3(t)$ are shown, for $\gamma=1$ and for different values of $a$, in Fig. \ref{fig:s2_s3_mixed}, where we also compare them with numerical simulations, finding excellent agreement. Similarly to what observed in Section \ref{sec:record_wait}, for $N\geq 2$ the function $S_N(t)$ assumes its maximal value at a characteristic time $t^*_N$, which can be shown to increase linearly with $N$ for large $N$.

\begin{figure}
\includegraphics[width = \linewidth]{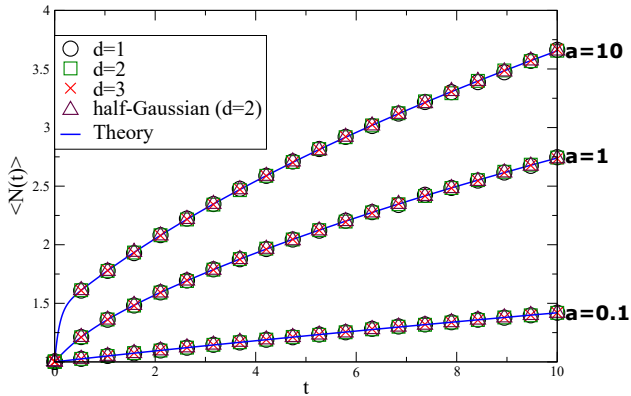}
\caption{Plot of the average number of records $\langle N(t) \rangle$ vs $t$ for the mixed model with $\gamma=1$ and $a=0.1,1,10$ (from bottom to top). The solid blue lines are given by the exact formula in Eq. (\ref{eq:avgN_final}). The symbols correspond to simulations with the choices $d = 1, 2, 3$ with $W(v)=\delta(v-1)$ and $d = 2$ with $W(v) =2/(\pi\left(1+v^2\right))$, for $v > 0$
(half-Cauchy).}\label{fig_av_record_mixed}
\end{figure}

Finally, let us consider the average number $\langle N(t)\rangle$ of records. Plugging $\tilde{P}_W(s)=\frac{a}{a+s}$ into Eq. (\ref{eq:Nt}), we obtain, after few steps of algebra,
\begin{equation}\label{eq:avgN_exp_LT}
\langle N(t)\rangle=S^{\rm Mixed}(t)+\int \frac{ds}{2\pi i}e^{st}\frac{a}{s}\sqrt{\frac{\gamma+s}{s(a+s)(a+\gamma+s)}}\,,
\end{equation}
where $S^{\rm Mixed}(t)$ is given in Eq. (\ref{eq:S_mixed}) for the case of exponential waiting times.
Performing the Laplace inversion in Eq. (\ref{eq:avgN_exp_LT}), we obtain
\begin{eqnarray}\label{eq:avgN_final}
&&\langle N(t)\rangle=S^{\rm Mixed}(t)\\
&&+a e^{-\gamma t/2}\int_{0}^{t} dt'\,e^{-a(t-t')}I_0\left(\frac{\gamma}{2}(t-t')\right) \nonumber\\
&\times &\left((1+\gamma t')I_0\left(\frac{\gamma}{2}t'\right)+\gamma t'I_1\left(\frac{\gamma}{2}t'\right)\right)\,.\nonumber
\end{eqnarray}
This result in Eq. (\ref{eq:avgN_final}) is plotted in Fig. \ref{fig_av_record_mixed} for $\gamma=1$ and $a=0.1,1,10$, where we also show the results of numerical simulations. The exact formula in Eq. (\ref{eq:avgN_final}) is in excellent agreement with simulations.

\section{Conclusion}\label{sec:conclusion}

To conclude, we have shown that there is a nontrivial mapping between the 
$x$-component of the RTP model (and its variants) in $d$ dimensions and the
discrete-time random walk in one-dimension with continuous and symmetric jump
distribution. Exploiting this mapping and using the Sparre Andersen theorem 
valid for discrete-time random walks, we have shown that several observables associated 
to the $x$-component of the RTP of duration $t$ in $d$ dimensions, such as
(i) the survival probability, (ii) the distribution of the time of the maximum and (iii) the record
statistics become universal at all time $t$, i.e. independent of the dimension $d$, as well as the speed
distribution $W(v)$ after each tumbling. Furthermore, we have shown that this universal 
behavior can be extended to two other variants of the basic RTP model.

It is relevant to note that our results are also valid  for an even larger class of RTP models. Indeed, focusing for simplicity on the IT model with exponentially distributed waiting times, we consider the joint probability $P(\{x_i\}|\{\tau_i\})$ of the $x$-component displacements $\{x_i\}=\{x_1,\ldots x_n\}$ conditioned on the running times $\{\tau_i\}=\{\tau_1,\ldots \tau_n\}$. If the joint probability factorizes as $P(\{x_i\}|\{\tau_i\})=\prod_{i=1}^{n}p(x_i|\tau_i)$, where the function $p(x|\tau)$ is symmetric in $x$ and is the same from run to run, then
it is easy to see that the expression for the joint probability $P(\{x_i\},n|t)$ in Eq.~(\ref{Pxn.1}) is still valid with 
\begin{equation}
\tilde{p}_s(x)=(\gamma+s)\int_{0}^{\infty}d\tau\,p(x|\tau)e^{-(\gamma+s)\tau}\,.
\end{equation}
Thus, if $p(x|\tau)$ is continuous in $x$ and symmetric around $x=0$, $\tilde{p}_s(x)$ is also continuous and symmetric and all the universal results presented in Section \ref{sec:IT} are still valid. For instance, let us consider a RTP which evolves in two dimensions according to the IT model but with an additional space-dependent force along the $y$ direction. In this case, it is easy to show that the distribution $p(x|\tau)$ is symmetric and continuous. Thus, the properties computed in Section \ref{sec:IT} turn out to be valid also for this generalized model.

It would be interesting also to investigate if the universality, e.g. the independence on the dimension $d$ extends to other observables of the $x$ component of the RTP. One example is the so-called occupation time, which denotes the fraction of time spent by the $x$-component on the positive side. The distribution of this occupation time in $d=1$ was computed recently using a generalisation of the Feynman-Kac method
in Ref. \cite{singh19}. We have checked numerically that the same result holds for all $d \geq 1$, indeed indicating the universality with respect to the dimension. However, proving this result analytically, using the mapping described in this paper looks challenging and therefore remains as an interesting open problem. 

In this paper, we have also studied the distribution of $t_{\max}$, i.e., the time at which the $x$-component of the RTP reaches its maximum. We have shown this distribution of $t_{\max}$ is also universal, i.e., independent of the dimension $d$ as well as the speed distribution $W(v)$. 
By symmetry, the distribution of $t_{\min}$ (denoting the time at which the minimum is reached) is also universal. One can also ask about the distribution of the time difference $\tau = t_{\min} - t_{\max}$ between the occurrence of the minimum and the one of the maximum. This distribution of $\tau$ was recently computed exactly for the one-dimensional Brownian motion \cite{mori_PRL_tau, mori_PRE_tau} and was found to be nontrivial (as well as related observables, see \cite{schehr10}). 
Therefore, it would be interesting to compute this distribution for the $x$-component of the RTP. 

Other interesting extensions of the present results would be to the case when the RTP is subjected to a constant drift in a certain direction. The problem of a single RTP in the presence of a drift has been studied in $d=1$ in Ref. \cite{GM_2019}. As stated above, if the drift is perpendicular to the $x$ direction, the results computed in this paper remain valid. On the other hand, if the drift involves also the $x$ component, the process can still be mapped to a discrete-time random walk model, though with a noise distribution $f(\eta)$ which is typically non-symmetric. Therefore the universality of the observables based on the standard Sparre Andersen theorem, as detailed in this
paper, will no longer hold. However, some observables of the random walk problem, such as the record statistics, has
been studied in the presence of a drift \cite{Record_review,Wergen, LDW09} and it was shown that, even though the Sparre Andersen universality does not hold, there are still vestiges of universal properties at late times. It would be interesting to investigate whether the record statistics of the $x$-component of the RTP model acquires a similar late time universality. Another way to deviate from the Sparre Andersen universality in the random walk problem is to introduce a walk on a lattice (but with arbitrary big jumps and not necessarily $\pm 1$ jumps).
Several interesting results for the survival probability, the distribution of the time of the maximum and the record statistics for this lattice model in one dimension have been derived recently \cite{mounaix20} and it would be interesting to consider a similar lattice model of RTP in higher dimensions.

\appendix
\section{Derivation of the formula in Eq. (\ref{fdz.1}) for the marginal distribution $P_d(x|l)$} \label{sec:f_d}

We consider a random vector $\vec l$ of fixed magnitude $l$ in $d$ dimensions and compute
the marginal distribution $P_d(x|L)$ of its $x$ component, given fixed $l$. The PDF of a random vector $\vec l$
of fixed magnitude $l$ is simply
\begin{equation}
P(\vec l)= \frac{1}{S_d\, l^{d-1}}\, \delta\left(|\vec l|-l\right)\,  ,
\label{pdf_vecl.A1}
\end{equation}
where 
\begin{equation}
S_d= \frac{2 \pi^{d/2}}{\Gamma(d/2)}\, .
\end{equation}
Note that $S_d$ is just the surface area of a $d$-dimensional sphere of unit radius.
It is convenient to rewrite Eq. (\ref{pdf_vecl.A1}) as
\begin{equation}
P(\vec l)= \frac{2}{S_d\, l^{d-2}}\, \delta\left(|\vec l|^2- l^2\right)\, .
\label{pdf_vecl.A2}
\end{equation}
Let $|\vec l|^2= z_1^2+z_2^2+\ldots z_d^2$ where $z_k$ denotes the component of the vector 
$\vec l$ along the $k$-th direction. Therefore, the marginal distribution $P_d(x|l)$, 
for instance along the $x$ direction,
is obtained by keeping $z_1=x$ fixed while integrating over the other components
\begin{eqnarray}
&& P_d(x|l)\\ & =&  \int P(\vec l)\, \delta(z_1-x)\, dz_1\, dz_2\, \ldots dz_d \nonumber
 = \frac{2}{S_d\, l^{d-2}}\, \\ &\times &
\int \delta\left(z_2^2+z_3^2+\ldots +z_d^2- (l^2-x^2)\right)\, dz_2\, dz_3\, \ldots dz_d\, ,\nonumber
\label{marg_dist.1}
\end{eqnarray}
where we used Eq. (\ref{pdf_vecl.A2}) in going from the first to the second line above.
Let $R^2= z_2^2+z_3^2+\ldots+z_d^2$. Then the $(d-1)$-dimensional integral in Eq. (\ref{marg_dist.1})
can be performed in the radial coordinate
\begin{equation}
P_d(x|L)  = \frac{2\, S_{d-1}}{S_d\, l^{d-2}}\, \int_0^{\infty} \delta\left(R^2- (l^2-x^2)\right)\, R^{d-2}\, dR  
\label{marg_dist.2}
\end{equation}
where we recall $S_{d-1}$ is the surface area of a $(d-1)$-dimensional unit sphere. The single
radial integral in Eq. (\ref{marg_dist.2}) can be trivially done by making a change of variable $R^2=u$
\begin{eqnarray}
P_d(x|l) &=& \frac{S_{d-1}}{S_d\, l^{d-2}}\, 
\int_0^{\infty} \delta\left(u-(l^2-x^2)\right)\, u^{(d-3)/2}\, du \nonumber \\
& = &  \frac{S_{d-1}}{S_d\, l^{d-2}}\, (l^2-x^2)^{(d-3)/2}\, \theta(l-|x|)\, .
\label{marg_dist.3}
\end{eqnarray}
Using the formula for $S_d$ in Eq. (\ref{pdf_vecl.A1}) and rearranging the terms, we get
\begin{equation}
P_d(x|l)= \frac{1}{l}\, f_d\left(\frac{x}{l}\right)\, ,
\label{fdz.A1}
\end{equation}
where
\begin{equation}
f_d(z)= \frac{\Gamma(d/2)}{\sqrt{\pi}\, \Gamma((d-1)/2)}\, (1-z^2)^{(d-3)/2}\, \theta(1-|z|)\,,
\end{equation}
as given in Eq. (\ref{fdz.1}).
One can check easily that $f_d(z)$ is normalized to unity over the support $z\in [-1,1]$.

\section{Distribution of the number $n_1$ of steps to reach the maximum}\label{app:n1}

\begin{figure}[t]
\includegraphics[width = \linewidth]{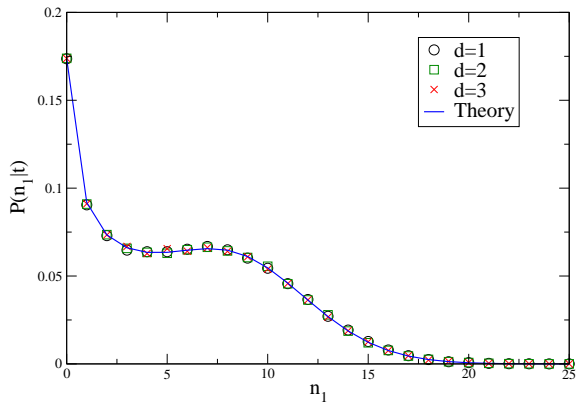} 
\caption{\label{fig:n_1} Probability distribution $P(n_1|t)$ of the number $n_1$ of running phases to reach the global maximum in the IT model, with $\gamma=1$ and $t=10$.
The continuous blue line corresponds to the exact result in Eq. (\ref{eq:P_n1|t}). The symbols correspond to simulations $d=1,2,3$ and with fixed velocity $v_0=1$. }
\end{figure}

In this Appendix, we compute the probability distribution of the number $n_1$ of running phases to reach the global maximum, fixing the total time $t>0$. We will perform the computation for the IT set-up but it is easy to generalize this result to the other models. We will first compute the joint distribution of $n_1$ and of time $t_{\max}$ of the maximum. Then, integrating over $t_{\max}$ we will obtain the marginal distribution for $n_1$. Similarly to the derivation described in \ref{sec:tmax}, we compute $P(t_{\max},n_1|t)$ as the product of the two weights $P_{\rm I}(t_{\max})$ and $P_{\rm II}(t-t_{\max})$, corresponding to the independent intervals $[0,t_{\max}]$ and $[t_{\max},t]$. In Section \ref{sec:tmax}, we have computed $P_{\rm I}(t_{\max})$ summing over $n_1\geq 1$ (see Eq. (\ref{eq_Qm})). Similarly, keeping $n_1$ fixed, one obtains
\begin{equation}\label{eq:P1_app}
P_{\rm I}(t_{\max})=\int \frac{ds}{2\pi i}e^{st_{\max}}\left(\frac{\gamma}{\gamma+s}\right)^{n_1}q_{n_1}\,,
\end{equation}
where $q_{n_1}$ is the survival probability of a symmetric RW of $n_1$ steps, given in Eq. (\ref{eq:SA}). The weight of the second interval $[t_{\max},t]$ can be written as (see Eq. (\ref{eq:Pn1}))
\begin{eqnarray}\label{eq:P2_app}
P_{\rm II}(t-t_{\max})&=&\frac{1}{\gamma}\int \frac{ds}{2\pi i}e^{s(t-t_{\max})}\sum_{n_2=0}^{\infty}\left(\frac{\gamma}{\gamma+s}\right)^{n_2}q_{n_2}\nonumber\\
&=&\frac{1}{\gamma}\int \frac{ds}{2\pi i}e^{s(t-t_{\max})}\sqrt{\frac{\gamma+s}{s}}\,,
\end{eqnarray}
where we have used Eq. (\ref{eq:gen_fun}). Note that, at variance with Eq. (\ref{eq:Pn1}), here we include also the term with $n_2=0$, which corresponds to the event $t_{\max}=t$. Then, the joint probability of $t_{\max}$ and $n_1$ can be written as
\begin{equation}
P(t_{\max},n_1|t)=P_{\rm I}(t_{\max})P_{\rm II}(t-t_{\max})\,,
\end{equation}
where $P_{\rm I}(t_{\max})$ and $P_{\rm II}(t-t_{\max})$ are given in Eqs.  (\ref{eq:P1_app}) and (\ref{eq:P2_app}). Integrating over $t_{\max}$, we obtain
\begin{equation}
P(n_1|t)=\int_{0}^{t}dt_{\max}\,P_{\rm I}(t_{\max})P_{\rm II}(t-t_{\max})\,.
\end{equation}
Taking a Laplace transform with respect to $t$, using the convolution theorem and Eqs. (\ref{eq:P1_app}) and (\ref{eq:P2_app}), we obtain
\begin{equation} \label{Eq_App}
\int_{0}^{\infty}dt\,e^{-st}P(n_1|t)=\gamma^{n_1-1}q_{n_1}\frac{1}{\sqrt{s}(\gamma+s)^{n_1-1/2}}\,.
\end{equation}
Performing the Laplace inversion one obtains that, 
\begin{equation}\label{eq:P_n1|t}
P(n_1|t)=\frac{(\gamma t)^{n_1-1}}{(n_1-1)!}q_{n_1}{}_{1}F_1\left(n_1-\frac{1}{2},n_1,-\gamma t\right)\,,
\end{equation}
where ${}_{1}F_1\left(n_1-\frac{1}{2},n_1,-\gamma t\right)$ is the Kummer's confluent hypergeometric function and $q_{n_1}$ is given in Eq. (\ref{eq:SA}). Note that Eq. (\ref{eq:P_n1|t}) is valid for $n_1\geq 1$. In the case $n_1=0$ the probability $P(n_1|t)$ reduces to the survival probability $S^{\rm IT}(t)$. The result in Eq. (\ref{eq:P_n1|t}) is shown in Fig. \ref{fig:n_1}, where we also compare it with numerical simulations, finding excellent agreement.

In the limit $n_1 \to \infty$, $t \to \infty$ but with the ratio $z = n_1/t$ fixed, we find from \eqref{Eq_App} that $P(n_1|t)$ approaches a scaling form 
\begin{equation}\label{scaling_Pn1}
P(n_1|t)  \to \frac{1}{\gamma t} F\left( \frac{n_1}{\gamma t}\right) \;, \; F(z) = \frac{1}{\pi \sqrt{z(1-z)}} \theta(1-z) \;,
\end{equation} 
where $\theta(x)$ is the Heaviside step function. The cumulative distribution of the scaled variable $n_1/(\gamma t)$ is thus given 
by the arcsine form. This reflects the fact that $\gamma t$ is the expected number of steps for the associated discrete time random walk and $n_1$ is just the number of steps till the maximum of this discrete walk. Indeed it is known \cite{SA_54} that for a discrete time random walk of $n$ steps with symmetric and continuous jump distribution, the cumulative distribution of $n_{\max}$ (the step at which the random walker reaches its maximum) is given by the arcsine law in the limit of $n$ large and $n_{\max}$ but keeping $n_{\max}/n$ fixed.

\end{document}